\documentclass[aps,prd,nofootinbib,superscriptaddress,twoside,twocolumn,floatfix,a4paper,reprint,preprintnumbers,showkeys]{revtex4-2}
\usepackage{xcolor}

\usepackage{epsfig,mathtools,bm,color,xcolor,graphicx,braket,adjustbox,esint,upgreek,dcolumn,psfrag}
\usepackage{amsmath,amssymb,amsfonts}
\usepackage[utf8]{inputenc}
\usepackage[english]{babel}
\usepackage{slashed}
\usepackage{orcidlink} 
\usepackage{multirow} 
\usepackage{soul,cancel}
\usepackage{tabularray}

\allowdisplaybreaks 
\usepackage{hyperref} 
\hypersetup{
colorlinks,
linkcolor={blue},
citecolor={green!70!black},
urlcolor={blue!80!black},
pdftitle={},
}
\usepackage{cleveref}

\newcommand{\beq}{\begin{equation}}
\newcommand{\eeq}[1]{\label{#1}\end{equation}}
\newcommand{\beqa}{\begin{eqnarray}}
\newcommand{\eeqa}{\end{eqnarray}}
\newcommand{\beas}{\begin{eqnarray*}}
\newcommand{\eeas}{\end{eqnarray*}}

\newcommand{\vek}{{\bm k}}
\newcommand{\veq}{{\bm q}}

\newcommand{\cf}{\textit{c.f.}}
\newcommand{\lhc}{\textit{lhc}}
\newcommand{\tbf}[1]{\textbf{#1}}
\newcommand{\tit}[1]{\textit{#1}}
\newcommand{\tgr}[1]{\textcolor{gray}{#1}}

\renewcommand{\arraystretch}{1.2}

\def\vec#1{\boldsymbol{#1}}

\newcommand{\eqn}[1]{Eq. (\ref{#1})}

\graphicspath{{Figs.dir/}}

\begin{document}

\title{Lattice study of $cc\bar u\bar s$ tetraquark channel in $D^{(*)}D^{(*)}_s$ scattering}

\def\UL{Faculty of Mathematics and Physics, University of Ljubljana, Ljubljana, Slovenia}
\def\IJS{Jozef Stefan Institute, Ljubljana, Slovenia}
\def\UR{Institut für Theoretische Physik, Universität Regensburg, 93040 Regensburg, Germany}
\def\IMSc{The Institute of Mathematical Sciences, CIT Campus, Chennai, 600113, India}
\def\HBNI{Homi Bhabha National Institute, Training School Complex, Anushaktinagar, Mumbai 400094, India}
\def\CYP{Department of Physics, University of Cyprus, Nicosia, Cyprus}
\def\CBST{Computation-based Science and Technology Research Center, The Cyprus Institute, Nicosia, Cyprus}

\author{Tanishk Shrimal\orcidlink{0009-0008-9568-0294}}
  \email{tanishks@imsc.res.in}
      \affiliation{\IMSc}
      \affiliation{\HBNI}
    
\author{Sara Collins\orcidlink{0000-0003-0979-7602}}
    \email{sara.collins@ur.de}
    \affiliation{ \UR}

\author{Priyajit Jana}
      \affiliation{\CYP}
      \affiliation{\CBST}
\author{M. Padmanath\orcidlink{0000-0001-6877-7578}}
    \email{padmanath@imsc.res.in}
    \affiliation{\IMSc}
    \affiliation{\HBNI}

\author{Sasa Prelovsek\orcidlink{0000-0002-7496-6188}}
  \email{sasa.prelovsek@ijs.si}
    \affiliation{\UL}
    \affiliation{\IJS}

\begin{abstract}
We present the first lattice QCD determination of coupled $DD_s^*$ and $D^*D_s$ scattering amplitudes in the $J^{P}=1^{+}$ channel and elastic $DD_s$ scattering amplitude in the $J^{P}=0^{+}$ channel. The aim is to investigate whether tetraquarks with flavor $cc\bar u\bar s$ exist in the region near threshold. Lattice QCD ensembles from the CLS consortium with $m_{\pi} \sim 280$ MeV, $a\sim0.09$ fm and $L/a = 24, 32$ are utilized. Finite-volume spectra are determined via variational analysis of two-point correlation matrices, computed using large bases of operators resembling bilocal two-meson structures within the {\it distillation} framework. The scattering matrix for partial wave $l=0$ is determined using lattice eigenenergies from multiple inertial frames following L{\"u}scher’s formalism as well as following the solutions of Lippmann-Schwinger Equation in the finite-volume on a plane-wave basis. We observe small nonzero energy shifts in the simulated spectra from the noninteracting scenario in both the channels studied, which points to  rather weak nontrivial  interactions between the mesons involved. Despite the nonzero energy shifts, the lattice-extracted $S$-wave amplitudes do not carry signatures of any hadron pole features in the physical amplitudes in the energy region near the threshold.
\end{abstract}

\maketitle

\section{Introduction\label{sec:intro}}
Several discoveries in the past two decades challenge the conventional picture of hadrons either as meson-like (quark-antiquark) or baryon-like (three-quark) objects. These newly discovered states, commonly referred to as exotics or XYZTs, have garnered significant theoretical interest in the recent years, thanks to all the experimental efforts across the globe. One particularly intriguing case is that of the $T_{cc}^+(3875)$ tetraquark, recently discovered in proton-proton collisions at the LHCb experiment \cite{LHCb:2021vvq,LHCb:2021auc}. This state was observed as a resonance peak in the mass spectrum of $D^0D^0\pi^+$ mesons, appearing immediately below the $D^{*+}D^0$ mass threshold. With the minimal quark content $cc\bar{u}\bar{d}$ and most likely quantum numbers $I(J^P) = 0(1^+)$, the $T_{cc}^+(3875)$ tetraquark is the longest-lived exotic state observed to this date. This discovery has sparked significant interest in exploring the potentially rich family of exotic hadrons containing two heavy quarks. 

The existence of doubly heavy tetraquarks at varying heavy quark masses have been theoretically proposed/predicted since early 1980s \cite{Ader:1981db,Carlson:1987hh}. Since then, numerous theoretical investigations have also been made over these decades, with a proliferating number of studies following the discovery of $T_{cc}^+$, \cf ~Ref. \cite{Chen:2022asf,Brambilla:2019esw,Brambilla:2022ura} for  detailed reviews. Although there is a general theoretical consensus that the doubly bottom tetraquark system could be stable under QCD interactions in the heavy-quark mass limit \cite{Carlson:1987hh,Manohar:1992nd,Eichten:2017ffp}, the experimental prospects for the discovery of such a deeply bound state in the near future is limited. From different theoretical studies, two strong candidates among the doubly heavy tetraquarks favoring bound state formation correspond to isoscalar axialvector $bb\bar u\bar d$ and isodoublet axialvector $bb\bar u\bar s$ tetraquarks. The discovered $cc\bar u\bar d$  tetraquark is considered to be the charm partner of the isoscalar axialvector $bb\bar u\bar d$ tetraquark, whereas the charm partner $cc\bar u\bar s$ of the isodoublet axialvector $bb\bar u\bar s$ tetraquark still lacks experimental evidence. Considering the experimental advancements in dicharmonium \cite{LHCb:2017iph,LHCb:2021vvq} and dibottomonium production \cite{CMS:2016liw}, the tetraquark with minimal quark content $cc\bar{u}\bar{s}$ and isospin $\frac{1}{2}$ could be a promising candidate for  experimental search in the near future. 

Following the doubly charm baryon discovery in 2017 \cite{LHCb:2017iph}, the theory community has been giving more attention to the studies of doubly heavy tetraquarks. Although majority of studies considered $QQ\bar u\bar d$ with heavy quarks $Q=c,b$, the strange tetraquarks $QQ\bar u\bar s$ have also been addressed by several phenomenological investigations.  These include those based on Heavy Quark Symmetry and mass relations \cite{Eichten:2017ffp,Karliner:2021wju}, relativistic quark model \cite{Lu:2020rog,Deng:2021gnb}, chiral quark soliton model \cite{Praszalowicz:2022sqx}, effective theory for the heavy meson interactions \cite{Gamermann:2006nm,Molina:2010tx,Dai:2021vgf,Gamermann:2007fi}\footnote{Ref. \cite{Gamermann:2007fi} considers axial vectors in hidden charm channels $\bar D^{(*)}D_s^{(*)}$, while authors expect those are related to the doubly charm channel.} and QCD sum rules \cite{Agaev:2022vhq}. Except \cite{Lu:2020rog}, generally predictions for $bb\bar u\bar s$ axialvector tetraquark points to the existence of stable configuration with respect to strong interaction. Phenomenological predictions for scalar $cc\bar u\bar s$ tetraquark are scattered 200 MeV above the $DD_s$ threshold and higher, whereas most predictions for axialvector $cc\bar u\bar s$ tetraquark lie at least 100 MeV above the $DD^*_s$ threshold. Note that despite this qualitative consistency in expectations for $cc\bar u\bar s$ tetraquarks, the mass estimates for isoscalar $cc\bar u\bar d$ tetraquark from these studies are scattered. 

A possible phenomenological reason why a $cc\bar u\bar s$ configuration is less likely to form a bound state than $cc\bar u\bar d$ lies in the relative binding strengths of the associated light antidiquarks. Specifically, the scalar light antidiquark $[\bar u\bar s]$ is expected to be less bound than the ``good" scalar light antidiquark $[\bar u\bar d]$  \cite{Jaffe:2004ph,Francis:2021vrr}. Considering effective theories of meson-meson interactions, the exchange of light mesons $\pi$ and $\rho$ in the system $D^{(*)}D^{(*)}$ might naturally lead to a stronger interaction than the exchange of heavier $K$ and $K^*$ in the system $D^{(*)}D_s^{(*)}$ \cite{Dai:2021vgf}. Phenomenological approaches such as those utilizing effective theories consider only a few of the binding mechanisms which may be important, and a thorough theoretical understanding would preferably rely directly on first principles investigations of QCD.

Lattice QCD offers a nonperturbative, first-principles framework for studying QCD in the hadronic regime, providing results with quantifiable systematic and statistical uncertainties. In recent years, several lattice studies have been performed on isoscalar doubly bottom tetraquarks $bb\bar u\bar d$ \cite{Bicudo:2017szl,Bicudo:2015kna,Francis:2016hui,Junnarkar:2018twb,Leskovec:2019ioa,Mohanta:2020eed,Hudspith:2020tdf,Hudspith:2023loy,Aoki:2023nzp,Alexandrou:2024iwi,Colquhoun:2024jzh,Tripathy:2025vao}, isodoublet doubly bottom tetraquarks $bb\bar u\bar s$  \cite{Francis:2016hui,Junnarkar:2018twb,Hudspith:2020tdf,Meinel:2022lzo,Alexandrou:2024iwi,Colquhoun:2024jzh}, isoscalar bottom-charm tetraquarks $bc\bar u\bar d$ \cite{Meinel:2022lzo,Padmanath:2023rdu,Alexandrou:2023cqg,Radhakrishnan:2024ihu} and isoscalar doubly charm tetraquarks $cc\bar u\bar d$ \cite{Cheung:2017tnt,Junnarkar:2018twb,Padmanath:2022cvl,Chen:2022vpo,Lyu:2023xro,Collins:2024sfi,Whyte:2024ihh}. A detailed review of various lattice investigations can be found in Refs. \cite{Bicudo:2022cqi,Francis:2024fwf}. More recently lattice investigations have been performed with increased rigor in terms of the extraction of the corresponding scattering amplitude from the finite-volume (FV) two-point correlation functions. Although there are some differences in the extraction procedure of amplitudes, a common feature among all these studies is the use of heavier-than-physical pion masses. In the calculations of $T_{cc}$ in $DD^*$ scattering, this leads to a strong interaction stable $D^*$ meson allowing the system to be treated with a two-body $DD^*$ scattering\footnote{A topical aspect in these calculations is the presence of logarithmic branch cuts arising from crossing channel process in the partial-wave projected amplitudes. We give a brief discussion on the relevant logarithmic branch cuts in our study and ignore such effects during the amplitude extraction.}. All lattice simulations of $DD^*$ scattering find significant attraction between $D$ and $D^*$, which leads to a shallow subthreshold pole on the second Riemann sheet, either at real or complex energy. 

Lattice calculations for the $bb\bar u\bar s$ system report the existence of a bound $T_{bb\bar u\bar s}$ tetraquark with $J^P\!=\!1^+$ and a binding energy in the interval [$30$, $100$] MeV with respect to the strong decay threshold $BB_s^*$ \cite{Francis:2016hui,Junnarkar:2018twb,Hudspith:2020tdf,Meinel:2022lzo,Alexandrou:2024iwi,Colquhoun:2024jzh}. The mass of this state $m$ has most often been taken as the finite volume ground state energy ($m=E_1$) in the rest frame $\vec P\!=\!\vec 0$,  which is a valid approach for states well below threshold. The majority of these lattice determinations  did not proceed beyond evaluating the rest frame energy spectrum and/or energy splittings. The heavy-quark mass dependence of the tetraquark $QQ\bar u\bar s$ from a recent lattice simulation \cite{Colquhoun:2024jzh} indicates that the mass of this tetraquark would match the energy of the strong decay threshold at about $m_b/m_Q\simeq 2$ (\footnote{Assuming that the relation $m=E_1(\vec P\!=\!\vec 0)$ still applies near threshold.}), so the bound state would dissolve roughly at and below $m_Q \simeq m_b/2$. This suggests that the $cc\bar u\bar s$  channel would not feature a strongly bound tetraquark.

Given that $cc\bar u\bar s$ tetraquarks, if they exist, are  expected near or above threshold, their existence has to be inferred from poles in the associated scattering amplitudes. The relevant amplitudes have not been previously determined from lattice QCD. The two existing lattice simulations \cite{Junnarkar:2018twb,Cheung:2017tnt} of the channel $cc\bar u\bar s$ extracted only the FV eigenenergies. The small energy shifts observed in the FV eigenenergies from noninteracting energies in Ref. \cite{Cheung:2017tnt} suggest nontrivial interactions in the scalar as well as axialvector channels. In Ref. \cite{Junnarkar:2018twb}, no conclusive signatures for nontrivial interactions were observed in the axialvector channel, with the shift of the ground state energy found to be consistent with zero after the continuum extrapolation.

In this work, we perform the first lattice QCD determination of the elastic $DD_s$ scattering amplitude in the $J^{P}=0^{+}$ channel and the coupled $DD_s^*$ and $D^*D_s$ scattering amplitudes in the $J^{P}=1^{+}$ channel, in search of signatures for potential tetraquarks with flavor content $cc\bar u\bar s$ in the near-threshold region. Variationally determined low energy FV spectra in two lattice QCD ensembles with different spatial volumes and in multiple moving frames for the $cc\bar u\bar s$ channels are utilized to extract the relevant scattering amplitudes. We employ two different prescriptions for the FV spectral quantization: one following the widely utilized L\"uscher's prescription \cite{Luscher:1990ux,Briceno:2014oea} and the second one following the solutions of Lippmann-Schwinger Equation defined in FV and on a plane-wave basis. 

The outline of this article is as follows. In Section \ref{sec:ScatChan}, we list various relevant channels in $S$-wave and their quantum numbers, along with the next few higher partial waves contributions appearing in the infinite-volume and are relevant in FV. A brief discussion on the logarithmic branch cuts from crossing channels in each of the channels studied is given in Section \ref{sec:lhc}. We discuss various technical details associated with this lattice QCD investigation in Section \ref{sec:LatSet}. These include the details of the lattice QCD ensembles used, the correlation measurement setup utilized, the operator basis employed, and the energy fits performed. The results and the discussions on the extracted FV spectra are presented in Section \ref{sec:FiniteVolume}. Our main results on the scattering amplitudes and the underlying hadronic pole information are discussed in Section \ref{sec:scatana}. Finally, we summarize and conclude our findings in Section \ref{sec:conclusions}.

\section{Channels and partial waves}
\label{sec:ScatChan}
There are four low lying two-meson channels that are relevant in a study of tetraquarks with explicit flavor content $cc\bar u\bar s$, that has isospin $(1/2)$ and strangeness $(+1)$. Suppressing the electric charge indices for brevity, they are $DD_s$, $DD_s^*$, $D^*D_s$, and $D^*D_s^*$. Limiting to the $S$-wave interactions within these two-meson channels, there are three sets of total angular momentum-parity $J^P$ quantum numbers that are interesting in the near-threshold regions: 
\begin{itemize}
 \item[(i)] scalar ($J^P=0^+$) in the $DD_s$ system, 
 \item[(ii)] axialvector ($J^P=1^+$) in the coupled  \\ 
 \quad $DD_s^*$-$D^*D_s$-$D^*D_s^*$ system, 
 \item[(iii)] tensor ($J^P=2^+$) in the $D^*D_s^*$ system. 
\end{itemize}
More generally, in Table \ref{tab:pw}, we list the set of $J^P$ (up to $J\le2$), the partial waves (up to $l\le2$) and different total intrinsic spins ($\tilde{s}$) that can lead to different $J^P$ values, including the ones above. 

Since there is only one light quark, there is no symmetry in the flavor sector under interchange of the $D^{(*)}_{(s)}$ mesons. Thus several combinations of $D^{(*)}_{(s)}$ are allowed, although similar combinations were disallowed in the case for the flavor content $cc\bar u\bar d$ (where $\bar s$ antiquark is replaced with $\bar d$ antiquark), that assumed isospin symmetry. In the case (i), the lowest inelastic threshold corresponds to $D^*D_s^*$ channel and opens at a significantly higher energy and hence we ignore this channel in this study. For the case (ii), that have nonzero total intrinsic spin, there could also be dynamical mixing contributions from $l=2 (D)$ partial waves in both $DD_s^*$ and $D^*D_s$ channels, forming a twin-coupled channel $^3S_1-{^3D_1}$ system, expressed in the standard spectroscopic notation $^{2\tilde{s}+1}l_J$. $D^*D_s^*$ scattering poses additional complication in this channel, as it also can contribute via both $S$ and $D$-wave interactions. We address the $S$-wave effects from $D^*D_s^*$ scattering up to the stage of FV spectrum extraction, whereas we ignore all nonzero partial waves components in the $D^*D_s^*$ channels starting from the stage of interpolators used. This approximation is justified as all $D^*D_s^*$-type interpolators with nonzero partial waves require internal meson momenta and hence lie higher in energy. In the case (iii), such dynamical mixing of partial waves is even more involved with the $S$-wave scenario.  At this point we suggest the readers, who are primarily interested in the extracted infinite-volume $S$-wave amplitudes (\{$l$,$\tilde{s}$\} combinations highlighted in bold faced text in Table \ref{tab:pw}), may direct themselves to Section \ref{sec:scatana}. 

\begin{table}[h]
\setlength{\tabcolsep}{4.6pt}
   \begin{tabular}{cccc}
   \hline \hline
   $J^P$ & \{$l$,$\tilde{s}$\}           &  channel            \\\hline\hline
   $0^+$ & \{\tbf{0,0}\}         &  $DD_s$ \\
         & \tgr{\{0,0\}}         &  $D^*D_s^*$ \\
         & \tgr{\{2,2\}}         &  $D^*D_s^*$         \\ \hline
   $1^-$ & \{\tit{1,0}\}         &  $DD_s$             \\
         & \tgr{\{1,[0,1,2]\}}   &  $D^*D_s^*$         \\ \hline 
   $0^-$ & \tit{\{1,1\}}         &  $DD_s^*$; $D^*D_s$ \\
         & \tgr{\{1,1\}}         &  $D^*D_s^*$         \\ \hline
   $1^+$ & \tbf{\{0,1\}}         &  $DD_s^*$; $D^*D_s$; $D^*D_s^*$  \\
         & \{2,1\}               &  $DD_s^*$; $D^*D_s$ \\
         & \tgr{\{2,[1,2]\}}   &  $D^*D_s^*$         \\ \hline
   $2^+$ & \{2,0\}               &  $DD_s$             \\
         & \{2,1\}               &  $DD_s^*$; $D^*D_s$ \\
         & \{0,2\}               &  $D^*D_s^*$         \\ 
         & \tgr{\{2,[1,2]\}} &  $D^*D_s^*$         \\ \hline
   $2^-$ & \{1,1\}               &  $DD_s^*$; $D^*D_s$ \\
         & \tgr{\{1,[1,2]\}}   &  $D^*D_s^*$         \\ \hline\hline
   \end{tabular}
   \caption{Different combinations \{$l$,$\tilde{s}$\} of partial wave $l$ and spin $\tilde s$ relevant for the $J^P$ values up until $l\le2$ and $J\le2$. The \{$l$,$\tilde{s}$\} that are addressed up to the stage of the amplitude analysis are shown in bold faced text, whereas the combinations denoted in gray are ignored even at the level of the correlation function measurements. The $P$-wave combinations that are considered in the amplitude fits, with an aim to filter out their contamination in the extracted $S$-wave amplitudes are identified with italicized text.}     \label{tab:pw}
\end{table}

\section{Meson exchange left-hand cuts}
\label{sec:lhc}

\begin{figure}[h]
\centering
\includegraphics[scale=0.2]{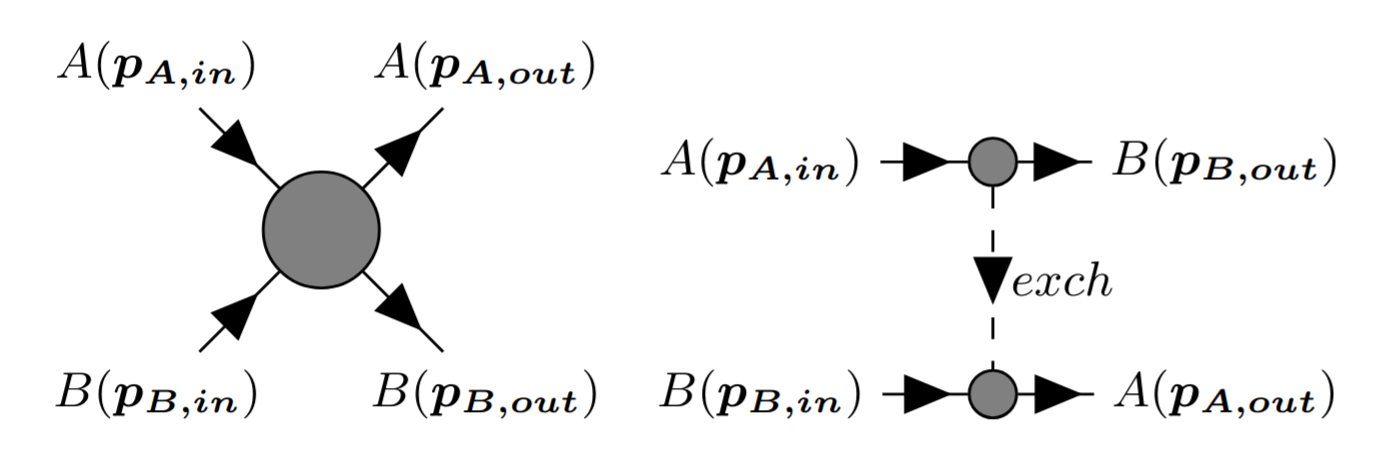}
 \caption{Left: s-channel diagram. Right: u-channel diagram }
   \label{diag:DDs(s-t)}
\end{figure}

Before we move on the details of the calculation, we remark on an important aspect in the study of two-hadron scattering in FV. Recently it was pointed out by several authors that the commonly used $2\rightarrow2$ L\"uscher's prescription to investigate hadron-hadron interactions in FV \cite{Luscher:1990ux,Briceno:2014oea} is not only blind to the three particle effects near and above the lowest three particle threshold, but also to the effects of logarithmic cut arising from a light meson exchange (or one-pion-exchange [OPE]) in the crossed channel, that can have a branch point on the left of the two particle threshold \cite{Du:2023hlu,Raposo:2023oru}. This means the range of validity of the $2\rightarrow2$ L\"uscher's FV scattering prescription is limited to the center-of-momentum energy $E_{cm}$, $E_{lhc}<E_{cm}<E_{ine}$, where $E_{ine}$ is the lowest three particle threshold and $E_{lhc}$ is the left-hand-cut (\lhc) branch point. Such \lhc ~effects could be critical in extracting the infinite volume physics, if the FV eigenenergies used to extract are close to or below the corresponding branch point \cite{Padmanath:2022cvl,Du:2023hlu, Collins:2024sfi}. For example, in our previous study of $DD^*$ scattering at a pion mass $m_\pi\sim 280$~MeV, the \lhc ~branch point due to pion exchange lies right next to the lowest energy levels (see Figs. 3 and 9 in Ref. \cite{Collins:2024sfi}). In this section, we briefly discuss the relevant light mesons that can lead to such \lhc s ~in the channels studied and their associated branch points. We observe that there are no eigenenergies close to the relevant \lhc ~branch points in this study (except for the one associated with $D^*D^*_s$ channel) and hence we ignore any related effects on the extracted amplitudes. 
 
Considering one-channel scattering of particles $A$ and $B$ in Figure \ref{diag:DDs(s-t)}, with $M_A > M_B$, let us derive the three-momentum-squared $p^2_{\lhc}$ of the scattering particles in the center-of-momentum frame at which the particle of mass $M_e$ exchanged in $u$-channel comes on-shell. Following Ref. \cite{Hansen:2024ffk}, we use the relation between Mandelstam variables 
\begin{align}
    s+t+u&=2(M_A^2+M_B^2) \nonumber \\
    s&=(p_{A,in}+p_{B,in})^2,~ \nonumber\\
    t&=(p_{A,in}-p_{A,out})^2,~ \nonumber\\
    u&=(p_{A,in}-p_{B,out})^2,
\end{align} 
The \lhc ~branch point of the partial-wave projected amplitude is given by $s_{lhc}=2M_A^2+2M_B^2-t-u$ with $u\!=\!M_e^2$ and minimal   $t\!\simeq\! 0$\footnote{Minimal $t\simeq -(\vec p_{A,in}\!-\!\vec p_{A,out})^2\!=\!-2p^2(1\!-\! \cos \theta)\!=\!0$ below threshold $p^2<0$ occurs at  $\cos \theta =1$, where the recoil of heavy mesons has been neglected.}:
\begin{align}
     s_{lhc}&=2M_A^2+2M_B^2-M_e^2\nonumber\\
            &=(M_A+M_B)^2+(M_A-M_B)^2-M_e^2
\end{align}
Using the energy-momentum dispersion relation, we extract the branch point in the momentum-squared as
\begin{align}
    (p^{1e}_{lhc})^2&=\frac{(s_{lhc}-(M_A+M_B)^2)(s_{lhc}-(M_A-M_B)^2)}{4s_{lhc}}\nonumber\\
            &=\frac{(M_A-M_B)^2-M_e^2}{4}\left(1-\frac{(M_A-M_B)^2}{s_{lhc}}\right).
\end{align}
Taking $M_A-M_B=\Delta M$ and assuming $\Delta M^2/s_{lhc}$ to be small, as is typically the case in scattering of heavy mesons scattering, the branch point appears below the threshold at.
\begin{align}
    (p^{1e}_{lhc})^2=-\frac{1}{4}[M_e^2-(M_A-M_B)^2].
    \label{psqlhc}
\end{align}
and extends to the left, therefore the name left-hand-cut \cite{Du:2023hlu}.
We utilize this formulae to evaluate the single meson exchange \lhc ~branch point in the respective channels in the next section. 

\section{Technical details  }
\label{sec:LatSet}
Conventionally, hadron spectroscopy using lattice methodology proceeds through the evaluation of two-point correlation functions
\begin{equation}  
\mathcal{C}_{ij}(t) = \sum_{\mathbf{x}}\left<O_i(\vec{P},t)O_j^{\dagger}(\vec{P},0)\right> = \sum_n \frac{Z_i^n Z_j^{n\dagger}}{2E^n} e^{-E^nt}
\label{twoptc}
\end{equation}
on Markov Chain Monte Carlo based importance sampled lattice QCD configurations. In \eqn{twoptc}, $O_i(\vec{P},t)$ are interpolators designed to carry the desired quantum numbers, with $\vec{P}$ being the total spatial momentum in the lab frame. $Z_i^n=\left<O_i|n\right>$ is the operator-state-overlap that carry the quantum numbers, while the exponential factor contains the energy $E^n$ of the state. The FV energy levels are extracted from asymptotic behavior of these two-point functions, which are then utilized to extract the infinite volume physics. In this section, we present the technical details of the configurations employed in this study, the interpolators implemented and  procedure followed in extracting the FV energies. 

\subsection{Lattice setup}\label{sec:setup}
We utilize two ensembles with dynamical $u/d, s$ quarks generated by the Coordinated Lattice Simulations (CLS) consortium \cite{Bruno:2014jqa,Bali:2016umi} with an inverse gauge coupling $\beta = 6/g^2 = 3.4$ (corresponding to a lattice spacing $a =$ 0.08636(98)(40) fm) \cite{Bruno:2016plf} and lattice extents $N_T \times N^3_L = 128\times24^3$ and $96\times32^3$ (referred to as H105 and U101, respectively). For our analysis, we utilize 494 (258) configurations of the large (small) volume ensemble. The dynamical quark fields are realized using a non-perturbatively improved Wilson-clover action, with the degenerate light quark masses ($m_{u/d}$) corresponding to a pion mass $m_{\pi}\sim280$ MeV and the strange quark mass corresponding to a $K$-meson mass $m_{K}\sim 467$ MeV. Note that lighter than physical $m_K$ is a result of the strategy, adopted by the CLS consortium, to approach the physical point along a trajectory of fixed average quark mass $2 m_{u/d} + m_s$. The gauge and the fermion fields fulfill open boundary conditions in the time direction and periodic boundary conditions in the spatial directions.

The valence charm quark is realized using the same relativistic action as for the sea quarks with the charm-quark hopping parameter $\kappa_c = 0.12315$, which leads to a slightly heavier than the physical spin-averaged 1S-charmonium mass $M_{av}$. Of the different $\kappa_c$-values  utilized in our previous investigations \cite{Prelovsek:2020eiw,Padmanath:2022cvl,Collins:2024sfi,Meng:2024kkp} this $\kappa_c$ gives $M_{av}$ closest to the experimental value. Correlator measurements are made employing Laplacian Heaviside smearing for the fermion fields, otherwise referred to as {\it distillation}, see Refs. \cite{Peardon:2009gh,Morningstar:2011ka,Piemonte:2019cbi} for details. We utilize 45 and 75 eigenvectors of the discretized gauge-covariant Laplacian on the small and large spatial volume ensemble, respectively. The sources are placed in the bulk far away from the time boundary such that any boundary effects have sufficiently died out \cite{Piemonte:2019cbi}. The correlation functions are averaged over multiple source timeslices and spin/momentum polarizations to improve the statistical precision. The entire study follows a bootstrap error analysis, in which the statistical uncertainties are extracted from the spread of the central 68\% samples in the bootstrap sample distribution \cite{Prelovsek:2020eiw}. 

The single meson masses and energies are obtained using a basis with two local operators ($\bar q_1\Gamma q_2$) with the following gamma structures ($\Gamma$) for pseudoscalars and vectors:  
\begin{equation}
\label{gammas}
J^P=0^-:\ \gamma_5,~\gamma_5\gamma_t~;\quad J^P=1^-: \gamma_i~,\gamma_i\gamma_t
\end{equation}
We list the lattice estimates for the masses of various relevant mesons in Table \ref{singmesonmases}. The resulting thresholds of the two-meson channels listed in Section \ref{sec:ScatChan} are collected in Table \ref{tmthresholds}. We remark that observables related to hadrons with valence charm quarks could be plagued with non-negligible discretization effects. In Ref. \cite{Piemonte:2019cbi}, we have compared the FV energies of the $D$ meson in nonzero momentum frames with the corresponding expectations from continuum dispersion relations. Small, but statistically significant, differences have been found, indicating non-negligible cut-off effects. We briefly revisit this point later in Section \ref{Efits}. 

\begin{table}[h]
\setlength{\tabcolsep}{4.6pt}
    \begin{tabular}{c|c|c}
    \hline 
    Meson &$N_L=32$ & $N_L=24$ \\     \hline
    $D$     &   $1926(^{+1}_{-1})$  &  $1931(^{+2}_{-2})$  \\
    $D_s$   &   $1980(^{+1}_{-1})$  &  $1979(^{+2}_{-1})$  \\
    $D^*$   &   $2048(^{+2}_{-2})$  &  $2051(^{+4}_{-4})$  \\
    $D_s^*$ &   $2097(^{+1}_{-1})$  &  $2098(^{+2}_{-2})$  \\
    $K$     &    $467(^{+3}_{-3})$  &    $464(^{+3}_{-3})$  \\
    $K^*$   &    $879(^{+11}_{-11})$&  $900(^{+9}_{-9})$    \\         
    \hline
    \end{tabular}
    \caption{Relevant single meson masses $m_H$ on both ensembles for the bare charm mass corresponding to $\kappa_c=0.12315$. The numbers in physical units [MeV] are obtained using the scale $a =$ 0.08636(98)(40) fm \cite{Bruno:2016plf}.}
    \label{singmesonmases}
\end{table}
\begin{table}[h]
    \centering
    \begin{tabular}{cc|ccc|cc}
    \hline
       channel   &&& $N_L\!=\!32$    &&& $N_L\!=\! 24$    \\   \hline
       $DD_s$    &&& $3906(^{+2}_{-2})~$ &&& $3911(^{+4}_{-3})~$ \\
       $DD_s^*$  &&& $4024(^{+2}_{-2})~$ &&& $4030(^{+4}_{-3})~$ \\
       $D^*D_s$  &&& $4027(^{+2}_{-3})~$ &&& $4031(^{+5}_{-5})~$ \\
       $D^*D_s^*$&&& $4145(^{+3}_{-3})~$ &&& $4150(^{+6}_{-5})~$ \\
       \hline
    \end{tabular}
    \caption{Energy of various relevant two-meson thresholds $E_{th}$ on each ensemble in physical units [MeV]. }
    \label{tmthresholds}
\end{table}

\begin{table}[h]
    \centering
    \begin{tabular}{c|c|c|c|c}
    \hline
        channel & $ex$ &  $E_{th}$  &   $E_{lhc}/E_{th}$ & $(p_{lhc}^{1M_{e}}/E_{th})^2$ \\ \hline
        $DD^*$  &$\pi$    &$DD^*$   &0.9979& -0.00100 \\ \hline
        $DD_s$  &$K^*$    &$DD_s$   &0.9789& -0.01044 \\ \hline
        $DD_s^*$&$K$      &$DD_s^*$ &0.9942& -0.00291 \\ \hline
        $D_sD^*$&$K$      &$DD_s^*$ &0.9945& -0.00267 \\ \hline
        $D_s^*D^*$&$K$      &$DD_s^*$ &1.0236& 0.01204 \\ \hline
        $D_s^*D^*$&$K$      &$D^*D_s^*$ &0.9937& -0.00314 \\ \hline
    \end{tabular}
    \caption{The \lhc ~branch points for different channels considered in this work evaluated for the large volume ensemble. The second column ($ex$) identifies the lightest allowed exchange particle, whereas the third column indicates the reference threshold in units of which the \lhc   ~ branch points are presented. The $\pi$-exchange \lhc ~branch point in the $DD^*$ channel is also given for comparison.}
    \label{tab:lhc}
\end{table}

Finally, we list the single-meson exchange \lhc ~branch points associated with different $D^{(*)}D^{(*)}_s$ scattering channels in Table \ref{tab:lhc}. In the case of $DD_s$ elastic scattering, we assign particles $A$ and $B$ in Section \ref{sec:lhc} to the $D_s$ and $D$ meson, respectively, and the lightest allowed exchange particle would be a $K^*$ meson. In the case of $DD_s^*$  scattering, one could make the assignment $A\rightarrow D_s^*$ and $B\rightarrow D$. Similarly, for the $D^*D_s$ channel, one can make the replacement $A\rightarrow D_s$ and $B\rightarrow D^*$. For both channels, the $K$-meson is the lightest allowed exchange particle. We list the \lhc ~branch points in terms of $E_{lhc}/E_{th}$ as well as $(p_{lhc}^{1e}/E_{th})^2$ for all these channels determined using \eqn{psqlhc}. We indicate these branch points in relevant figures with dotted lines. In Table \ref{tab:lhc}, we also present the \lhc ~branch point in the $DD^*$ channel arising from a $\pi$-exchange for comparison.

\subsection{Interpolators}

In this study, our focus is on the first two channels (with $J^P=0^+$ and $1^+$) listed in Section \ref{sec:ScatChan}. Due to the reduced symmetries in FV, it is essential to address the contributions from other continuum quantum channels that could populate the spectrum in relevant FV irreducible representations (irreps) together with the $S$-wave channels of interest \cite{Johnson:1982yq,Thomas:2011rh,Padmanath:2018tuc}. To this end, we utilize two-meson interpolators of type 
 \begin{align}
     O^{D^{(*)}D^{(*)}_s}(\vec{P}) &= \sum_{k,(i),(j)} A_{k,(i),(j)}D_{(i)}^{(*)}(\vec{p}_{1k})D_{s(j)}^{(*)}(\vec{p}_{2k}), \nonumber \\
     \mbox{ with  } ~~\vec{P} &= \vec{p}_{1k}+\vec{p}_{2k},
 \end{align}
following the same projection formulae used in Refs. \cite{Piemonte:2019cbi,Prelovsek:2020eiw,Padmanath:2022cvl}. Operators with $\vec P=\vec 0$ are constructed with partial-wave method  \cite{Prelovsek:2016iyo}, while others  are constructed with projection method. Further details of the operator construction, partial-wave and lattice symmetry group projection can be found in Ref. \cite{Prelovsek:2016iyo}. The individual charmed mesons are realized using quark bilinears $\bar q\Gamma c$ that are projected to a definite momentum following the {\it distillation} framework. The pseudoscalar and  vector quark bilinears are realized with two gamma structures listed in Eq. (\ref{gammas}). With the two gamma structures utilized for single meson components, we adopt the strategy (like in our previous publications \cite{Padmanath:2022cvl,Collins:2024sfi,Meng:2024kkp,Prelovsek:2025vbr}) of constructing two-meson interpolators with or without $\gamma_t$ factors in both the single meson components. In this investigation, we build correlation matrices for several irreps in different inertial frames with total momenta $|\vec{P}|L/2\pi=0,1,\sqrt{2}$ and 2. 
Table \ref{tab:op_elastic} is a compilation of all the irreps and operator information relevant for the study of the scalar channel, whereas Table \ref{tab:op_inelastic} carries the same information for the study of axialvector channel.
\begin{table*}[tbh!]
\setlength{\tabcolsep}{10pt}
    \begin{tabular}{c|c|c|c|c|c|c}
    \hline\hline
        $\vec{P}L/2\pi$ & LG    & $\Lambda^{[P]}$ & $J^P$        &Partial Wave ($l$)& interpolators: $M_1(\vec{p_1})M_2(\vec{p_2})$&$N_{ops}$   \\ \hline
        $(0,0,0)$         &$O_h$  &  $A_1^+$    & $0^+$        & $0$            & $D[000]D_s[000]; D[001]D_s[001]$                 & $2\times2$ \\ \cline{3-7}
                          &       &  $T_1^-$    & $1^-$        & $1$            & $D[001]D_s[001]$                                 & $1\times2$ \\ \cline{3-7}                   
                          &       &  $E^+$      & $2^+$        & $2$            & $D[001]D_s[001]$                                 & $1\times2$ \\ \hline
        $(0,0,1)$         &$Dic_4$&  $A_1$      & $0^+,1^-,2^+$& $0,1,2$        & $D[000]D_s[001];$                             &            \\  
                          &       &             &              &                & $D[001]D_s[000];D[001]D_s[011];$                 & $4\times2$ \\  
                          &       &             &              &                & $D[011]D_s[001]$                                 &            \\ \hline                  
        $(0,1,1)$         &$Dic_2$&  $A_1$      & $0^+,1^-,2^+$& $0,1,2$        & $D[000]D_s[011];$                                & $3\times2$ \\ 
                          &       &             &              &                & $D[011]D_s[000];D[001]D_s[010]$                  &            \\ \cline{3-7}
                          &              &  $B_2$      &$1^{-},2^{+}$ & $1,2$           & $D[001]D_s[010],$                        & $1\times2$ \\  \hline
        $(0,0,2)$         &$Dic_4$&  $A_1$      & $0^+,1^-,2^+$& $0,1,2$        & $D[001]D_s[001]$                                 & $1\times2$ \\ \hline\hline
    \end{tabular}
    \caption{Compiled list of all the irreps and interpolators used in the study of scalar $cc\bar u\bar s$ tetraquarks. LG refers to the spatial lattice symmetry group and $\Lambda^{[P]}$ indicates the FV irrep, and parity, if applicable. $J^P$ is the total angular momentum -parity that can contribute to the FV irreps, and $l$ refers to the two-meson partial wave that can lead to the indicated $J^P$ values. The sixth column lists all the operators used in the respective irreps, while the numbers in the last column indicate the total number of two meson interpolators $N_{ops}$ used; the second factor $2$ accounts for two different choices of gamma structures listed in \eqn{gammas}.}
    \label{tab:op_elastic}
\end{table*}

\begin{table*}[tbh!]
\setlength{\tabcolsep}{10pt}
   \begin{tblr}{c|c|c|c|c|c|c}
   \hline\hline
       $\vec{P}L/2\pi$ & LG    & $\Lambda^P$ & $J^P$        &Partial Wave ($l$)& interpolators: $M_1(\vec{p_1})M_2(\vec{p_2})$  & $N_{ops}$\\
       \hline
       $(0,0,0)$         &$O_h$  &  $T_1^+$    & $1^+$        & $0$            & $D[000]D_s^*[000]; D[001]D_s^*[001]$           &            \\
                         &       &             &              &                & $D^*[000]D_s[000]; D^*[001]D_s[001]$          & $7\times2$ \\
                         &       &             &              &                & $D^*[000]D_s^*[000]$                           &            \\\cline[dashed]{5-6}
                         &       &             &              & $2$            & $D[001]D_s^*[001];D^*[001]D_s[001]$            &            \\\cline{3-7}         
                         &       &  $A_1^-$    & $0^-$        & $1$            & $D[001]D_s^*[001]; D^*[001]D_s[001]$           & $2\times2$ \\\cline{3-7}
                         &       &  $E^-$      & $2^-$        & $1$            & $D[001]D_s^*[001]; D^*[001]D_s[001]$           & $2\times2$ \\\cline{3-7}
                         &       &  $T_2^+$    & $2^+$        & $2$            & $D[001]D_s^*[001]; D^*[001]D_s[001];$          &            \\\cline[dashed]{5-6}
                         &       &             &              & $0$            & $D^*[000]D_s^*[000]$;                          & $4\times2$ \\\cline[dashed]{5-6}
                         &       &             &              & $2$            & $D[011]D_s[011]$                               &            \\ \hline
       $(0,0,1)$         &$Dic_4$&  $A_2$      & $0^-,1^+,2^-$& $0,1,2$        & $D[000]D_s^*[001];D[001]D_s^*[000]$      & $4\times2$ \\
                         &       &             &              &                & $D^*[000]D_s[001];D^*[001]D_s[000]$            &            \\ \hline
      $(0,1,1)$         &$Dic_2$&  $A_2$      & $0^-,1^+,2^-$& $0,1,2$        & $D[000]D_s^*[011];D[011]D_s^*[000]$            &            \\
                         &       &             &              &                & $D[001]D_s^*[010]\times2$                      & $8\times2$ \\
                         &       &             &              &                & $D^*[000]D_s[011];D^*[011]D_s[000]$            &            \\
                        &       &             &              &                & $D^*[001]D_s[010]\times2$                      &            \\ \hline
       $(0,0,2)$         &$Dic_4$&  $A_2$      & $0^-,1^+,2^-$& $0,1,2$        & $D[001]D_s^*[001];D^*[001]D_s[001]$            & $2\times2$ \\
       \hline\hline
   \end{tblr}
   \caption{Similar to Table \ref{tab:op_elastic}, but for the case of axial vector $cc\bar u\bar s$ tetraquarks. The horizontal dashed lines in the tables separate partial wave projected operator sets across scattering channels and provide respective partial wave information. This classification is made only in the rest frame irreps as such a projection is allowed only in the rest frame, where parity is a good quantum number. In the moving frames, we list the operators corresponding to the lowest shell of noninteracting levels in the $DD_s^*$ and $D^*D_s$ channels, where partial wave projection is invalid and multiple partial waves can in principle contribute.}
   \label{tab:op_inelastic}
\end{table*}

We investigate the $DD_s$ correlator data in the $A_1$ irreps in all the momentum frames considered to access the scalar $cc\bar u\bar s$ tetraquarks.  The $DD_s^*$ and $D^*D_s$ channels cannot contribute to the scalar quantum numbers, whereas the $D^*D_s^*$ channel opens significantly higher up in the spectrum (see Table \ref{tmthresholds}). Hence $DD_s$ scattering analysis can be treated within an elastic approximation near the $DD_s$ threshold. Even within this approximation, once we include the moving frame data in the analysis, contributions from $J^P=1^-$ and $2^+$ quantum numbers could arise out of $l=1$ and $l=2$ partial waves in the $DD_s$ scattering (see Table \ref{tab:pw}). To gauge their contamination, we additionally include $T_1^-$, $T_2^+$, and $E^+$ irreps in the rest frame and $B_2$ irrep \footnote{We follow the naming convention for the FV irreps consistent with that used in the {\it TwoHadronsInBox} package made public by the authors of Ref. \cite{Morningstar:2017spu}.} in the moving frame with $|\vec{P}|L/2\pi=\sqrt{2}$. We provide a detailed account of different irreps and operators studied in this regard, the contributing total angular momentum and the related partial wave information in Table \ref{tab:op_elastic}. For the $T_2^+$ irrep, one additionally has two-meson operators of type $DD_s^*$, $D^*D_s$, and $D^*D_s^*$ in the same energy range studied. Hence the information related to the $T_2^+$ irrep is presented in Table \ref{tab:op_inelastic} together with other irreps utilized for $DD_s^*$-$D^*D_s$ inelastic system.

The axialvector channel is interesting from an experimental perspective as the discovered $T_{cc}$ tetraquark favors these quantum numbers and from phenomenology as most of the predictions in the doubly heavy tetraquark systems are with $J^P=1^+$. As considered in the previous works for axial-vector channel in doubly heavy tetraquarks \cite{Whyte:2024ihh,Collins:2024sfi,Alexandrou:2023cqg,Padmanath:2023rdu,Lyu:2023xro,Chen:2022vpo,Meinel:2022lzo,Padmanath:2022cvl}, we utilize the correlation data in the $T_1^+$ irrep in the rest frame and $A_2$ irreps in the moving frame to study about axialvector $cc\bar u\bar s$ tetraquarks. The instances of multiple operators associated with the same noninteracting levels are indicated with ``$\times~2$" in the sixth column of Table \ref{tab:op_inelastic}. Given that the lowest two relevant two-meson scattering channels $DD_s^*$ and $D^*D_s$ in this case are close to each other (see Table \ref{tmthresholds}), one has to treat this case as an inelastic system. The $D^*D_s^*$ threshold lies relatively close to some of the excited levels in the rest frame irreps $T_1^+$ and $T_2^+$, and hence the associated $D^*D_s^*$-like operators can influence the energy estimates for these levels. To address this, we include these relevant low-lying $D^*D_s^*$-like operators into the basis for $T_1^+$ and $T_2^+$ irreps. Similar to the case for scalar channel, there are higher partial wave contributions that appear inevitably in the moving frame irreps considered. The channels $J^P=0^-$ and $2^{\pm}$ arise from $l=1$ and 2 partial waves (see Table \ref{tab:pw}). To constrain these undesired contributions, we additionally include $A_1^-$, $E^-$, and $T_2^+$ rest frame irreps in our study. 

Note that with the flavor content $cc\bar u\bar s$, one cannot have quark bilinear operators of the form quark-antiquark. One can indeed consider diquark-antidiquark type interpolators such as those used in Refs. \cite{Cheung:2017tnt,Junnarkar:2018twb}. We have omitted such interpolators from the basis, as operator  $[c\Gamma c][\bar u C\gamma_5 \bar s]$ ($\Gamma=C\gamma_i$ for axialvector channel and $\Gamma=C\gamma_5$ for scalar channel) was found to have negligible effects on eigenenergies in either of the $cc\bar u\bar s$ channels studied in Ref. \cite{Cheung:2017tnt} and the diquark $[\bar u C\gamma_5 \bar s]$ is known to be less bound than  the good diquark $[\bar u C\gamma_5 \bar d]$ \cite{Francis:2021vrr}.


\subsection{Energy fits\label{Efits}}
\begin{figure*}[t]
    \centering
    \includegraphics[width=\linewidth]{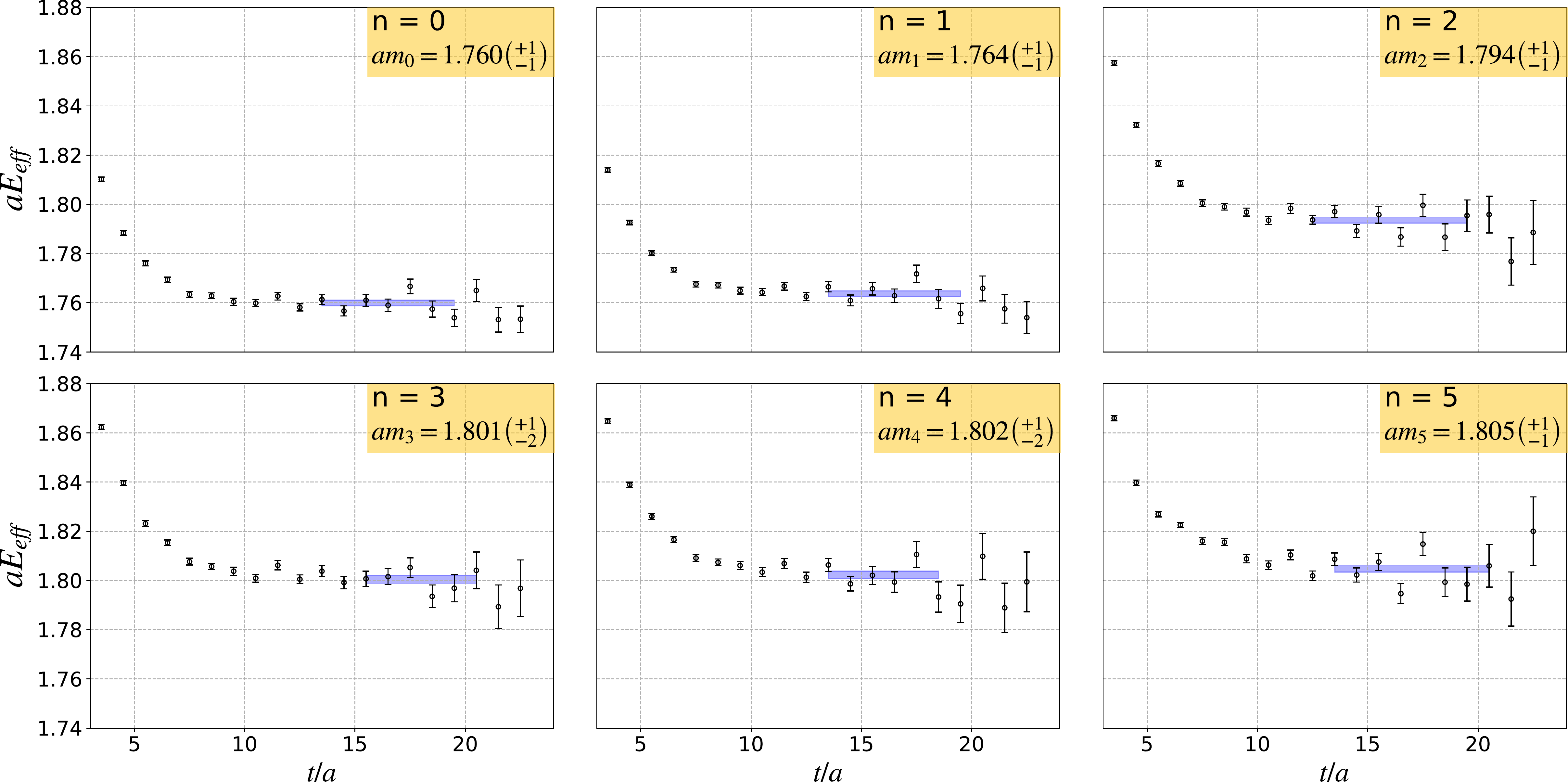}
    \caption{Effective energies as a function of the time interval for the GEVP eigenvalue correlators $\lambda^n(t)$. The plots are presented for the lowest six states in the $T_1^+$ irrep on the large volume ensemble. The bands represent the single-exponential fit estimates for the respective energies and their errors. } 
    \label{fig:effective_energy}
\end{figure*}

Correlation matrices $\mathcal{C}$ (\ref{twoptc}) are evaluated for all the irreps and utilizing all the operators listed in Tables \ref{tab:op_elastic} and \ref{tab:op_inelastic} and are variationally analyzed \cite{Michael:1985ne} following the solutions of generalized eigenvalue problem (GEVP)
\begin{equation}
\mathcal{C}(t)v^{(n)}(t) = \lambda^{(n)}(t) \mathcal{C}(t_0)v^{(n)}(t),
\label{gevp}
\end{equation}
where $\lambda^{(n)}(t)$ and $v^{(n)}(t)$ are eigenvalues and the eigenvector representing the $n^{th}$ eigenlevel with energy $E_n$. $t_0$ is the reference timeslice for the GEVP, which in our case is chosen to be $t_0=4$, in general. The energy extraction proceeds through single-exponential fits to $\lambda^{(n)}(t)$ with their asymptotic form, $\lim_{t\rightarrow\infty}\lambda^{(n)}(t)\sim A_ne^{-E_nt}$. The operator-state overlap factors 
\begin{equation}
Z_i^{n}=\left<O_i|n\right> = \sqrt{2E^n}(V^{-1})_i^n e^{E^{n}t_0/2},
\label{Zfacs}
\end{equation}
are related to the eigenvectors $V=\{v^{(n)}\}$ and provide information on the nature of the eigenlevel extracted. The signal quality and the large time saturation of the ground state signal can be assessed from the effective energy defined as $aE_{eff}^n = ln[\lambda^{(n)}(t)/\lambda^{(n)}(t+1)]$. In Figure \ref{fig:effective_energy}, we present the $aE_{eff}^n$ for the lowest six states in the $T_1^+$ irrep for the large volume ensemble to demonstrate the signal quality. The band represents the final chosen fit estimate and the fitting time interval [$t_{min}$, $t_{max}$].  Considering the plateau observed in $aE_{eff}^n$, we make conservative choices for the fitting time intervals. Our main results for eigenenergies are extracted from single exponential fits to $\lambda^{(n)}(t)$ in these fitting time intervals. We additionally perform similar fits to the ratios $R^n(t)=\lambda^n(t)/\mathcal{C}_{m_1,p_1}(t) \mathcal{C}_{m_2,p_2}(t)$, where, $\mathcal{C}_{m_i,p_i}$ are reference single meson correlation functions with definite momentum $p_i$, which are chosen based on the normalized $\tilde Z_i^{n}$ factors, as practiced widely in the literature. The normalization of the $Z_i^{n}$ factor is such that for any given operator $i$, the largest normalized overlap factor across all the extracted states \{$n$\} is unity. The fits to the ratio correlators $R^n(t)$ directly render the energy difference estimate 
\beq
\Delta E^n = E^n-E_{m_1}(p_1)-E_{m_2}(p_2).
\eeq{eq:ratiofit}

\begin{figure*}[t]
    \centering
    \includegraphics[width=\linewidth]{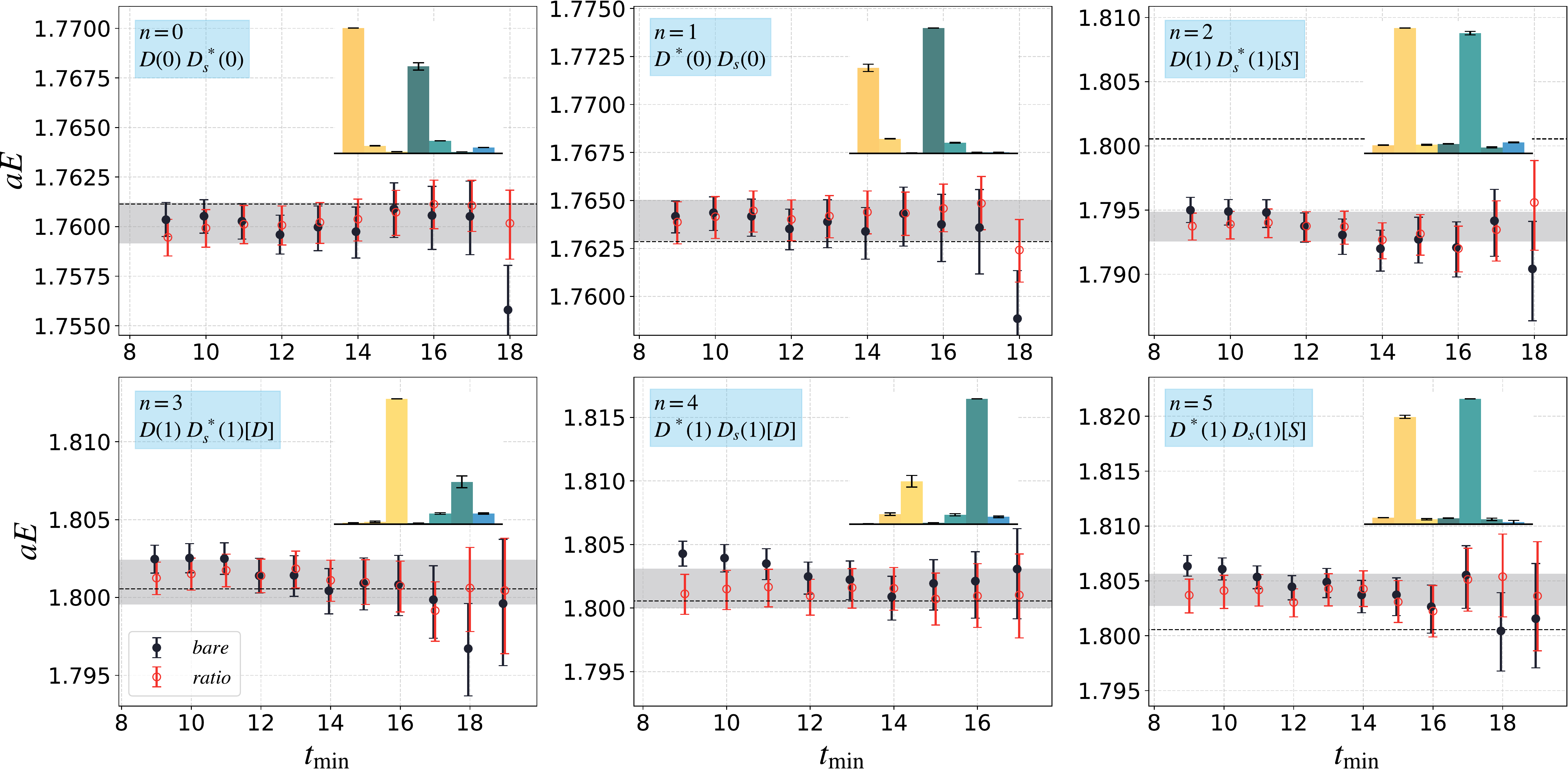}
    \caption{Dependence of fitted energy estimates on the choice of $t_{min}$ for different levels in the $T_1^+$ irrep for the large volume ensemble. The red markers are determined from the ratio of correlators ($R^n(t)$ defined in the text) followed by adding the associated noninteracting level energy (indicated within each block) evaluated using continuum dispersion relation for single mesons. The black markers are based on fits to the $\lambda^{(n)}(t)$. The dashed horizontal black line in each subplot represents the reference noninteracting energy, calculated using the individual meson masses and continuum dispersion relation. The inset figures showcase the normalized operator state overlaps ($\tilde{Z}_n^i$), such that the largest value of $\tilde{Z}_n^i$ for a given operator $i$ across all states $\{n\}$ is unity. The $x$-axis in the inset figures denotes the operator index (only for the cases with single meson components having gamma structures $\gamma_5$ or $\gamma_i$) in the order provided in Table \ref{tab:op_inelastic}.  The red, green and pink colored bars represent operators of type $DD_s^*$, $D^*D_s$, and $D^*D_s^*$. The operator ordering presented in the inset are listed in the legend appended above the figure.} 
    \label{fig:tmin_dependence}
\end{figure*}

The fit quality is influenced primarily by the early time boundary $t_{min}$ of the fitting time interval, whereas $t_{max}$ is less influential to $\chi^2$ cost functions used in the energy fits, owing to the decreased signal-to-noise ratio. To this end, we perform a comparative study of $t_{min}$ dependence of $\Delta E^n$ in \eqn{eq:ratiofit} evaluated from ratios $R^n(t)$ with energy differences that are evaluated from separate fits to the $\lambda^{(n)}(t)$ and single-meson correlators $\mathcal{C}_{m_i}$ for fixed late time boundary $t_{max}$ appropriately chosen for each correlator data. In Figure \ref{fig:tmin_dependence}, we demonstrate the $t_{min}$ dependence of the six lowest levels in the $T_1^+$ irrep for the large volume ensemble, along with the $\tilde Z_i^{n}$ factors in the insets. The final choice of the fitting time interval or equivalently $t_{min}$ is arrived at based on the consistency between these two procedures in the region of ground state plateauing. The bands indicate the final choice of the fit estimates, whereas the reference two-meson noninteracting levels are indicated in the legend. 

The observation of small, but statistically significant deviations of the lattice dispersion relation from that in the continuum for charmed mesons, presented in Ref. \cite{Piemonte:2019cbi}, suggests caution in the FV scattering analysis that generally relies on continuum dispersion relations. To this end, we have adopted the strategy, proposed and described in Section IV.B of Ref. \cite{Piemonte:2019cbi}, for the two-meson spectra extracted in this study. In this approach, the final eigenenergies for the scattering analysis are built from the energy splittings $\Delta E^n$ by adding back the continuum energy $E^{cont}_{i}(p_i)=\sqrt{m_{i}^2+p_i^2}$ of both the mesons as 
\begin{equation}
E^{calc}_n = \Delta E^n + E^{cont}_{1}(p_1) + E^{cont}_{2}(p_2).
\label{Ecalc}
\end{equation}
We observe that a different choice of noninteracting reference level considering other near-degenerate energy levels leads to consistent estimates for $E^{calc}_n$. Similar procedures are also followed by other groups, {\it c.f.} Ref. \cite{Green:2021qol}. For brevity, we drop the superscript ``$calc$" from $E^{calc}_n$ in the following discussions.  

The above discussed procedure can introduce a systematic uncertainty due to the difference in the single meson masses ($m_{i}$) measured on different volumes. Such systematic variation in single meson masses across multiple volumes were also indicated and addressed in Refs. \cite{Wilson:2023anv,Whyte:2024ihh}. Single-meson mass difference can be observed to be statistically significant for the $D$ mesons, whereas other charmed meson masses are consistent within the statistical uncertainties. To account for this systematic, we investigate the variation in the extracted amplitudes when choosing different choices for $m_{m_i}$ used in \eqn{Ecalc}. We consider three scenarios, where the single meson masses used in \eqn{Ecalc} are either chosen from respective ensembles or from the large or small volume ensembles. We also have investigated the variation in extracted amplitudes while using the reference single meson masses in the FV quantization conditions (discussed later on in Section \ref{sec:scatana}) by choosing those either from large or small volume ensembles. We observe the extracted amplitudes are robust to such variation in conventions for single meson masses in our setup. Additionally, we have performed another check by introducing a extra factor 2 in the statistical uncertainties for single meson masses, in which cases the variation in $m_{m_i}$ across different volumes becomes statistically consistent. In all these analysis, we find that the resulting variation in the extracted amplitudes are negligible compared to the respective statistical uncertainties. 

\section{Finite Volume Energy Eigenvalues}
\label{sec:FiniteVolume}
\begin{figure*}[tbh]
    \centering
    \includegraphics[width=\linewidth]{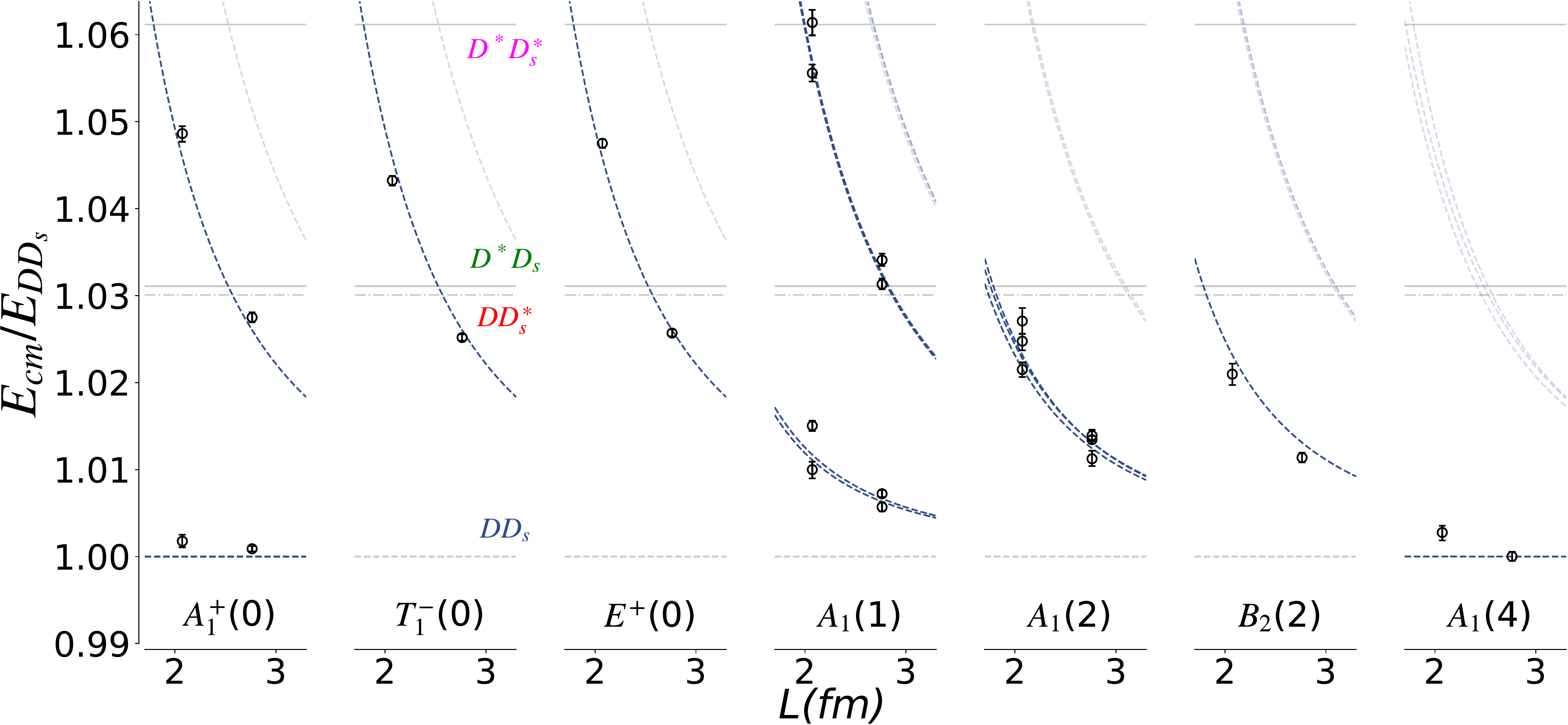}
    \includegraphics[width=\linewidth]{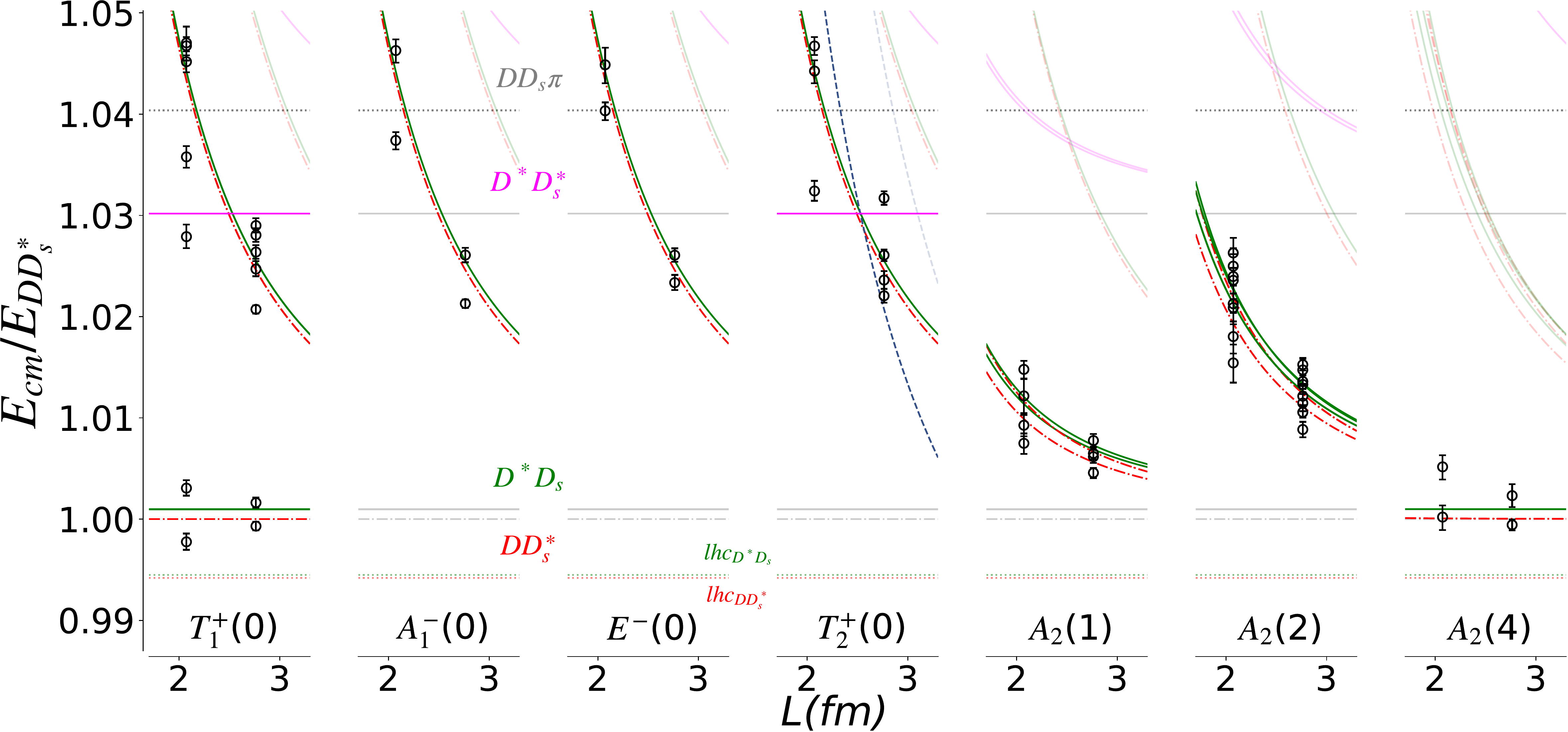}
    \caption{The FV energy eigenvalues for the coupled two-meson system involving channels $DD_s$(blue dashed), $DD_s^*$(red dot-dashed), $D^*D_s$(green solid), $ D^*D_s^*$(purple solid). The $y$-axis represents the energies in the center-of-momentum frame in units of $DD_s$ (top panes) and $DD_s^*$ (bottom panes) threshold for varying spatial extent, of the boxes used, along the x-axis. The panes presented in the top are relevant for the study of scalar channel, whereas the ones in the bottom are relevant for the axialvector channel. Different panes correspond to FV irreps which are also labeled at the bottom with the corresponding $P^2$ value in brackets. The black markers are the lattice-extracted eigenenergies on the two ensembles utilized. The curves represent the energies of relevant low lying non-interacting levels as a function of volume. The two-meson thresholds are presented in gray solid lines, whereas the faded colored curves in the high energy regions within each irreps, correspond to the lowest shell of noninteracting two-meson levels that are ignored in the entire analysis. The lowest three particle threshold $DD_s\pi$ is the gray dotted horizontal line above the $D^*D_s^*$ threshold. } 
    \label{fig:full_spectum}
\end{figure*}

In Figure \ref{fig:full_spectum}, we present the extracted FV eigenenergies in all the irreps and utilizing all the operators listed in Tables \ref{tab:op_elastic} and \ref{tab:op_inelastic}.  The eigenenergies are presented in the center-of-momentum frame $E_{cm}=(E^2-\vec P^2)^{1/2}$ and in units of the energy of the $DD_s$ (top) and $DD_s^*$ (bottom) threshold. The markers represent the simulated eigenenergies, evaluated using \eqn{Ecalc}, whereas the continuous curves are the noninteracting levels in the different relevant two-meson scattering channels labeled within the figure. The noninteracting levels in lattice units are built using continuous dispersion relation $aE=\sqrt{(am_1)^2+(a\vec{p_1})^2}+\sqrt{(am_2)^2+(a\vec{(P-p_1)})^2}$, where the meson masses $am_i$ are taken from the large volume ensemble. A first-hand information on the underlying interactions can be arrived at from the observed deviations of the simulated energy levels from the noninteracting scenario. The first observation is that the rest frame ground state energy splittings in the $A_1^+$ and $T_1^+$ irreps extracted in this study are qualitatively consistent with the observed patterns in the study reported in Ref. \cite{Cheung:2017tnt}. In the rest of this section, we discuss a first impression of the nature of interactions that can be immediately assessed from these energy deviations. 

\ul{\it Scalar channel}: As one can see in the top pane of Figure \ref{fig:full_spectum}, where the FV eigenenergies in irreps relevant for the scalar $cc\bar u\bar s$ tetraquarks are presented, all the relevant low lying noninteracting levels correspond to the $DD_s$ channel. This justifies an elastic approximation in this channel in the low energy regime. The most relevant irrep, in this case, is $A_1^+$, where the simulated energy levels are slightly positively shifted with respect to the nearest noninteracting energies, suggesting a repulsive nature of the interaction between the $D$ and $D_s$ mesons at the length scales corresponding to the momentum involved. This repulsive nature can also be inferred from the positive energy shifts in the energy levels within $A_1(4)$ (where the number inside the bracket is $|\vec{P}^2|$). However, note that in the irreps $T_1^-(0)$ and $B_2(2)$ irreps, where the lowest contributing partial wave is $l=1$, a negative energy shift is evident pointing to the attractive nature of interactions between $D$ and $D_s$ mesons in $P$-wave. Interestingly, in $A_1(1)$ and $A_1(2)$ irreps, where the $l=0$ and 1 partial waves can contribute, the simulated energy levels are distributed symmetrically about the noninteracting curves, suggesting a mixture of contributions from $S$ and $P$ wave scattering. In the $E^+(0)$ irrep, where $l=2$ partial wave can contribute, the simulated levels can be observed to be consistent with noninteracting level. Given this observation, we assume negligible interactions between the $D$ and $D_s$ mesons in the $l=2$ partial wave. Another irrep, where $l=2$ partial wave in $DD_s$ scattering becomes relevant is $T_2^+(0)$ at an energy range about the $D^*D_s^*$ threshold, where we assume that this contribution is going to be negligible and would hardly affect the rest of the spectrum in the relevant energy range.

\ul{\it Axialvector channel}: In this case, there are two two-meson decay channels ($DD_s^*$ and $D^*D_s$) that lie very close to each other in the low energy regime. Thus scattering analysis in this case should involve an inelastic treatment of the amplitude. An interesting observation that is evident is the nature of energy shifts in the lowest two levels in the $T_1^+(0)$ irrep, which is the most relevant irrep in the study of axial-vector channel. While the lowest energy gets a negative energy shift, the first excited state displays a positive energy shift. Such a behavior is apparent in both the ensembles studied, with a larger shift for the small volume ensemble. Such a pattern of energy shift can also be observed in the $A_2(4)$ irrep, with an exception of the ground state in the small volume ensemble. We observe that in both ensembles, the negatively shifted near-threshold level in the $T_1^+(0)$ irrep is dominated by the overlap with $D(0)D_s^*(0)$ operator, despite a non-negligible Fock component of the  $D^*(0)D_s(0)$ operator. Similarly the positive shifted near-threshold level in the $T_1^+(0)$ irrep is dominantly determined by the $D^*(0)D_s(0)$ operators, although it has a sizable overlap with $D(0)D_s^*(0)$ operator. See for reference, the insets of Figure \ref{fig:tmin_dependence}, where we have presented the normalized overlap factors $\tilde Z_i^{n}$ for the six lowest levels in the $T_1^+(0)$ irrep to the operators with the spin structure of single meson components purely build out of $\gamma_{5/i}$. Similar observation can also be made on the negative and positive shifted levels ($n=2$ and 5) around the first excited FV noninteracting $DD_s^*$ and $D^*D_s$ levels. The levels with dominant overlaps with operators projected to the $D$-wave ($n=3$ and 4) can be seen to consistent with the $DD^*_s$ and the $D^*D_s$ noninteracting levels, which is evident from Figure \ref{fig:tmin_dependence}. Leaving aside the exceptional case of the ground state in the $A_2(4)$ irrep and small volume ensemble, we expect that the observed behavior in energy shifts and the overlap factors could arise from two scenarios: either from independent intrinsic $S$-wave interactions within each channel that are either attractive and repulsive in nature or due to coupled channel effects. A third scenario could be a mixture of both. 

\begin{figure}[tbh!]
    \centering
    \includegraphics[width=0.97\linewidth]{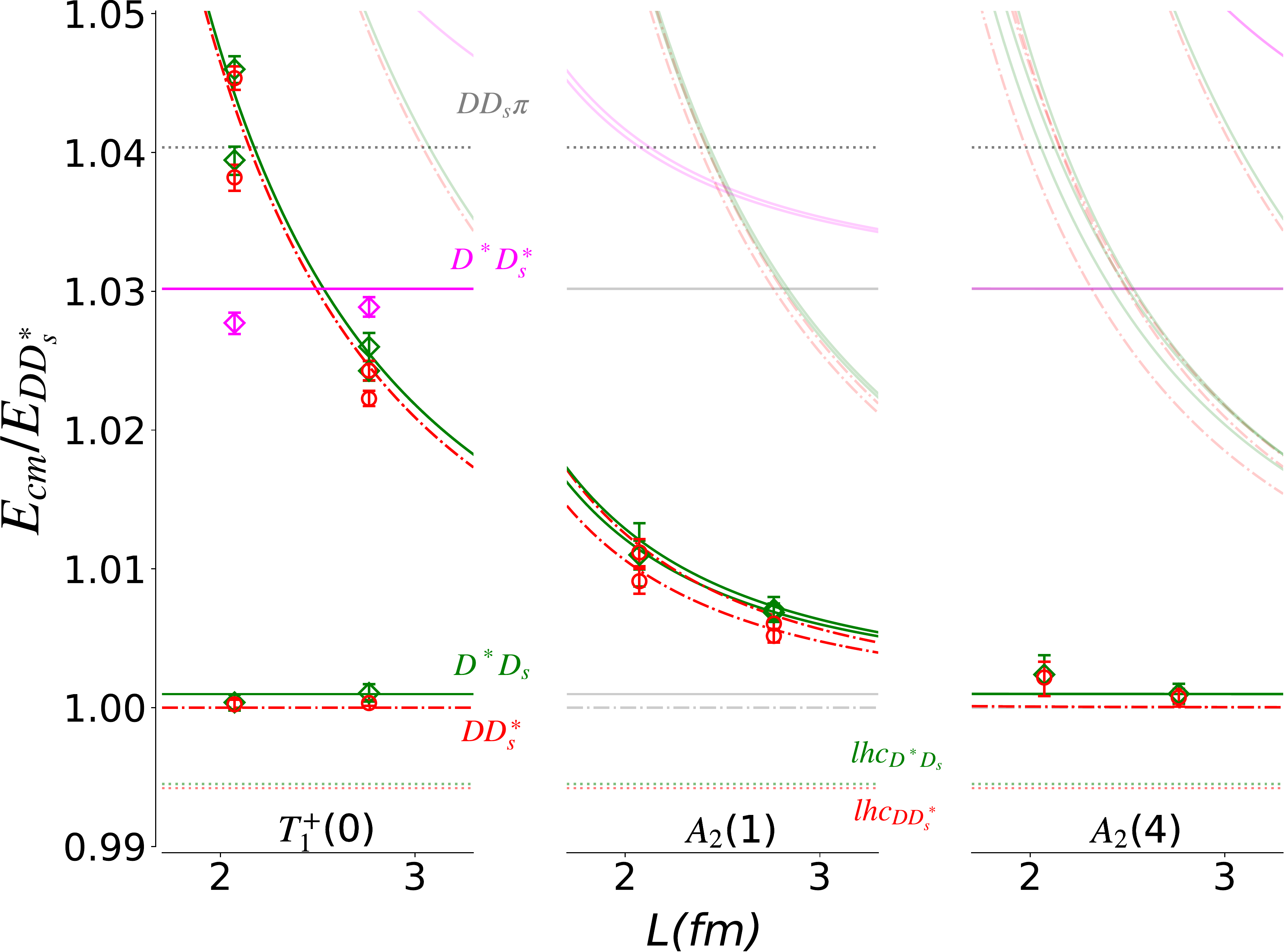}
    \caption{FV eigenenergies in $T_1^+(0)$, $A_2(1)$, and $A_2(4)$ irreps determined with two operator bases, first excluding $D^*D_s$ and second excluding $DD_s^*$-type operators. We plot the energies in units of $DD_s^*$ threshold along the y-axis for different volumes of the box along the x-axis.  The red circles / green diamonds are the lattice energy levels obtained with operator basis omitting $D^*D_s$ / $DD_s^*$ type operators. The magenta diamonds are levels determined by $D^*D_s^*$-like interpolators, and are unaffected by the other operators in the basis used. The curves represent the noninteracting two-meson energy levels as a function of volume.}
    \label{fig:decoupled_spectum}
\end{figure}

To assess the intrinsic interactions within each of the channels ($DD_s^*$ and $D^*D_s$), we follow an elastic approach in each case, where we consider pruned operator basis omitting either $D^*D_s$-type or $DD_s^*$-type interpolators. From the resultant correlation matrices, we extract the respective FV energy eigenvalues in either channel. In Figure \ref{fig:decoupled_spectum}, we present the energy eigenvalues thus extracted in the three most relevant FV irreps ($T_1^+(0)$, $A_2(1)$, and $A_2(4)$) within elastic $DD_s^*$ [red markers] and elastic $D^*D_s$ [green markers]. Clearly, the low energy eigenvalues are consistent with the noninteracting levels, although small deviations can be observed in the levels that are higher up in energies close to the $D^*D^*_s$ threshold. Except for these small deviations at high energies, this consistency of energy levels in individual channels within an elastic approach suggests an intrinsically noninteracting scenario with the individual channels near the thresholds. This implies that the observed energy shifts in the full spectrum presented in Figure \ref{fig:full_spectum} have originated from coupled channel effects. 

\begin{figure}
    \centering
    \includegraphics[width=\linewidth]{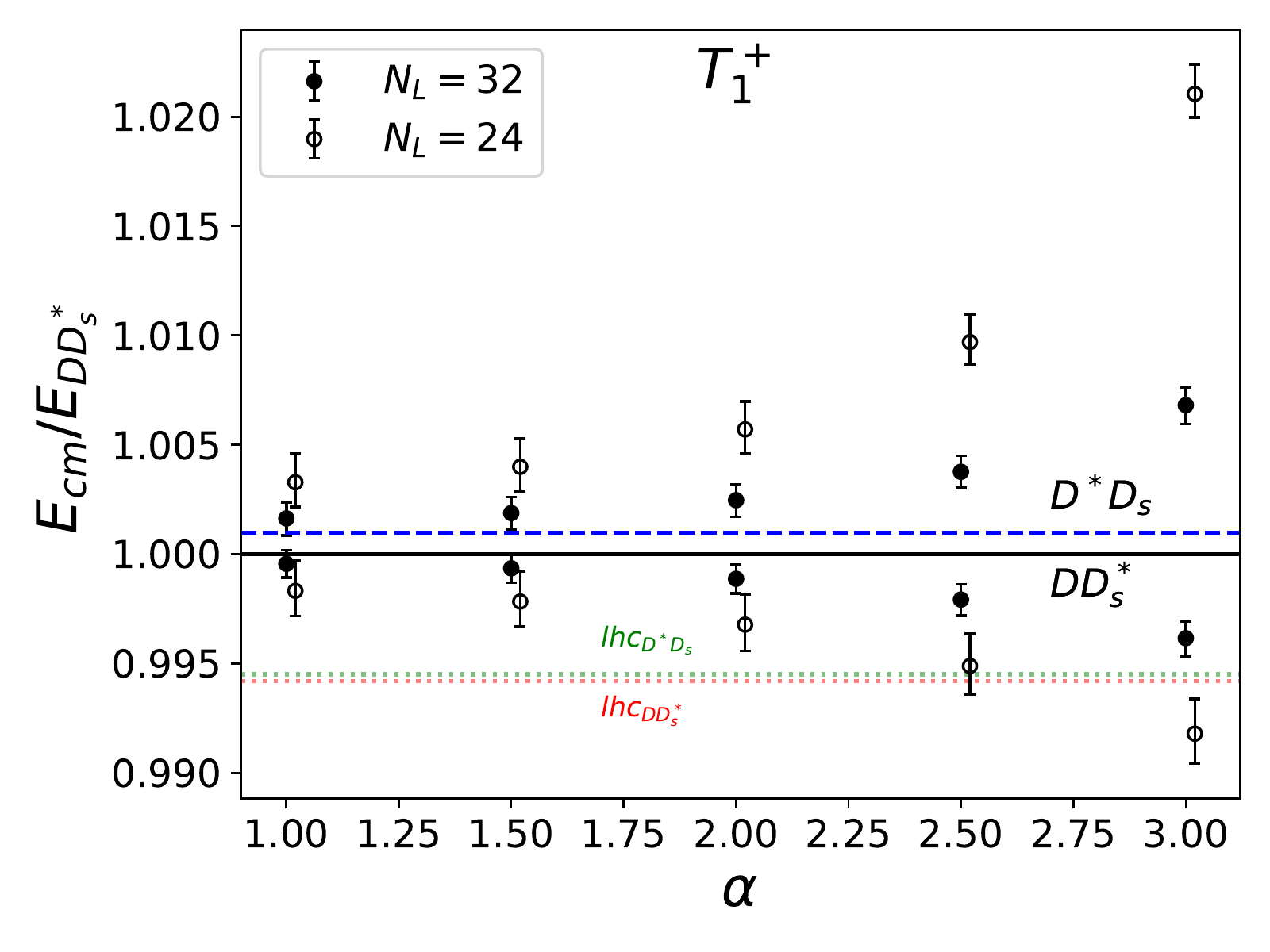}
    \caption{$DD_s^*-D^*D_s$ coupling dependence of lattice eigenenergies in the cm frame. We plot the FV energies in $T_1^+$ irrep after artificially enhancing the cross-correlations between the two channels (see \eqn{eq:couplingenhance}). The y-axis represents the eigenenergies in the cm frame in units of $DD_s^*$ threshold with the coupling factor $\alpha$ on the x-axis. The filled(unfilled) markers are the energy levels for the large(small) volume ensemble. We observe that the splitting behavior of energy levels is indeed a result of cross-correlations between the two channels.}
    \label{fig:coupling_dependence}
\end{figure}

To further assess this inference on coupled channel effects, we make an additional check on the effects of cross-correlations between the channels on the lowest two energy eigenvalues in the $T_1^+(0)$ irrep. Focusing only on the lowest two energy levels, we consider pruned correlation matrices 
\begin{equation}
\tilde{\mathcal{C}}(t)=\left[
\begin{array}{cc}
 \mathcal{C}_{DD_s^*,DD_s^*}(t) & \alpha~\mathcal{C}_{DD_s^*,D^*D_s}(t) \\
 \alpha~\mathcal{C}_{D^*D_s,DD_s^*}(t) & \mathcal{C}_{D^*D_s,D^*D_s}(t)
\end{array}
\right], \label{eq:couplingenhance}
\end{equation}
in the $T_1^+(0)$ irrep, where the operators of type $D^*D_s^*$ are omitted for simplicity. Here we refer to $\alpha$ as the strength of coupling between the channels $DD_s^*$ and $D^*D_s$, with $\alpha=1$ in the real simulated data. We vary this parameter $\alpha$ and assess its effects on the low-lying FV eigenenergies hoping to gain a better understanding of the origin of the observed energy shifts. In Figure \ref{fig:coupling_dependence}, we present the resulting low lying eigenenergies. It is evident from the figure that energy shifts get enhanced with increasing $\alpha$ indicating the role of cross correlations between the channels and thus the coupled channel effects. Note that this enhancement is more rapid in the small volume ensemble. 

Now we briefly address the contributions from the $D^*D^*_s$ to the axialvector quantum channel. In contrast to the scalar case, in this case $D^*D^*_s$ threshold appears relatively closer to the $DD_s^*$ and $D^*D_s$ thresholds. Particularly, $D^*D^*_s$ threshold appears close to the first excitation shell in the $T_1^+$ irrep, which is the most relevant irrep for the analysis of axialvector quantum numbers. Between the two volumes used, the noninteracting level energies of the first excitation shell in the $DD_s^*/D^*D_s$ system and the $D^*D^*_s$ threshold are in opposite hierarchical order as evident from Figure \ref{fig:full_spectum}. The levels dominantly overlapping with the $D^*D^*_s$-like interpolators are also observed to be consistently negatively shifted in both volumes, suggesting an attractive interaction between $D^*$ and $D^*_s$ mesons within an elastic assumption. This feature is mostly unaffected when assuming decoupled channels as evident from the corresponding levels in Figure \ref{fig:decoupled_spectum}. Despite signatures for an attractive nature of $D^*D^*_s$ interactions, we refrain from making any amplitude extraction due to limited lattice eigenenergy levels to constrain the associated amplitude.

Before we move on to the discussion of the scattering analysis and results, we briefly discuss the higher partial wave effects in the axial-vector study. In the $A_1^-(0)$ irrep, where $l=1$ is the lowest contributing partial wave in either $DD_s^*$ and $D^*D_s$ channels leading to $J^P=0^-$, we observe nontrivial negative energy shift in the ground state energy with respect to the lowest relevant noninteracting level, whereas there is only a mild positive shift in the first excitation. Another quantum channel where $l=1$ partial wave can contribute is $J^P=2^-$, which appears in the $E^-(0)$ irrep. Although small energy shifts are observed, they are relatively less significant with respect to those observed in the $A_1^-(0)$ irrep. This observation suggests that it is crucial to gauge the $P$-wave contributions to the moving frame FV irreps, which could influence the $S$-wave amplitude fits. We address this issue further in the section on scattering analysis, to the extent allowed by the FV data we have. 

The next higher partial wave is the $D$-wave with $l=2$, which can lead to $J^P=1^+$ as well as $2^+$ quantum numbers appearing in the $T_1^+(0)$ and $T_2^+(0)$ irreps, respectively. $D$-wave could contribute to $J^P=1^+$, and hence it could naturally lead to a physical partial wave mixing. Like in the previous investigations of similar systems \cite{Padmanath:2022cvl,Chen:2022vpo,Whyte:2024ihh}, we observe the eigenenergies in the $T_1^+(0)$ irrep with dominant overlap to the operators projected to the $D$-wave are consistent with the respective noninteracting energies. This is also supported by phenomenological expectations in $T_{cc}^+$ channel \cite{Du:2021zzh}, which we assume should not be qualitatively different in $T_{cc\bar u\bar s}$ channel studied here. As for the $J^P=2^+$ channel, the energy levels in $T_2^+(0)$ irrep that dominantly overlap with the $DD_s^*$ and $D^*D_s$ interpolators are consistent with the respective noninteracting levels. Thus in the subsequent scattering analysis, we ignore the $D$-wave amplitudes and their contribution to the low-lying FV eigenenergies. 

\ul{\it Tensor channel}: This is the third case presented in Section \ref{sec:ScatChan}, where $D^*D^*_s$ can contribute in $S$-wave leading to $J^P=2^+$, whereas other $D^{(*)}D^{(*)}_s$ channels can contribute in nonzero partial waves. It can be seen that the level dominantly determined by the $DD_s$ operator that appear in the $T_2^+(0)$ irrep is consistent with the associated $DD_s$ noninteracting level, suggesting no nontrivial interactions between the $D$ and $D_s$ mesons in $D$-wave. Similar consistency is also be observed in the levels dominantly overlapping with $DD^*_s$ and $D^*D_s$ channels in $D$-wave, suggesting effectively no interactions. However, significant nonzero energy shifts are observed in the levels with dominant overlap to $D^*D_s^*$-type operators. Note that the $D^*D_s^*$ operators included in the analysis are $S$-wave projected, and the near-threshold $D^*D_s^*$ amplitudes should be dominantly $S$-wave, due to the phase space suppression in non-zero partial waves. This collectively implies that the observed positive energy shifts in the $D^*D_s^*$ levels in the $T_2^+$ irrep are most probably a result of repulsive interactions between the $D^*$ and $D_s^*$ mesons in $S$-wave\footnote{Despite the meson to baryon difference in the scattering channel, a system of two spin $3/2$ $\Omega$ baryons in $S$-wave leading to a total angular momentum of 2 was also observed to show similar positive energy shifts with respect to the respective elastic threshold, possibly suggesting similar repulsive interactions \cite{Dhindsa:2024yrj}.}. We refrain from performing any further analysis or study of this observation, considering the lack of sufficient degrees of freedom to make robust inferences. 

\section{Scattering Analysis}
\label{sec:scatana}
In this section, we present the scattering amplitudes/matrices extracted from the FV spectra. We follow two procedures of the FV quantization: the standard approach that was first proposed in the context of QCD on the lattice by L\"uscher \cite{Luscher:1990ux} and later generalized by several authors, {\it c.f.} Refs. \cite{Briceno:2014oea,Briceno:2017max}. The second approach follows the proposals in Refs. \cite{Meng:2021uhz,Meng:2023bmz} that uses a Lippmann-Schwinger Equation (LSE) defined on FV in a plane wave basis. In subsection \ref{ScatConventions}, we briefly discuss our conventions for scattering physics in the infinite volume and the procedure we follow in extracting them from the FV spectrum. Following this we discuss the main results and the related details of the elastic $DD_s$ system in subsections \ref{elasticDDs} and \ref{More_elasticDDs} and inelastic $DD_s^*$-$D^*D_s$ system in subsection \ref{inelasticDDs}. We conclude this section with brief remarks on our observations and inferences in contrast to the existing phenomenological expectations as well as lattice determinations in Section \ref{discussions}. 

\subsection{Extracting the scattering amplitudes\label{ScatConventions}}
\ul{\textit{Infinite-volume}}- The information on scattering is encoded in the unitary scattering matrix 
\begin{equation}
\label{Smatrix}
S = 1 + i2\tilde{\rho}~T\tilde{\rho} = \frac{1 + i\tilde{\rho}K\tilde{\rho}}{1 - i\tilde{\rho}K\tilde{\rho}}, 
\end{equation}
where $S$, $T$, and $K$ are matrices in the space of relevant open scattering channels ($c$), their orbital angular momentum ($l$), and the total intrinsic spin ($\tilde{s}$), such that the total angular momentum $J\!=\!|\tilde{s}-l|,...,|\tilde{s}+l|$.  $\tilde{\rho}$ is a diagonal matrix with the nonzero diagonal elements given by $\rho_c=\tilde{\rho}^2_c=\frac{2k_c}{E_{cm}}$, where $k_c$ is the three-momentum of the scattering particles within the channel $c$ in the center-of-momentum frame and is related to $E_{cm}$ and the Mandelstam $s=E_{cm}^2$ through the dispersion relation
\begin{equation}
    4sk^2_c=(s-(m_1+m_2)^2)(s-(m_1-m_2)^2),
\end{equation}
where $m_1$ and $m_2$ are the mass of the scattering particles in the channel $c$. For brevity, we omit this subscript $c$ in the single channel case discussions. 

In the elastic scenario, when only a single partial wave contributes, the scattering amplitude can be defined in terms of a single variable as $S_{l}=e^{2i\delta_{l}}$, where the phase shift $\delta_{l}$ depends on energy. In terms of $\delta_{l}$, $T_{l}$ could be expressed as
\begin{align}
T_{l}&= \frac{E_{cm}}{2}\frac{1}{k\cot \delta_{l} - ik}~, \nonumber \\ \mbox{ with }\qquad  K_l^{-1} &= 2k\cot\delta_l/E_{cm} \label{Tmatrix},
\end{align}
where $k=|\vec k|$. In the two scattering channel scenario ($DD_s^*$ and $D^*D_s$), the $S$-matrix could be defined as 
\begin{align}
 \label{Scoupled}
 S_{ij} &= 
 \begin{cases} 
  \eta \, e^{2i\delta_i} & \text{if } i = j, \\
  i\sqrt{1-\eta^2} \, e^{i(\delta_i+\delta_j)} & \text{if } i \neq j,
 \end{cases}
\end{align}
where $i, j$ run over the two channels, and $\eta$ is the associated inelasticity. In terms of $\delta$ and $\eta$, the $T$-matrix is given by:
\begin{align}
 \label{Tcoupled}
 T_{ij} &= 
 \begin{cases} 
  \frac{\eta \, e^{2i\delta_i} - 1}{2i\rho_i} & \text{if } i = j, \\
  \frac{\sqrt{1-\eta^2} \, e^{i(\delta_i+\delta_j)}}{2\tilde{\rho}_i\tilde{\rho}_j} & \text{if } i \neq j.
 \end{cases}
\end{align}

The unitarity constraint on the $S$-matrix implies the $T$-matrix to be a multi-valued function in Mandelstam $s$ with a square-root branch cut opening at each threshold leading to a doubling of complex-valued sheets \cite{osti_4169822,Briceno:2017max}. In the elastic case, the two sheets are identified by the signature of the $Im(\rho)$: conventionally referred to as ``physical" sheet for $Im(\rho)>0$ and ``unphysical" sheet for $Im(\rho)<0$. Above the threshold along real energies, the sheets are connected along the square-root branch cut and moving from above the cut to below the ``physical" sheet is connected to the ``unphysical" sheet. 

In an inelastic system involving $n$ channels, the sheets are identified with an array of $n$ elements carrying values $\pm1$, indicating the signature of $Im(\rho_c)$ with the subscript $c$ referring to channels running from 1,...,$n$. The sheet with all of the values $+1$ is referred to as a ``physical" sheet, whereas all others are generally referred to as ``unphysical" sheets. At any point along the cut above $m$ thresholds, the ``physical" sheet is connected to that ``unphysical" sheet that has the corresponding $m$ values to be $-$1. In the two channel case of $DD_s^*$ and $D^*D_s$, there are four sheets labeled as (++) [or ``physical"], ($-+$), ($--$), and ($+-$). Above the inelastic threshold, ($++$) is connected to ($--$), whereas in between the elastic and inelastic thresholds ($++$) is connected to ($-+$). The value of $T$-matrix elements in various sheets can appropriately be built once the real-valued $K$-matrix elements are constrained from the lattice data. 

Such an evaluation of the $T$-matrix is possible from the $K$-matrix through the relation 
\begin{equation}
\label{Rkn0}
(T^{-1})_{cl\tilde{s};c'l'\tilde{s}'} = (K^{-1})_{cl\tilde{s},c'l'\tilde{s}'} - i \rho_c\delta_{cc'}\delta_{ll'}\delta_{\tilde{s}\tilde{s}'}.
\end{equation}
Here the primed and unprimed indices refer to the outgoing and incoming particle configurations. The unitarity of $S$-matrix imposes $K$-matrix to be hermitian, whereas the time-reversal symmetry of QCD says it should be a real symmetric matrix \cite{osti_4169822,Briceno:2017max}. In the FV amplitude analysis based on L\"uscher's prescription \cite{Luscher:1990ux,Briceno:2014oea}, one may choose a symmetric matrix of real functions (of $s$ or $k^2_c$ for real energies above the scattering threshold) to parametrize the energy dependence of the $K$-matrix that respects unitarity. By extracting the best parameter values for the chosen $K$-matrix parametrization that can reproduce the simulated FV energy spectra, one can extract the $T$-matrix elements adapting the signatures within $Im(\rho)$ to represent the sheet being studied \cite{osti_4169822,Briceno:2017max}. In an alternative procedure \cite{Mai:2017bge,Meng:2021uhz}, one utilizes a parametrized potential that reproduces the FV spectra, and then extracts the  scattering matrix from this potential by solving LSE.  We briefly discuss these two procedures below. 

\ul{\textit{FV analysis \'a la L\"uscher}}- First we briefly discuss the generalized form of the L\"uscher's prescription utilized to extract the set of best parameters $\{a\}$ for a given $K$-matrix functional form, that would reproduce the simulated FV spectra. For the case of elastic scattering of two spinless particles in $S$-wave, this prescription looks like 
\beq
k~cot \delta_0(k) = 2Z_{00}[1;(\frac{kL}{2\pi})^2]/(L\sqrt{\pi}),
\eeq{eLQC} 
where $Z_{00}$ is the L\"uscher's zeta function described in Ref. \cite{Luscher:1990ux} and $L$ is the spatial extent of the cubic box. For $DD_s$ elastic scattering in $S$-wave we could utilize this relation to determine the real part of the inverse amplitude $K_0^{-1}$ in \eqn{Tmatrix} for each FV energy levels available. The energy dependence of the amplitude can then be studied by parameterizing these $K_0^{-1}$ estimates as a real function of $s$ or $k^2$. 

For a general case, involving multiple partial waves and/or multiple scattering channels, one has to consider a more generalized form of the quantization prescription presented in \eqn{eLQC} \cite{Briceno:2014oea}. To this end, we utilize the generalized L\"uscher-based quantization condition given by 
\begin{align}
\label{QCn}
\det[ (\tilde K^{(J)}_{cls;c'l's'}(E_{cm},\{a\}))^{-1}&\delta_{JJ'} \nonumber \\ -\delta_{ss'}\delta_{cc'}&B^{\vec P,\Lambda}_{lJ;l^\prime J^\prime}(E_{cm})]=0
\end{align}
such that the $K$-matrix parametrization with the best parameter values can faithfully reconstruct the simulated FV spectra \cite{Morningstar:2017spu}. Here, $\Lambda$ is the FV irrep and $B^{\vec P,\Lambda}$ is referred to as box matrix that can be expressed in terms of the generalized L\"uscher's zeta functions, which we evaluate using the {\it TwoHadronsInBox} package \cite{Morningstar:2017spu}. The $\tilde K$-matrix is related to the $K$-matrix defined in \eqn{Smatrix} through $K^{(J)} = \tilde{k} ~\tilde{K}^{(J)}~\tilde{k}/\rho$, where $\tilde{k}$ is a diagonal matrix composed of $k_c^{l+1/2}$ as the diagonal elements. We work with the definition of $T$-matrix by the {\it HadSpec} collaboration ({\it c.f.} Ref. \cite{Dudek:2014qha}), whereas the definition $\tilde K$ is according to Ref. \cite{Morningstar:2017spu}.

\begin{figure}[tbh!]
    \centering
    \includegraphics[width=\linewidth]{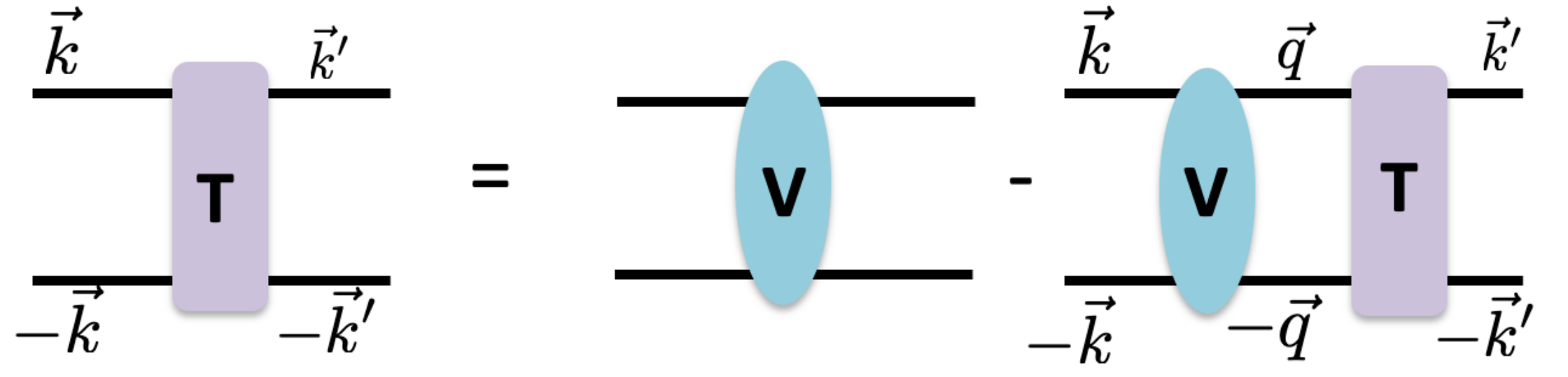}  
    \caption{Diagrammatic representation of the Lippmann Schwinger equation. }
    \label{fig:lse}
\end{figure}

\ul{\textit{FV analysis via LSE}}- Extracting the energy dependence of hadron-hadron scattering amplitude following the solutions of LSE is an alternative \cite{Mai:2017bge,Meng:2021uhz} to the L\"uscher-based approach. The LSE-based procedure also has the advantage that one can incorporate the long-range interactions arising out of OPE discussed in Section \ref{sec:lhc} \cite{Du:2023hlu,Meng:2023bmz}. In this procedure, the associated off-shell transfer matrix $T(\vek,\vek',E)$ is related to the hadron-hadron potential $V(\vek,\vek')$ involved through a transcendental equation 
\beqa
T(\vek,\vek';E) &=& V(\vek,\vek') \label{LSEif}\\ 
&& -\int \frac{d^3q}{(2\pi)^3}V(\vek,\veq)G(\veq;E)T(\veq,\vek';E)\nonumber 
\eeqa 
illustrated diagrammatically in Figure \ref{fig:lse}. Here $\vek$, $\vek'$ and $\veq$ are momenta in cm frame and $G(\veq;E)$ is the energy dependent loop function. The on-shell partial-wave projected amplitude $T_l(k,E)=-T_l(\vek,\vek;E)$~\footnote{The negative sign accounts for the convention difference between the L\"uscher-based and the LSE-based analysis. In the latter, a notion of positive potential ($V>0$) implying repulsive interaction is assumed leading to a relation $S=1-2i\tilde{\rho}~T\tilde{\rho}$, unlike in \eqn{Smatrix}.} is related to the scattering phase shifts and inelasticities as indicated in Eqs. (\ref{Tmatrix}) and (\ref{Tcoupled}). We follow the procedure proposed in Ref. \cite{Meng:2023bmz}, where the LSE presented in \eqn{LSEif} can be recasted into an appropriate FV form by replacing integration over $\veq$ with a summation to yield 
\beq
\mathbb{T} = \mathbb{V} - \mathbb{V}.\mathbb{G}.\mathbb{T}, 
\eeq{LSEfv}
where $\mathbb{G}(\veq_n;+E)=\frac{\mathcal{J}(\veq_n)}{L^3}G(\veq_n;E)\delta_{n,n'}$ and $\mathbb{V}=V(\veq_n,\veq'_n,)$.  $\mathcal{J}(\veq_n)$ is the Jacobian determinant associated with change in coordinates from lab frame to center of momentum frame, which will be trivial in the nonrelativistic limit. Then the FV energy levels for any given potential $V$ can be obtained from the solutions of the determinant equation 
\beq
\mbox{det}[\mathbb{G}^{-1}+\mathbb{V}] = 0,
\eeq{planewaveQC}
as they would represent the poles in $\mathbb{T}=\mathbb{G}^{-1}(\mathbb{G}^{-1}+\mathbb{V})^{-1}\mathbb{V}$ derived from \eqn{LSEfv}. 

All the channels we consider in this study forbid any pion exchange diagrams like in the case of $DD^*$ system \cite{Du:2023hlu}. The allowed lightest meson exchange is either involving a $K$ or a $K^*$, which induce the left-hand cut further below the threshold, \cf ~Section \ref{sec:lhc} for details. Unlike in Refs. \cite{Du:2023hlu,Meng:2023bmz,Collins:2024sfi}, we ignore explicit realization of such a long range potential arising out of $K^{(*)}$-meson exchange diagrams, aiming at extraction of the scattering amplitude in the near-threshold region, where $E>E_{lhc}$. The potential $V(\vek,\vek')$ is approximated with purely contact potentials with terms up to $\mathcal{O}(k^2)$. The contact interactions are regularized with exponential form $e^{\frac{-(k^n+k'^n)}{\Lambda^n}}$. $\Lambda$ is the scale beyond which the contact potential is cut off, and the value of $n$ determines how rapid is this cut off. The parametrized form of the contact potentials utilized in each cases are discussed in the respective subsections below and in the appendices referred to later on. More details of the formalism and our implementation can be found in Refs. \cite{Oller:2019opk,Meng:2021uhz,Meng:2023bmz,Prelovsek:2025vbr}.

\ul{Amplitude fits}- The extraction of best-fit parameter values, in $\tilde{K}$-matrix or $V({\vek,\vek'})$, that can faithfully reconstruct the simulated FV spectra follows a minimization program involving a cost function $\chi^2$ defined as 
\begin{align}
\chi^2(\{a\}) = \sum_{L} dE_{L, i}(\{a\})~\mathbb{C}^{-1}_{i,i'}~dE_{L, i'}(\{a\})~.
\label{chisq}
\end{align}
Here $dE_{L, i}(\{a\}) = E_{cm}(L, i) - E_{cm}^{an.}(L, i; \{a\})$ is the difference between a simulated lattice energy level $E_{cm}(L, i)$ and the analytically calculated energy level $E_{cm}^{an.}(L, i; \{a\})$ that satisfies the quantization conditions given in \eqn{QCn} or \eqn{planewaveQC} for the set of parameter values $\{a\}$ \cite{Briceno:2014oea, Morningstar:2017spu}. $\mathbb{C}$ refers to the data covariance matrix, which we evaluate following the procedure outlined in Appendix A of Ref. \cite{Prelovsek:2020eiw}. 

In the elastic cases, one could alternatively utilize a definition of the $\chi^2$ function in terms of the momentum of scattering particles $k^2$ given by 
\begin{equation}
    \chi^2= \sum_{L}dk^2_{L,i}~\mathbb{C}^{-1}(k)_{ii'}~dk^2_{L,i'}
    \label{kchi2}
\end{equation}
where, $dk^2_{L,i}=k^2(L, i)-k^2(L, i;\{a\})$ is the difference between momentum-squared of the scattering particles at the simulated energy values ($k^2(L, i)$) and at the analytically calculated energy values ($k^2(L, i;\{a\})$). This formulation of the cost function could be particularly convenient in the case of elastic systems \cite{Green:2021qol}. We utilize either of the cost functions defined in Eqs. (\ref{chisq}) and (\ref{kchi2}) in our amplitude determinations.

The energy or momentum solutions of L\"uscher-based quantization conditions in \eqn{QCn} are identified from the zeros in eigenvalues (as a function of $E_{cm}$ for each lattice QCD ensemble, lab frame momenta, and FV irrep) of the matrix
\begin{equation}
\label{Omega}
\tilde A(E_{cm}) = \frac{A}{\det((\mu^2+AA^{\dagger})^{1/2})},
\end{equation}
by following an eigenvalue decomposition program along the lines as discussed in Ref. \cite{Woss:2020cmp}. Here $A$ is chosen to be the argument of determinant in \eqn{QCn} and $\mu=2.0$ is chosen throughout this study. Our results remain statistically unaffected with the variation in the value of $\mu$ across a wide range [0.1, 10.]. The LSE-based quantization condition is also applied following $\chi^2$ functions defined in Eq. (\ref{chisq}), where the energy solutions are identified following the solutions of a reformulated version of the relation in \eqn{planewaveQC}, as discussed in Appendix A1 of Ref. \cite{Meng:2023bmz}.

This problem of determining the correct energy dependence of the amplitude that results in the simulated FV spectra is a famous inverse problem in hadron spectroscopy using lattice techniques \cite{Briceno:2017max}. A typical functional form used in capturing the energy dependence of the amplitude can introduce a model dependent systematic in the extracted infinite volume physics. This model dependent systematic can be ameliorated by considering different functional forms to parametrize the energy dependence of the amplitude and then demonstrate the model (in)dependence in the extracted infinite volume results \cite{Dudek:2014qha,Wilson:2014cna}. We utilize different functional forms to investigate the  variations in our results with our choice of parametrization. The inelastic amplitudes we extract following different parameterizations suggest qualitatively similar conclusions. However, in the elastic sector such a conclusion turned out to be a challenging one, due to lack of constraints on analytic properties of the L\"uscher-based amplitudes. We discuss more on this in the following subsections.


\subsection{Elastic $DD_s$ scattering} \label{elasticDDs}

\begin{figure*}[tbh!]
    \centering
    \includegraphics[width=\linewidth]{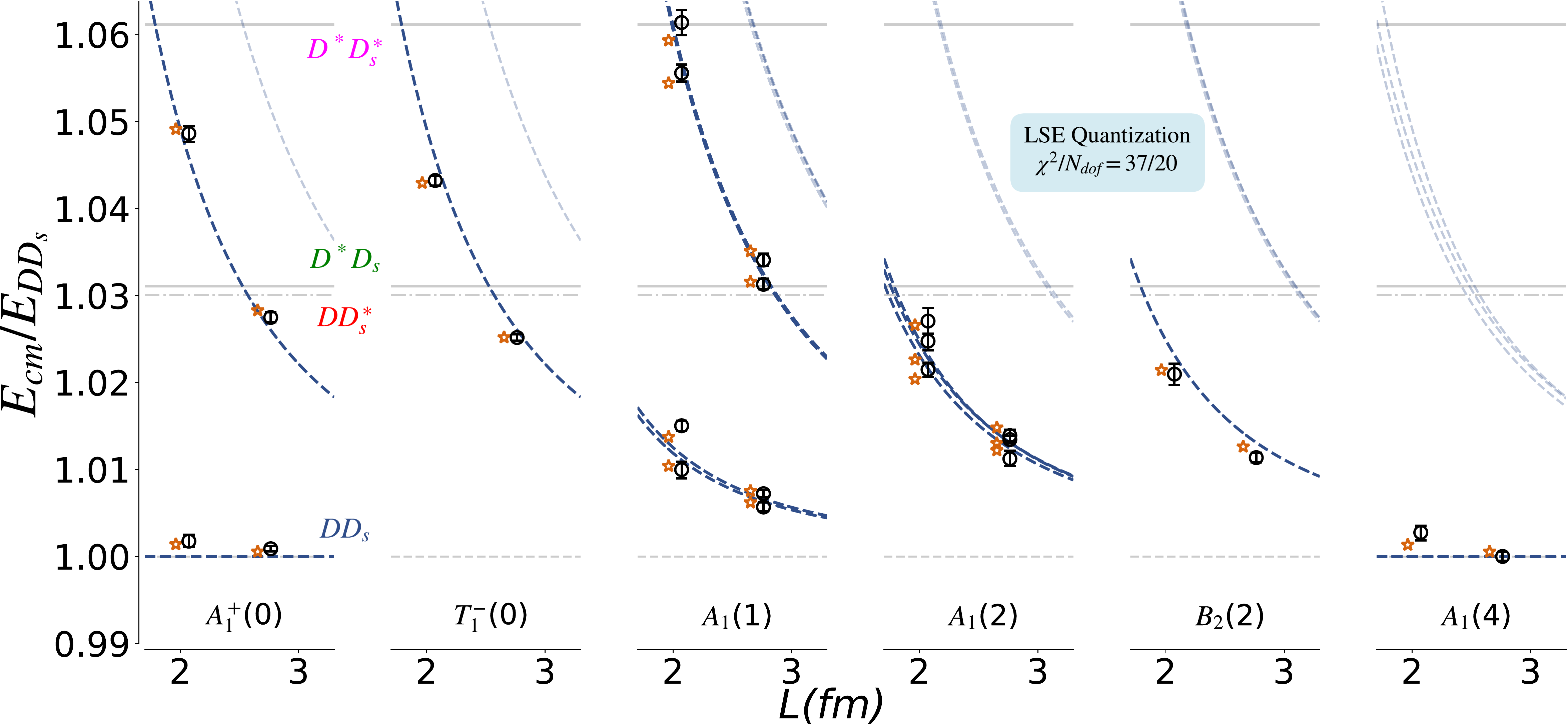} \\
    \includegraphics[width=\linewidth]{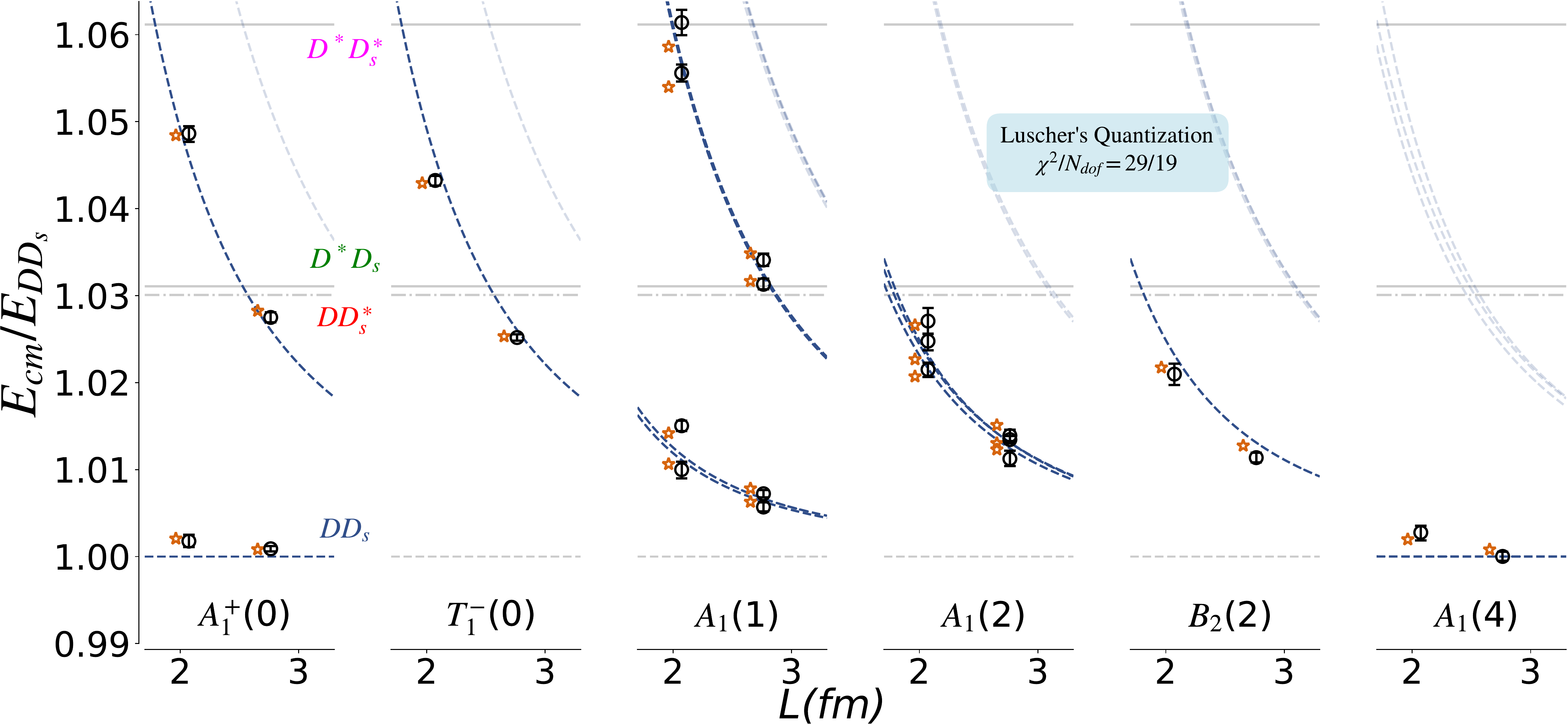}
    \caption{The simulated FV eigenenergies (circles) plotted together with the analytically reconstructed energy levels (stars, slightly shifted to the left), in the irreps used for the analysis $DD_s$ scattering in $S$- and $P$-waves, as a function of the spatial lattice extent studied. Top row shows the spectral reconstruction following LSE-based quantization condition, whereas the bottom row shows similar reconstruction using L\"uscher's quantization condition assuming a Breit-Wigner parametrization for the amplitudes. See Appendix \ref{app:elastic_fits} for details of parameterizations used within the L\"uscher prescription for amplitude extraction. The eigenenegies are presented in units of the energy of the elastic threshold $E_{DD_s}$. The black circles are the lattice energies included in the amplitude fit. }
    \label{fig:scalar_chosen_spectrum}
\end{figure*}

\begin{table}[tbh!]
    \centering
    \renewcommand{\arraystretch}{1.3}
    \begin{tabular}{ccccc}
    \hline
       $c_0$[S]               & $c_2$[S]             & $c_2$[P]           & $c_4$[P]          & $\chi^2/d.o.f$ \\
       $[$GeV$^{-2}]$        & $[$GeV$^{-4}]$      & $[$GeV$^{-4}]$    & $[$GeV$^{-6}]$   &                \\ \hline
       $3.6(^{+5}_{-5})$     & $-3.2(^{+6}_{-6})$  & $-12(^{+5}_{-4})$ & $5(^{+5}_{-5})$  & 36.7/20        \\ \hline\hline
       $a_0$ [fm]          & $r_0$ [fm]        & $a_1$ [fm$^3$]  & $r_1$ [fm$^{-1}$] &                \\ \hline
       $-0.157(^{+16}_{-15})$& $-0.08(^{+7}_{-8})$ & $0.086(^{+44}_{-36})$ & $-1.5(^{+0.4}_{-1.6})$&                \\ \hline           \end{tabular}
    \caption{Best fit parameter values for the elastic $DD_s$ scattering in $S$ and $P$ wave using solutions of LSE with parameterizations of the contact potential presented in Eqs. (\ref{VelasticS_LSE}) and (\ref{VelasticP_LSE}). The effective range parameter values from this fit are also presented.}
    \label{tab:Elastic_fits_final}
\end{table}

In this case, our main focus is on the $S$-wave interactions in the $DD_s$ scattering within an elastic assumption leading to scalar quantum numbers. To this end, we could utilize the lowest two levels in the $A_1^+(0)$ irrep from both ensembles and also the ground states from the $A_1(4)$ irrep presented in Figure \ref{fig:full_spectum}, where the higher partial wave contributions are assumed to be suppressed near the threshold. Assuming negligible mixed contributions from different partial waves on the energy levels, one could further consider the energy levels in the lowest shell of $A_1(1)$ and $A_1(2)$ irreps in the pure $S$-wave fits. A repulsive nature of interaction will drive the interacting FV energy levels to acquire positive energy shifts. One could further extend the fits to address $S$ and $P$-wave amplitudes combinedly, utilizing the remaining excited levels from $A_1(1)$ irrep and those from $T_1^-(0)$ and $B_2(2)$ irreps. 

In Figure \ref{fig:scalar_chosen_spectrum}, we present the analytically reconstructed energy levels (orange stars, slightly shifted to the left for clarity) using the best-fit parameters determined based on the combined $S$- and $P$-wave fit along with the simulated energy levels (circles) to demonstrate the quality of fits. The energy spectra are all presented in units of the elastic threshold energy $E_{DD_s}$. The top row demonstrates the quality of fits from the quantization program following solutions of LSE, whereas the bottom row shows the same for fits based on L\"uscher-based FV analysis. The parameterizations utilized in these fits are discussed in the next paragraph  and in the next subsection. The reduced $\chi^2$ values can be seen to be slightly larger than one suggesting mild tension in the reconstructed spectrum, reflected as some deviations of order 1.5$\sigma$ in a few of the moving frame levels. Despite slightly large reduced $\chi^2$ values observed, it is evident that the analytically reconstructed energy levels from the fits  represent the simulated energy spectrum very well in either cases.

\begin{figure}[h!]
    \centering
    \includegraphics[width=1.0\linewidth]{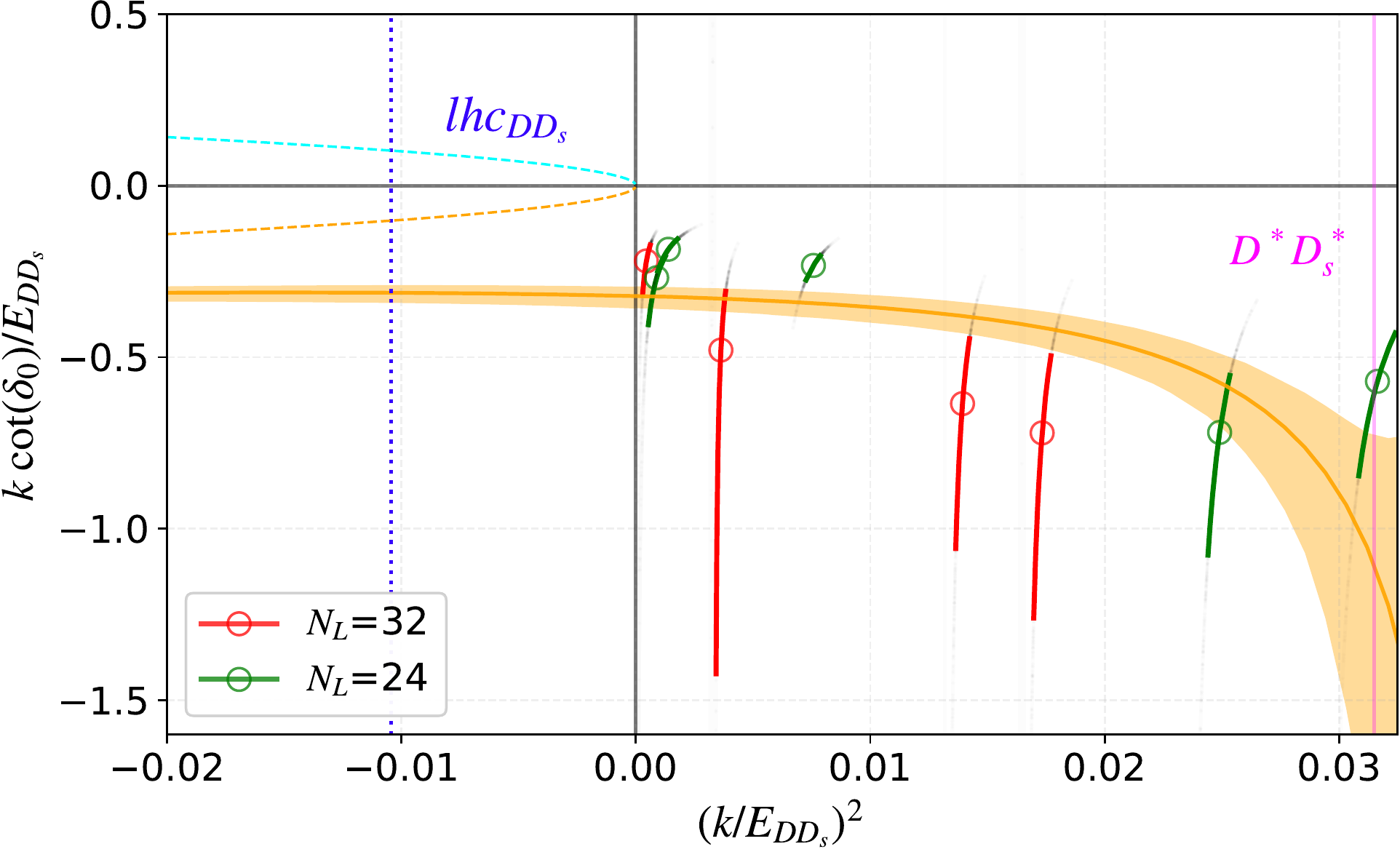}
    \caption{$k\cot{\delta}/E_{DD_s}$ vs. $(k/E_{DD_s})^2$ plot for the elastic $DD_s$ scattering in $S$-wave as estimated from the solutions of LSE (\eqn{planewaveQC}). The red and green markers correspond to the simulated FV eigenenergies, where the $k\cot{\delta_0}$ values are determined following L\"uscher's prescription. The solid orange curve and the associated band represent the fitted energy dependence of  amplitude determined from the parametrized contact potential given in \eqn{VelasticS_LSE}.  The cyan and orange curves represent the unitary parabola $\pm i\sqrt{-k^2}/E_{DD_s}$, whereas the vertical blue dotted line is the \lhc ~branch point associated with a $K^*$ meson exchange in the $u$-channel process. The vertical magenta line represents the $D^*D_s^*$ threshold, close to and above which inelastic effects could be important.}
    \label{fig:scalar_chosen_pcotdel}
\end{figure}

Our main results in elastic $DD_s$ scattering are based on the fits following the solutions of LSE (\eqn{planewaveQC}), where we have used the following parametrization for the contact potential 
\beqa
\mathbb{V}^{[S]}(\vek,\vek') &=& 2c_0[S]+2c_2[S](k^2+k'^2) \label{VelasticS_LSE} \mbox{ ~and} \\
\mathbb{V}^{[P]}(\vek,\vek') &=& (2c_2[P]+2c_4[P](k^2+k'^2))~\vek.\vek' \label{VelasticP_LSE}
\eeqa
for the $S$ and $P$-wave, respectively. The regularization scale $\Lambda$ and degree $n$ is chosen to be 0.9 GeV and 40, respectively, partly motivated by our recent study of $DD^*$ scattering with similar procedure \cite{Prelovsek:2025vbr}. The solutions are statistically robust to varying $\Lambda$ in the interval [0.6,1.0] GeV, where the highest utilized eigenlevel appear in the extracted FV spectrum. The analytically reconstructed spectrum presented in the top row of Figure \ref{fig:scalar_chosen_spectrum} is following this parametrization. We present the extracted momentum-squared (energy) dependence of the on-shell $S$-wave amplitude $k\cot(\delta_{0})$ along with the simulated energy levels in Figure \ref{fig:scalar_chosen_pcotdel}, in which either of the quantities are presented in units of $E_{DD_s}$. The constraint curves $\pm i\sqrt{-k^2}$  for the presence of a real (virtual) bound pole are indicated by the orange (cyan) dashed line. The corresponding best fit parameters featuring in the contact potential (Eqs. (\ref{VelasticS_LSE}) and (\ref{VelasticP_LSE})) are summarized in Table \ref{tab:Elastic_fits_final}. We present the fitted parameter values in Table \ref{tab:Elastic_fits_final} and $k\cot(\delta_{0})$ in Figure \ref{fig:scalar_chosen_pcotdel} with nonrelativistic normalization, where $\frac{2\pi}{\mu}T^{-1}=k\cot(\delta_{0})-ik$ and $\mu$ being the reduced mass of the channel \cite{Oller:2019opk}. This is consistent with the conventions that we have followed in our previous studies of isoscalar $DD^*$ scattering \cite{Collins:2024sfi,Prelovsek:2025vbr}.

From Figure \ref{fig:scalar_chosen_pcotdel}, it is evident that there are no crossings of the fitted amplitude (the orange curve and the band) with the unitary parabola $\pm i\sqrt{-k^2}/E_{DDs}$ suggesting no shallow subthreshold poles along the real energy axis. There is also no crossing of the real axis above the threshold, suggesting no $S$-wave resonance in the low energy region. We also reaffirm the absence of any physically sensible poles in the $DD_s$ $S$-wave amplitude across the threshold by investigating the analytic structure of the amplitude across the complex energy plane. The negative sign of  $k\cot\delta_{0}$  and positive energy shifts indicate slightly repulsive interaction of $DD_s$ in S-wave. 

\begin{figure}[h!]
    \centering
    \includegraphics[width=1.0\linewidth]{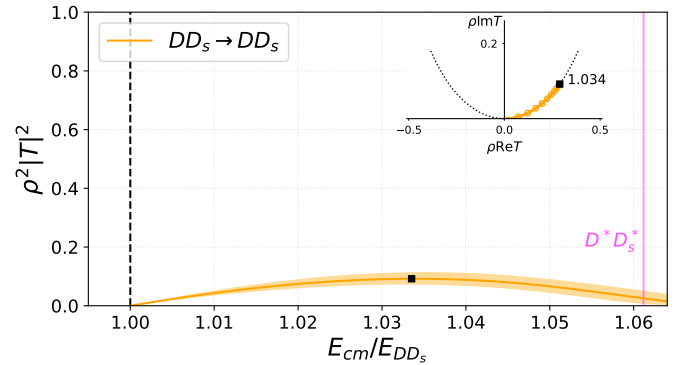}
    \caption{The energy dependence of the $DD_s$ amplitude: $\rho^2|T|^2$ proportional to the experimental $DD_s$ cross-section plotted as a function of the $E_{cm}/E_{DDs}$. The magenta vertical line indicates the $D^*D^*_s$ threshold. The inset figure: Variation of $\rho T=e^{i \delta}\sin{\delta}$ for $DD_s$ scattering in the complex plane, where the dashed line represents part of the Argand circle. The solid orange curve represents the rising feature, the orange unfilled circles represents the falling feature, and the black square indicating the energy at which the cross section peaks. }
    \label{fig:Cross_section_elastic}
\end{figure}

We present the energy dependence of the extracted amplitude above $DD_s$ threshold in Figure \ref{fig:Cross_section_elastic}, where the plotted quantity $\rho^2|T|^2$ is proportional to the experimental cross-section in $DD_s$ scattering. The peak in $\rho^2|T|^2$ reaching a value of unity is a quintessential feature that points to the presence of an elastic $DD_s$ resonance, in which case the $DD_s$ scattering phase crosses the value $\pi/2$. To demonstrate the $DD_s$ phase variation further, we also present the Argand diagram representation of variation in the elastic $DD_s$ amplitude ($T$) as a function of $E_{cm}/E_{DD_s}$, where the solid orange curve represents the rising feature in the amplitude, the unfilled orange circles trace the falling feature beyond the peak at $E_{cm}/E_{DD_s}\sim1.034$, indicated by the black square. The absence of a prominent peak rising to the maximum peak value of unity and equivalently the limited variation in the elastic $DD_s$ phase (presented in the inset figure) suggest no resonance features in the extracted amplitude across the energy region constrained. Similarly, the lack of threshold enhancement in this quantity suggests the absence of any shallow subthreshold poles. The falling feature in $\rho^2|T|^2$ at higher energies is driven by the zero in $\rho^2|T|^2$ that is related to the pole in $k\cot\delta_{0}$, which moves away to higher energies with increasing cutoff $\Lambda$ \cite{Meng:2024kkp}.

Various fits based on conventional L\"uscher's  formalism  also lead to consistent behavior of the amplitude in the region constrained by the lattice eigenenergies. They also provide a faithful analytical reconstruction of the FV spectra (bottom row in Figure \ref{fig:scalar_chosen_spectrum}), yet some of them lead to shallow subthreshold poles. Given the lack of data in the subthreshold regions, and potential effects of the \lhc ~branch point that cannot be accounted in the standard $2\rightarrow2$ L\"uscher-based FV quantization formalism, it is inappropriate to give any physical significance to such subthreshold poles. We provide additional details on these fits and our observations in the next subsection as well as in the Appendix \ref{app:elastic_fits}.  

\subsection{More on elastic $DD_s$ scattering}
\label{More_elasticDDs}

In this subsection, we provide further details on the various fits we make to the $S$-wave $DD_s$ amplitudes and the higher partial wave effects we observe in the extracted $S$-wave amplitude. Appendix \ref{app:elastic_fits} presents the list of general functional forms used to parametrize the amplitudes in the L\"uscher-based FV analysis. Considering only $l=0(S)$ and $1(P)$ partial waves, the $\tilde{K}^{-1}$-matrix is a two-dimensional diagonal matrix, with one dimension each for each partial wave. The parametrization of the contact potential in LSE was presented in  Eqs. (\ref{VelasticS_LSE}) and (\ref{VelasticP_LSE}). For a pure $S$-wave or $P$-wave fit, one utilizes only the relevant diagonal element in the interaction matrix to parametrize the respective energy dependence. 

\begin{figure}[h!]
    \centering
    \includegraphics[width=1.0\linewidth]{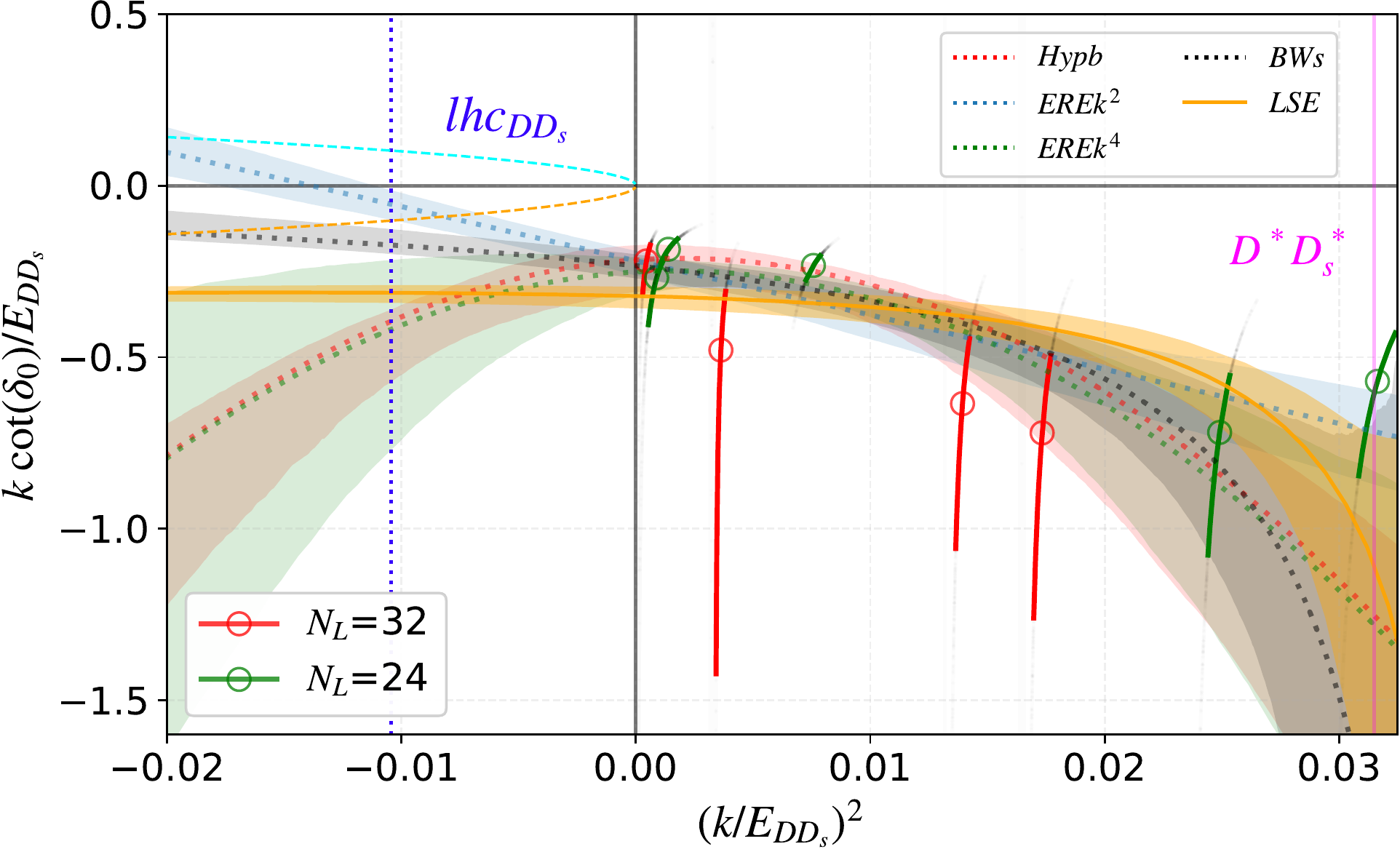}
    \caption{Same as in Figure \ref{fig:scalar_chosen_pcotdel}, but with different functional forms used to parametrize the energy dependence. Solid orange curve is based on solutions of LSE, whereas other dotted curves with associated bands are those following L\"uscher-based analysis. See Appendix~\ref{app:elastic_fits} for the definitions of the fit forms for the L\"uscher-based analyses.
    \label{fig:scalar_pcotdel}}
\end{figure}

\ul{\textit{$S$-wave scattering}}- A compilation of the fit results for elastic $DD_s$ scattering in $S$-wave with different parameterizations is presented in Appendix \ref{app:elastic_fits}. The corresponding momentum-squared dependence of a subset of parametrized amplitudes are presented in terms of the extracted $k\cot{\delta_0}$ in Figure \ref{fig:scalar_pcotdel}. As expected from good and sensible fits, the energy or equivalently momentum-squared dependence of all fitted amplitudes indeed follow the behavior followed by the spread of FV eigenenergies in the region immediately above the threshold. However, below the threshold and at energies well above the threshold, where there is no lattice data to constrain the dependence, the fits demonstrate differences in behavior. Using two-parameter effective range expansion (ERE) with terms up to the $k^2$ term, the best-fit results would have suggested a crossing with the bound state constraint (blue dotted curve) close to the \lhc ~branch point. Such a crossing would suggest a bound state solution in the $DD_s$ amplitude that should lead to an unobserved FV eigenenergy, similar to what was argued in the case of isovector $DD^*$ scattering in Ref. \cite{Meng:2024kkp}. It is also observed that all the different parameterizations used with at most two parameters and following L\"uscher's FV amplitude analysis lead to similar bound state solutions at energies close to or towards the left of the \lhc ~branch point. Since these solutions are close to the \lhc ~branch point or on the left of it, we refrain from giving it any physical significance. Note that such bound state solutions does not appear in the amplitudes based on the solutions of LSE. 

One way to get rid of such bound state solutions arising in L\"uscher-based amplitudes is by considering parameterizations that include more terms incorporating the shape parameters. Such parameterizations of  $\tilde{K}_0^{-1}$, including up to $k^4$ term or $s^2$ terms or with more parameters such as a hyperbolic~(Hyp) fit form (see Appendix \ref{app:elastic_fits} for details), circumvent the issue of bound state solutions. However they lead to subthreshold poles away from the real energy axis in the physical Riemann sheet that violate causality \cite{Gribov:2009cfk,Briceno:2017max}. Such acausal solutions are evident if one keeps track of the zeros of the denominator of $T_{l}$ expressed in \eqn{Tmatrix}. It is  observed that the complex energy location of such subthreshold poles also vary, yet remain acausal, across different parameterizations. In contrast, the solutions of LSE do not lead to any such acausal poles in the extracted amplitudes, which is also the case in the isovector $DD^*$ scattering studied in Ref. \cite{Meng:2024kkp}. We also observe the Breit-Wigner type parametrization of the $\tilde{K}_0$, while following L\"uscher's prescription, also does not lead to any physical sheet subthreshold resonance poles (see gray curve/band in Figure \ref{fig:scalar_pcotdel}). {The reconstructed FV spectra presented in Figure \ref{fig:scalar_chosen_spectrum} based on L\"uscher's prescription is assuming a Breit-Wigner type parametrization for the amplitude. In the next paragraph, we give a brief remark on the unphysical features we observe, while using L\"uscher's prescription for extracting the elastic $DD_s$ amplitudes in $S$-wave.

Physical amplitudes are expected to respect unitarity, analyticity and crossing symmetry \cite{Peskin:1995ev,Mizera:2023tfe}. Analyticity of the amplitudes is a consequence of its causal nature. In most of the lattice investigations these days, the discrete FV eigenenergies are used to constrain partial wave projected amplitudes that respects unitarity, yet there are no constraints imposed from analyticity or crossing symmetry. One could impose these symmetries following dispersion relations that are required by analytic physical amplitudes, given one has a rich collection of all relevant lattice-extracted amplitudes. Such an exercise has been performed for the case of $\pi\pi\rightarrow\pi\pi$ scattering using FV data in Refs. \cite{Mai:2019pqr,Rodas:2023gma,Rodas:2023nec,Cao:2023ntr}. In the absence of such a rich collection of lattice data (as is the case in this study), the amplitude behavior in subthreshold regions and highly above threshold regions that are not constrained by lattice-data relies on naive extrapolations, which need not be free of unphysical features, as observed in some of the L\"uscher-based amplitudes we extract. We remark that this is not the first observation of such behavior and such acausal poles have been reported in literature, \cf ~footnote 7 in Ref. \cite{Briceno:2017qmb}. Note that the fact that solutions of LSE does not lead to any such acausal poles is related to the inherent causality constraints within the LSE formulation of the scattering problem \cite{Mengprivate}. This is the main reason for projecting the solutions of LSE as our main result for the elastic $DD_s$ scattering discussed in the previous section. 

\begin{figure}[h]
    \centering
    \includegraphics[width=1.0\linewidth]{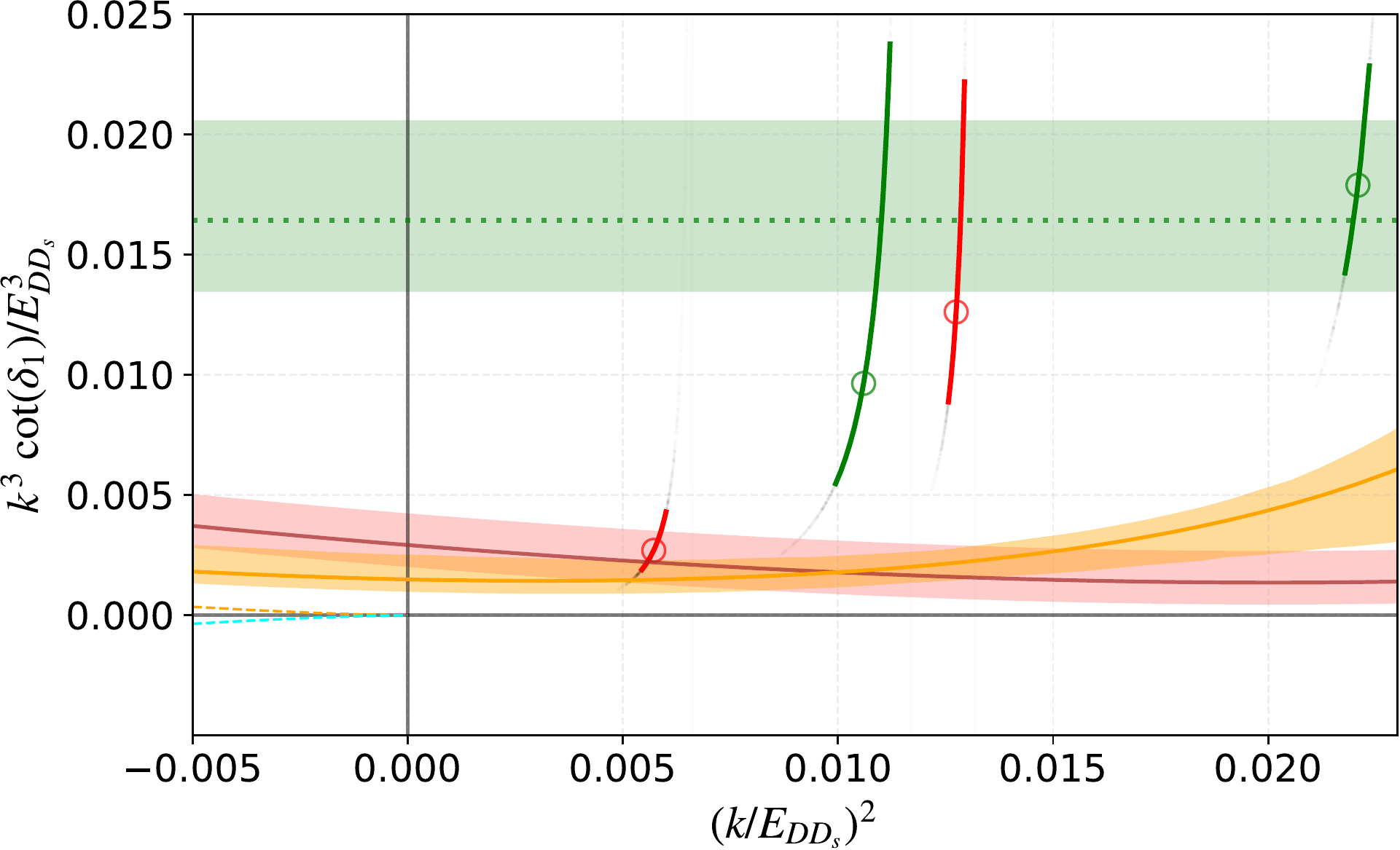}
    \caption{The momentum-squared dependence of the $P$-wave amplitude in $DD_s$ scattering. The circle markers are obtained via L\"uscher's formalism from irreps where $l=1$ is the lowest contributing partial wave. The solid~(orange and red) and dotted (green) curves represent fits with LSE and L\"uscher's approach, respectively.}
    \label{fig:Pwave_amplitude}
\end{figure}

\ul{\textit{Higher partial wave effects}}- Although the focus of this investigation is on $S$-wave interactions in the $D^{(*)}D_s^{(*)}$ channels, the use of moving frame irreps calls for analysis considering any contributions from higher partial waves. As argued in Section \ref{sec:FiniteVolume}, the rest frame FV eigenenergies with dominant overlaps to $D$-wave projected interpolators are consistent with the noninteracting scenario, and hence the related contributions are assumed to be negligible to the rest of the spectrum. However, in FV irreps with $P$-wave as the lowest contributing partial wave, we observe nonzero energy shifts in eigenenergies pointing to nontrivial interactions. This suggests the need to reliably constrain the associated $P$-wave amplitude, both in the interest of identifying potentially nontrivial features in the $P$-wave amplitude as well as in filtering out any contamination from moving frame irreps in the lattice-extracted $S$-wave amplitudes. Below we briefly discuss the $P$-wave amplitudes in $DD_s$ scattering and its effects on the extracted $S$-wave interactions. 

In Figure \ref{fig:Pwave_amplitude}, we present $P$-wave amplitude in $DD_s$ scattering in terms of $k^3\cot{\delta_1}$ as a function of $k^2$ in dimensionless units built out of $E_{DD_s}$. The amplitudes presented are extracted following a pure $P$-wave fit to the eigenenergies in FV irreps $T_1^-(0)$ and $B_2(2)$ that has $P$-wave as the lowest contributing partial wave. The $y$-axis values of the markers are evaluated following L\"uscher's formalism, whereas the solid (dotted) curves with bands represent extracted amplitudes following LSE-based (L\"uscher-based) analysis. The cyan and orange dashed curves ($\pm i(\sqrt{-k^2})^3/E^3_{DDs}$) on the second and third quadrant represents the $P$-wave constraint curves for the existence of real(virtual) bound states. 

First we discuss the analysis of $P$-wave contributions following the LSE-based analysis. The two solid curves shows the energy dependence of the amplitude as determined using the contact potential presented in \eqn{VelasticP_LSE}, with single parameter ($c^p_2$; red) and both parameters (orange). Note that the fit quality is very well reflected in the reconstructed energy spectrum demonstrated in Figure \ref{fig:scalar_chosen_spectrum}. Although one and two parameter fit reliably captures the amplitude at the lowest eigenenergies, they seem to fail in capturing the range parameter of the amplitude. In view of the large errors in $k^3cot\delta_1$ and the fewer degrees of freedom with lowest contributing partial wave as the $P$-wave, it would be interesting to investigate this further with higher-statistics data and a richer spectrum of energy levels. Alternatively, including the \lhc ~effects associated with a $K^*$-meson exchange might mitigate the need of higher order terms in the contact potential as argued in Ref. \cite{Meng:2023bmz}. However, either of these exercises go beyond the scope of the present work. Most importantly, the LSE-based amplitudes reliably reproduce the FV spectra and do not cross the constraint curves ($\pm i(\sqrt{-k^2})^3/E^3_{DDs}$) in the near-below threshold region suggesting no shallow real (virtual) bound states. They also do not cross the $x$-axis above threshold and hence does not carry any signature for resonances. 

Now we move on to the $P$-wave amplitudes extracted using L\"uscher-based analysis. The dotted green curve represents the energy independent amplitude extracted based on a constant parametrization of $\tilde K_1$ element. Clearly $\tilde{K}_l=$constant does not capture the amplitudes in the low energy and its energy dependence reliably. Including higher order terms in the ERE to accommodate leads to amplitudes that suggest real-axis crossing above threshold with positive slope, pointing to existence of off-axis poles in the physical sheet. We provide an extended account of this in Appendix \ref{app:elasticP}. Despite the unphysical features in relation to the $P$-wave amplitudes, we observe the lattice-extracted $S$-wave amplitude remains unaffected in the L\"uscher-based analysis as well. 

Following the pure $P$-wave fits, we utilize these parameters as initial parameters for combined $S$ and $P$-wave fits on larger set of data that includes eigenenergies from other FV irreps. The faithful reconstruction of FV spectra with these combined fits has been demonstrated in Figure \ref{fig:scalar_chosen_spectrum}. We observe that the resulting $S$-wave amplitudes are consistent with those obtained from pure $S$-wave fits within the statistical uncertainty, as evident from the listed best fit parameter values in Table \ref{tab:Elastic_fits}. More importantly, our main result for the elastic $DD_s$ scattering following the LSE-based fits leads to amplitude that does not carry any unphysical features and the lattice-extracted $S$-wave amplitude is left unaffected by the $P$-wave contributions in FV.

\begin{figure*}[tbh]
    \centering
    \includegraphics[height=7cm,width=0.9\linewidth]{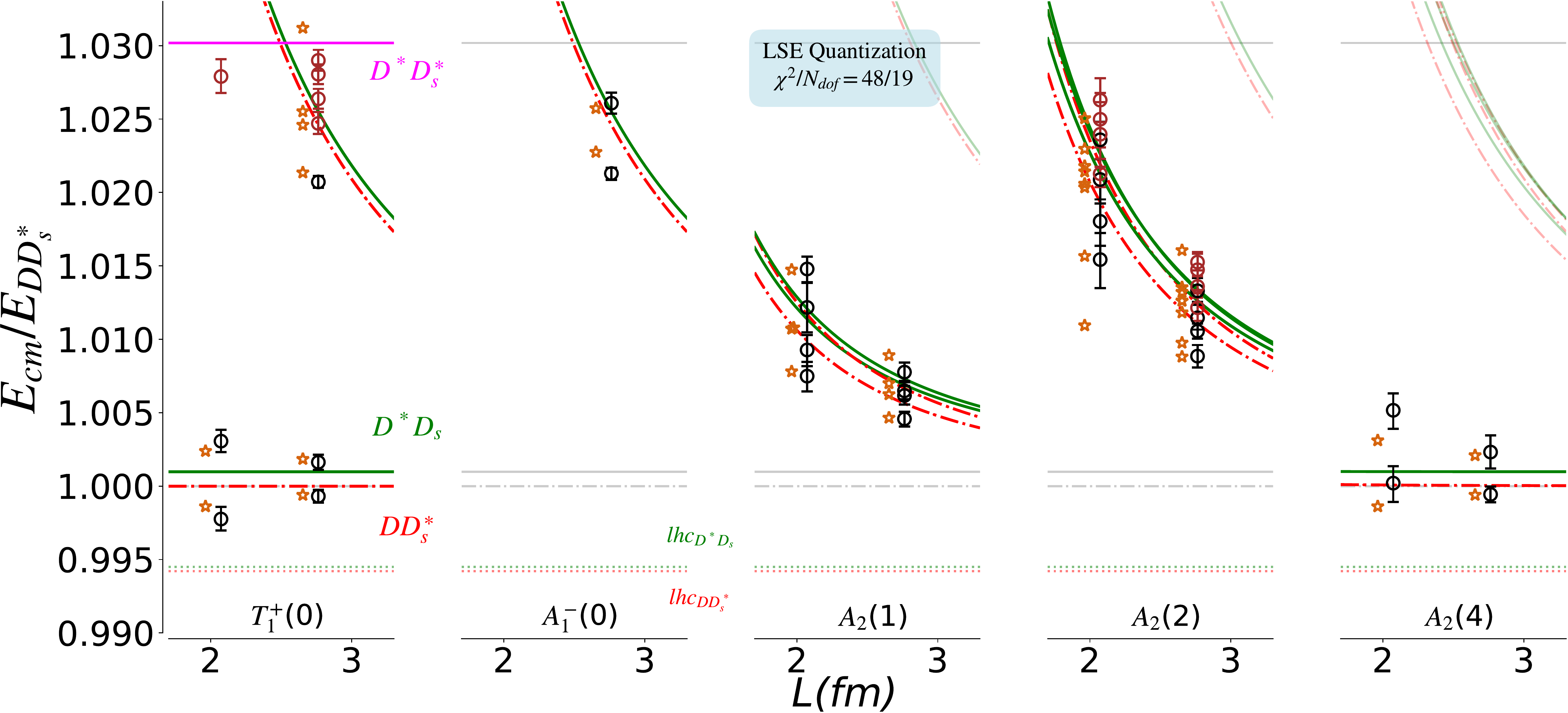}
    \includegraphics[height=7cm,width=0.9\linewidth]{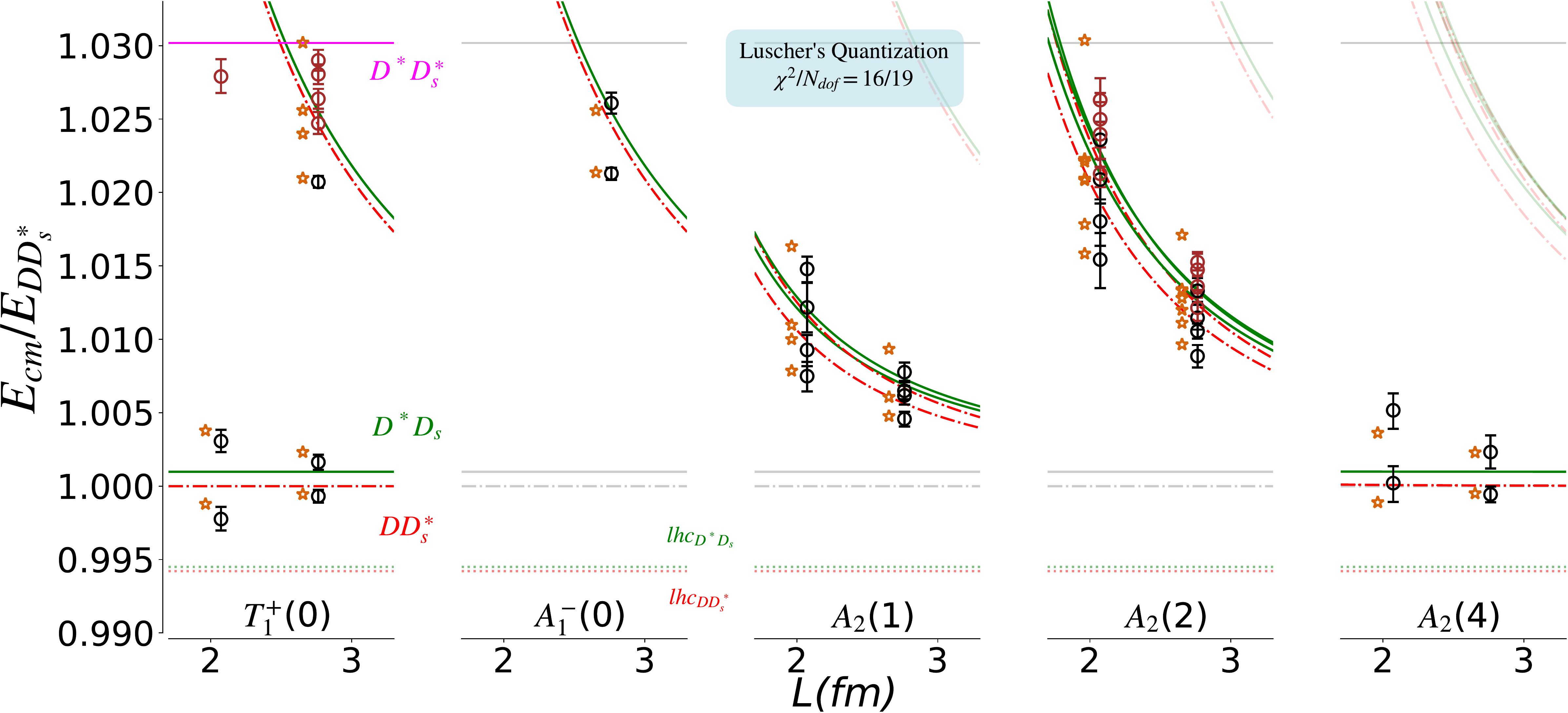}
    \caption{Same as in Figure \ref{fig:scalar_chosen_spectrum}, but for the case of $DD_s^*$-$D^*D_s$ coupled channel system. Reconstructed spectrum for channels with $J^P=1^+$ in $S$-wave. The eigenenergies are presented in units of $E_{DD_s^*}$ and the lattice spatial extent in fermi along the x-axis. The curves in red and green are the non-interacting levels. The black circles are those eigenenergies included in the amplitude fits, whereas the brown circles are those ignored. Top row presents the analytically reconstructed spectra following solutions of LSE in \eqn{LSEfv}, whereas bottom row presents the reconstructed spectra from L\"uscher-based fits.}
    \label{fig:avector_chosen_spectrum}
\end{figure*}

\subsection{Coupled $DD_s^*-D^*D_s$ scattering \label{inelasticDDs}} 

In this section, we discuss the results for $S$-wave interactions in the coupled $DD_s^*-D^*D_s$ scattering leading to axialvector quantum numbers. The lower panes for irreps $T_1^+(0)$, $A_2(1)$, and $A_2(4)$ in Figure \ref{fig:full_spectum} carry the relevant information for the $S$-wave scattering. The $DD_s^*$ and $D^*D_s$ channels with spin $\tilde{s}=1$ allow possibility of a physical mixing between partial waves $l=0$ and $2$. Since the simulated FV levels that have dominant overlap with $l=2$ partial waves are observed to be consistent with the noninteracting scenario, we assume the contributions from $l=2$ partial wave are negligible in the systems studied. However, one can observe nonnegligible deviations in irreps with $P$-wave as the lowest contributing partial wave. Thus we consider parameterizations of the amplitudes for pure $S$-wave as well as combined $S$ and $P$ wave scattering for the inelastic system, in order to assess the robustness in the extracted $S$-wave amplitudes. We additionally include levels from the irreps $A_1^-(0)$ and $A_2(2)$  to further constrain the energy dependence $P$-wave amplitudes. 

In Figure \ref{fig:avector_chosen_spectrum}, we present the simulated FV eigenenergies (circles) of the $DD_s^*-D^*D_s$ system along with the analytically reconstructed spectrum (stars) using the best-fit parameters from combined $S$- and $P$-wave fits. The eigenenergies are presented in units of the elastic threshold energy $E_{DD_s^*}$. The top row corresponds to the reconstructed spectrum from the solutions of LSE, whereas in the bottom row, we present similar results extracted following a L\"uscher-based FV analysis. We discuss more on the details on the fit forms and results below. 

Being a two-channel inelastic system, the interaction ($\tilde{K}$ and $V$) is a matrix assuming $S$ and $P$-wave contributions to the lowest allowed total angular momentum ($J^P=1^+$ and $0^-$, respectively), with no physical mixing of even and odd partial waves. Yet, one requires at least six parameters each to perform a fit assuming an energy independent amplitude. Given the limited nature of FV irreps available and the limited number of energy levels within them, the amplitude fits are less well constrained compared to the elastic scenario we have discussed earlier. Yet, the quality of spectral reconstruction following either procedures is evident from the figure, and is also reflected as reasonable $\chi^2/d.o.f.$ ($\sim 48/19$ and $\sim 16/19$ for the LSE-based and L\"uscher-based fits, respectively). The relatively large reduced $\chi^2$ for the LSE-based fits is observed to be arising from relatively large tension in the $A_1^-$ irrep, which serves in determining the $P$-wave contributions. Given the limited degrees of freedom in constraining the $P$-wave contributions, our main focus here is the extraction of $S$-wave amplitudes and argue their robustness with/without accommodating the $P$-wave contributions in the FV spectra. 

Before we proceed to discuss our main results, we briefly comment on the LSE-based fits in this inelastic system. Unlike in the elastic $DD_s$ scattering, a relatively closer $D^*D^*_s$ threshold limits the potential regularization to be at a lower value of $\Lambda\sim0.65$ GeV or less. Although the LSE-based fits lead to a qualitatively good reconstruction of the FV spectra (see Figure \ref{fig:avector_chosen_spectrum}), the resultant fits are generally more unstable and lead to relatively large $\chi^2$ values, particularly when accommodating higher partial waves. With results that are sensitive to the cutoff parameters and the input values, extracting robust conclusions employing LSE-based fits turns out to be challenging and involves both technical as well as conceptual difficulties. We consider such an extensive analysis is a work for the future. While such LSE-based multi-channel analyses have been addressed in the infinite-volume, inelastic two-particle scattering data in the finite-volume has seldom been studied following LSE-based formulations in the past and we have only attempted an exploratory coupled-channel LSE analysis in this work. We provide the potential parameterizations utilized in our fits based on LSE in Appendix \ref{app:inelastic_fits}. In this work, the main observation from LSE-based fits to the inelastic system is that the FV spectra are qualitatively reproduced and that $S$-wave amplitudes $k\cot(\delta_0)$ obtained  subsequently assuming a decoupled scenario in the respective channels agree with the estimates from L\"uscher-based fits near threshold, which we discuss later in this section. From now on, we discuss the fit results based on Luscher's FV formulation. 

Our main results for the inelastic $DD_s^*-D^*D_s$ scattering amplitudes are based on an effective range expansion in $S$ and $P$-wave. Choosing the notation 1 for $DD^*_s$ and 2 for $D^*D_s$ channels, the ERE parametrization used for the $S$-wave amplitude is 
\begin{align}
\tilde{K}^{-1}=\left[
\begin{array}{cc}
 a_{11}+b_{11}k_1^2 & a_{12}+b_{12}(k_1^2+k_2^2) \\
                    & a_{22}+b_{22}k_2^2
\label{ERE_inelastic}
\end{array}
\right].
\end{align}
The parametrization for the $P$-wave amplitudes are limited to a form with only the leading $k^2$ independent terms in the above equation. Note that there is no physical partial wave mixing that can happen between the $S$ and $P$-waves, and thus the parametrized $\tilde K^{-1}$ matrices are always block diagonal in the partial wave space. The reconstructed levels presented in the bottom panes of Figure \ref{fig:avector_chosen_spectrum} are based on the above proposed form of the $\tilde K^{-1}$ matrix. The best fit parameter values with this form of the $\tilde{K}^{-1}$-matrix in units of $\beta=E_{DD^*_s}$ are 
\begin{align}
a_{11}[S] &= 0.202(^{+21}_{-19})\cdot\beta, &  b_{11}[S] &= 0.0026(^{+1}_{-4})/\beta, \nonumber \\
a_{12}[S] &= -0.0012(^{+1}_{-1})\cdot\beta, & b_{12}[S] &= 0.096(^{+16}_{-4})/\beta, \nonumber \\
a_{22}[S] &= -0.15(^{+1}_{-2})\cdot\beta, & b_{22}[S] &= -0.24(^{+3}_{-2})/\beta, \nonumber \\
a_{33}[P] &= 0.0025(^{+4}_{-3})\cdot\beta^3, && \nonumber \\
a_{34}[P] &= 0.0001(^{+40}_{-1})\cdot\beta^3, && \nonumber  \\
a_{44}[P] &= -0.14(^{+1}_{-9})\cdot\beta^3. \label{eqn:ERE_spwave_inelastic_chosen}
\end{align}
The corresponding scattering length ($a_{0,c}=(\beta a_{cc}[S])^{-1}$) and the effective range ($r_{0,c}=2b_{cc}[S]/\beta$) in the individual channels ($c=1,2$) are
\begin{align}
a_{0,1} &= 0.24(^{+3}_{-2})~\mbox{fm}, &  r_{0,1} &= 0.00025(^{+1}_{-4})~\mbox{fm}, \nonumber \\
a_{0,2} &= -0.33(^{+4}_{-2})~\mbox{fm}, & r_{0,2} &= -0.023(^{+3}_{-1})~\mbox{fm} 
\label{eqn:ERE_spwave_inelastic_a0r0}
\end{align}
in physical units. 

\begin{figure}[tbh]
    \centering
    \includegraphics[width=\linewidth]{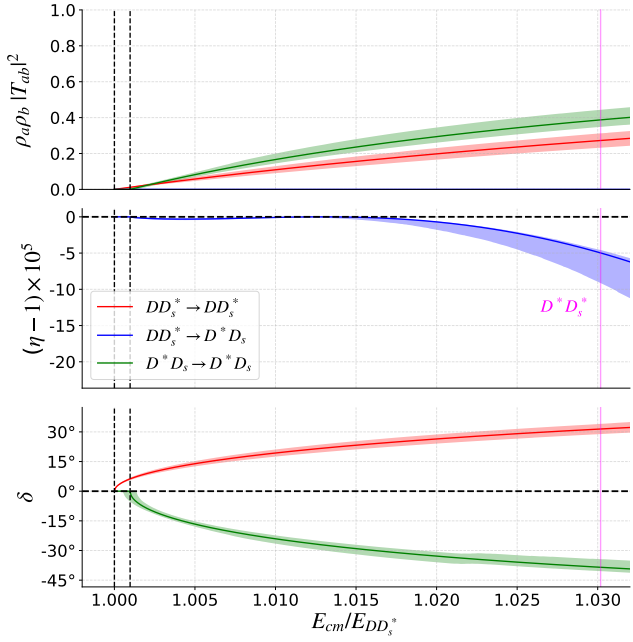}
    \caption{$DD_s^*-D^*D_s$ scattering amplitudes in $S$-wave. Top: The quantity $\rho_a\rho_b|T_{ab}|^2$ in the top pane, representing the experimental cross section, where $a$ and $b$ runs over the channels $DD_s^*$ and $D^*D_s$, respectively. See the text for definitions and details. Middle: The inelasticity between the channels, as defined in Eq. (\ref{Tcoupled}). Bottom: The phase shifts $\delta_0$ in the $DD_s^*$ and $D^*D_s$ channels presented in degrees.}
    \label{fig:Cross_section_inelastic}
\end{figure}

In Figure \ref{fig:Cross_section_inelastic}, we present the energy dependence of the extracted amplitude based on Eq. (\ref{ERE_inelastic}). In the top pane, we present the quantity $\rho_a\rho_b|T_{ab}|^2$, which is proportional to the experimental cross-section of different scattering processes considered. In the bottom and the middle panes, we present the same information in the amplitudes ($S_{aa}=\eta e^{i2\delta_a}$) in terms of the phases ($\delta_a$) and the associated (in)elasticity ($\eta$), respectively (See \eqn{Scoupled} for definition). Here $a$ and $b$ assume the channels $DD_s^*$ and $D^*D_s$, respectively. The color convention followed in the figure is listed in the legend.  

From the amplitudes presented in Figure \ref{fig:Cross_section_inelastic}, it is evident that the off-diagonal component is substantially smaller than the diagonal components in the scattering matrix. This suggests negligible coupling between the channels $DD_s^*$ and $D^*D_s$, essentially indicating the elastic nature of scattering in the individual channels. It can also be observed that the diagonal amplitudes shows a gradual monotonic rise suggesting no signatures for any resonance. The inelasticity presented in the middle pane indicates mild deviations with increasing energy, yet remains significantly small. The phase shifts in the individual channels shows a gradual variation reaching out to values $\pi/4$ at very large energies. The pole content of these amplitudes also suggest no interesting features across all different Riemann sheets in the energy regions constrained by the lattice data, as will be detailed below.

\begin{figure}[tbh]
    \centering
    \includegraphics[width=\linewidth]{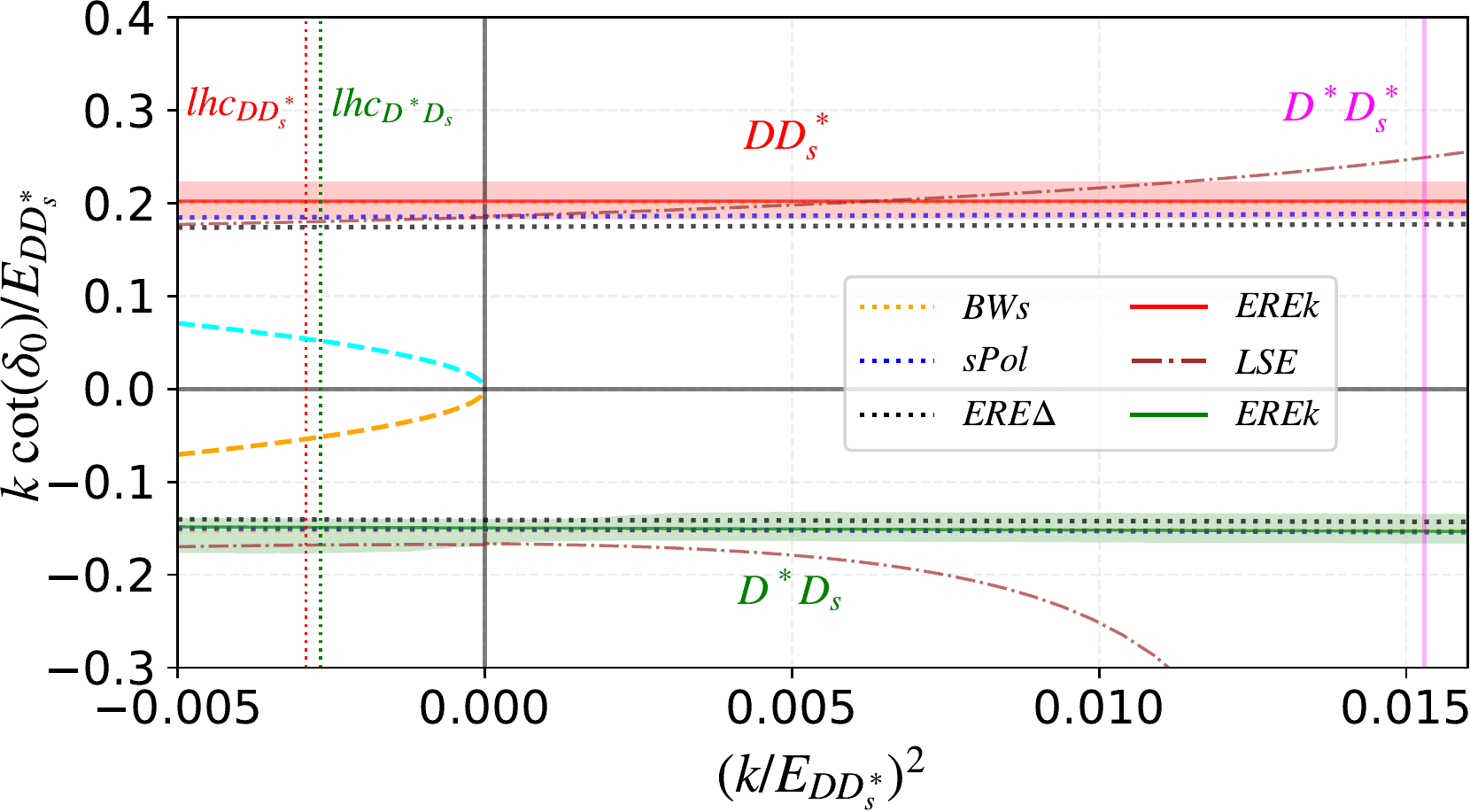}
    \caption{$k\cot(\delta_{0,a})$ vs. $(k)^2$ in units of $E_{DD_s^*}$ for individual channels in the inelastic coupled channel $DD_s^*$-$D^*D_s$ scattering in $S$-wave. The cyan and orange curves represent the unitary parabola $\pm i\sqrt{-k^2}/E_{DD_s^*}$, whereas the vertical dot-dashed lines represent the \lhc ~branch points associated $K$-meson exchange in the respective channels. The solid green and red curves/bands represent the $D^*D_s$ and $DD_s^*$ amplitudes, respectively, assuming an ERE parametrization and using L\"uscher's prescription. The other dashed and dot-dashed curves represent amplitude with other parametrizations and procedures as indicated in the legend. See Appendix~\ref{app:inelastic_fits} for further details on the fit forms.}
    \label{fig:inelastic_pcotdel}
\end{figure}

Given the nearly elastic nature of the individual scattering channels, one could investigate the amplitudes in terms of $k\cot(\delta)$ to assess the pole content in the constrained region. In Figure \ref{fig:inelastic_pcotdel}, we present the quantity $k\cot(\delta_{0,a})$ as a function of $k^2$, where the index $a$ refers to the scattering channels $DD_s^*$ and $D^*D_s$. The orange curve/band represents the elastic amplitudes from the diagonal components of our main result, whereas other curves represent equivalent amplitudes determined using other different parameterizations. The negative shifted energy levels in Figure \ref{fig:avector_chosen_spectrum} determine the $k^2$ dependence of the $DD_s^*$ phase shift leading to a positive $k\cot(\delta_0)$ and the positive shifted levels constrain the $k^2$ dependence of the $D^*D_s$ phase shift leading to a negative $k\cot(\delta_0)$. This observation is consistent with the observed operator-state-overlaps, primarily for the ground states, presented in Figure \ref{fig:tmin_dependence}, which suggests the negatively (positively) shifted level is dominantly overlapping with $DD_s^*$-like ($D^*D_s$-like) operator. 

It is evident that the amplitudes in the elastic assumption are nearly energy independent. This energy independence is also evident from best fit parameters presented in Eq. (\ref{eqn:ERE_spwave_inelastic_chosen}), where the leading energy independent terms in the parametrization of the diagonal elements can be observed to be dominant compared to subleading parameters and consistent across different fits performed. The fact that either of the amplitudes do not cross the $x$-axis anywhere in the constrained region or immediately below the threshold indicates no evidence for hadronic poles at the energies studied. The leading off-diagonal parameter in the coupled-channel system within the ERE assumption can be seen to be significantly small compared to the diagonal elements suggesting no striking surprises in this conclusion. 

To reaffirm the absence of any hadronic poles in the $DD_s^*$-$D^*D_s$ $S$-wave amplitude in the lattice-constrained energy regions, we investigate the analytic structure of the amplitude across the complex energy plane. To this end, using the formulae in \eqn{Rkn0} for the $T$-matrix (using the $K$-matrix determined from the amplitude fits) and adapting the signature of $Im(\rho)$ for the desired sheet, we evaluate the $T$-matrix. The resulting $T$-matrix is then examined for zeros in $|det(T^{-1})|$ leading to a pole singularity in $T$. As expected, there were no pole singularities observed in the extracted amplitudes. The monotonic rise in the cross section and the gradual variation observed in the phase shifts in the energy range of interest suggest the absence of any near poles in $S$-wave coupled channel $DD_s^*$-$D^*D_s$ scattering. 

\begin{figure}[tbh]
    \centering
    \includegraphics[width=\linewidth]{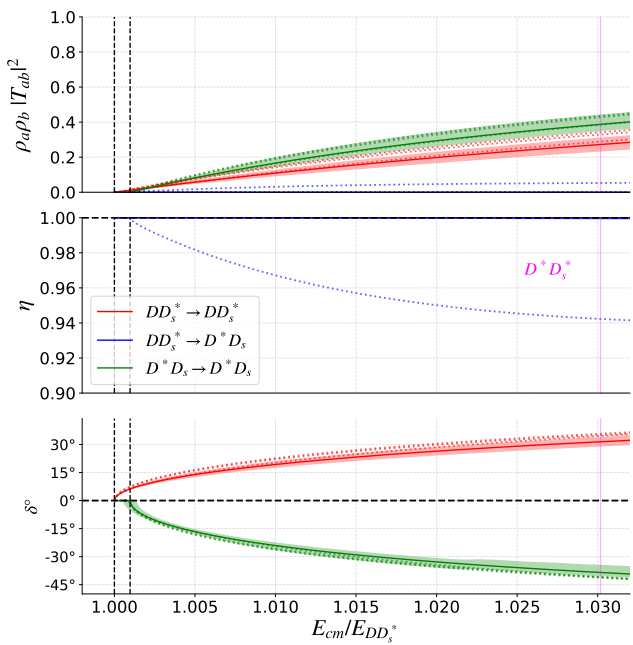}
    \caption{Same as in Figure \ref{fig:Cross_section_inelastic} (except the middle pane), but for a set of selected parameterizations studied following the L\"uscher-based amplitude fits. A limited set is plotted for clarity, and other omitted parameterizations also show similar patterns. In the middle pane, we present the elasticity ($\eta$) of the channels, in contrast to the inelasticity, which was presented in Figure \ref{fig:Cross_section_inelastic}. See Appendix~\ref{app:inelastic_fits} for further details on the fit forms. }
    \label{fig:Cross-section-inelastic-summary}
\end{figure}

Now we address the parametrization dependence of the observation we make in this case. In Figure \ref{fig:Cross-section-inelastic-summary}, we present the similar quantities as presented in Figure \ref{fig:Cross_section_inelastic}, except for the middle pane, for a set of different parameterizations utilized. It is evident that the quantity in the top pane representing the event distribution in different reactions shows qualitatively similar behavior, largely independent on the parametrization. Similar inference can also be arrived at from the energy dependence of the phase shifts presented in the bottom pane. In the middle pane, we present the channel elasticity ($\eta$) as defined in \eqn{Scoupled}. In the case of Breit-Wigner type parametrization, the extracted amplitude indicates a rise in inelasticity up to a few percent, yet leading to no interesting poles in the amplitudes across the constrained energy regions and all four Riemann sheets. We observe all parameterizations used in this case suggest qualitatively similar features and do not carry any poles across the complex energy plane in the region constrained by the lattice eigenlevels and immediately below the threshold. We present the form of parameterizations used and the summary of different fits made in Appendix \ref{app:inelastic_fits}. 

Constraining $P$-wave contributions in this coupled channel study is a more complicated challenge due to the limited nature of lattice data serving to constrain the $P$-wave amplitudes. With two channels involved in the $P$-wave scattering leading to the same quantum numbers, one requires at least three parameters assuming an energy independent amplitude. Even with a three parameter form for the $P$-wave effects, the amplitude fits turns out to be a formidable task with the available lattice data.  Accounting for corrections by incorporating higher energy dependent terms further reduces the stability of the fits and mostly lead to results with unacceptable $\chi^2$ values. Considering this we limit ourselves to the pragmatic case of the three parameter forms for the $P$-wave amplitudes, with each parameter representing the leading energy independent term in the three different reactions possible. The $P$-wave parameters are fitted for with constrained $S$-wave parameters based on pure $S$-wave fits, following which the quality of reconstructed FV spectra is investigated for. 

As already demonstrated in Figure \ref{fig:avector_chosen_spectrum}, with the above assumption on the $P$-wave amplitudes, the low energy spectrum is reconstructed faithfully independent of the procedure utilized, despite different degrees of fit quality. Other parameterizations also show similar features confirming the reliability of the fits. We summarize the $P$-wave amplitudes also in Appendix \ref{app:inelastic_fits}. Due to the limited degrees of freedom available for the $P$-wave, we refrain from any further investigation of the corresponding amplitude in this study. 


\subsection{Discussion \label{discussions}} 
In this subsection, we briefly discuss our results in the context of phenomenological expectations and results from previous lattice calculations. Phenomenologically, the favorite candidates for bound doubly heavy tetraquarks are $bb\bar u\bar d$ and $bb\bar u\bar s$ with $I(J^P)=0(1^+)$ and $I(J^P)=1/2(1^+)$, and binding energy $\sim$100 MeV and $\sim$50 MeV, respectively. The lowest state in the charm counterpart of the latter configuration is expected to be $\gtrsim100$ MeV above the $DD^*_s$ threshold based on nonlattice investigations \cite{Karliner:2021wju,Eichten:2017ffp,Lu:2020rog,Deng:2021gnb,Praszalowicz:2022sqx}, whereas the phenomenological predictions for $cc\bar u\bar d$ are quite scattered and only a few agree with the experimentally observed mass. The $cc\bar u\bar d$ configuration has been studied extensively in the recent times on the lattice following its discovery by LHCb in 2021 \cite{LHCb:2021vvq,LHCb:2021auc}. However, there exists only two relatively old lattice QCD studies of the isodoublet $cc\bar u\bar s$ system \cite{Cheung:2017tnt,Junnarkar:2018twb}, wherein only the FV spectrum was determined in the rest frame.

In Ref. \cite{Junnarkar:2018twb}, the authors utilized local meson-meson and local diquark-antidiquark interpolators and observed that the ground state energy in $J^P=1^+$ channel is statistically consistent with the threshold after the continuum extrapolation. In Ref. \cite{Cheung:2017tnt}, the authors utilized large bases of interpolators that include local diquark-antidiquark interpolators as well as bilocal meson-meson interpolators to extract the low lying FV spectrum in the rest frame. The calculation  was performed in a single lattice QCD ensemble with $m_{\pi}\sim 400$ MeV. Despite the difference in the pion mass used, this study  finds a  pattern of small energy shifts (shown in Figure 7 of Ref. \cite{Cheung:2017tnt}) similar to our results. In particular, the positive energy shifts observed in the scalar channel $DD_s$ scattering is consistent with our observations. A mixture of both positive as well as (mild) negative energy splitting can be observed in the axialvector channel, consistent with our observations. In our work, we made several steps beyond \cite{Cheung:2017tnt} by investigating these channels using two volumes and using moving frame irreps in extracting the near-threshold $S$-wave amplitudes. We observe that these extracted amplitudes do not carry any signs of near-threshold physical hadronic poles in the isodoublet $cc\bar u\bar s$ system. This is consistent with conclusions based on the FV energies extracted in Refs. \cite{Cheung:2017tnt,Junnarkar:2018twb} and also with various phenomenological expectations \cite{Eichten:2017ffp,Karliner:2021wju,Lu:2020rog,Deng:2021gnb,Praszalowicz:2022sqx,Dai:2021vgf,Agaev:2022vhq}.

\section{Summary and conclusions}
\label{sec:conclusions}

We present the first lattice determination of scattering amplitudes of relevant two-meson channels in search for doubly charm strange tetraquarks with flavor content $cc\bar u\bar s$. The focus is given for the $S$-wave amplitudes in the elastic $DD_s$ scattering and coupled $DD_s^*$-$D^*D_s$ scattering leading to scalar ($J^P=0^+$) and axialvector ($1^+$) quantum numbers. The simulated finite-volume (FV) energy spectrum reveals small nonzero energy shifts from the corresponding noninteracting expectations, suggesting nontrivial interactions between the mesons involved. An amplitude analysis of the extracted FV eigenenergies suggests no nontrivial pole features in the physical amplitudes that could be associated with an exotic hadron in the energy region constrained and immediately below the threshold. A conservative upper bound on the energy regions constrained are about 160 MeV (80 MeV) above the $DD_s$ ($D^*D_s$) threshold in the scalar and axialvector channels, respectively. In these energy regions, our observations are consistent with most of the phenomenological expectations in these channels and with the existing lattice results. Followup lattice investigations are required to address the unattended systematic uncertainties, such as cutoff effects, and affirm our findings. 

This study is performed on two lattice QCD ensembles at a single scale ($a\sim0.086$ fm) and with $N_f=2+1$ dynamical Wilson-clover fermions generated by the CLS consortium. The light and strange mass corresponds to $m_{\pi}\sim280$ MeV and $m_{K}\sim467$ MeV, whereas the charm quark mass corresponds to a value of the spin averaged 1S charmonium that is slightly larger than its physical value. The correlation measurements are made for basis composed of purely bilocal two-meson interpolators and are performed following the {\it distillation} procedure. Further details of the setup can be found in Section \ref{sec:setup}.

The FV eigenenergies related to the elastic $DD_s$ scattering suggests repulsive interactions between the $D$ and $D_s$ mesons in $S$-wave. We constrain the $S$-wave amplitude following pure $S$-wave fits as well as combined $S$ and $P$-wave to accommodate potential $P$-wave effects on the FV eigenenergies. In Figures \ref{fig:scalar_chosen_spectrum}, \ref{fig:scalar_chosen_pcotdel} and Table \ref{tab:Elastic_fits_final}, we present our main results, which are based on amplitude fits following the solutions of the Lipmann Schwinger equation in \eqn{LSEfv}. The extracted $S$-wave amplitude indicates no hadronic poles in the vicinity of the $DD_s$ threshold. Amplitude fits following L\"uscher's prescription indicates either shallow bound poles or acausal off-axis physical sheet poles in this sector, which we believe is a result of lack of constraints on the analytic properties of the constrained amplitude parameterizations used. We observe that the extracted $S$-wave amplitude for $DD_s$ scattering is unperturbed by the $P$-wave contributions to the FV irreps utilized.  

The amplitudes extracted from the energy shifts in the coupled channel $DD^*_s$-$D^*D_s$ system suggest a nearly decoupled system of $DD^*_s$ and $D^*D_s$ scattering in $S$-wave. The near-threshold eigenenergies with small negative energy shifts overlap dominantly with the $DD_s^*$ channel, suggesting a rather weakly attractive interaction. On the other hand, the levels with positive energy shifts overlap dominantly with the $D^*D_s$ channel, suggesting  a repulsive nature of interaction. Our main results are based on L\"uscher-based amplitude fits assuming an effective range expansion approach and are presented in Figures \ref{fig:avector_chosen_spectrum}, \ref{fig:Cross_section_inelastic} and in Eq. (\ref{eqn:ERE_spwave_inelastic_chosen}). The $S$-wave amplitudes do not indicate presence of any hadronic pole in the energy region constrained that would lead to any nontrivial features in the experimental cross section. In this case, we find the lattice extracted amplitudes based on different parameterizations point to qualitatively similar conclusions. Additionally we ensure the amplitudes faithfully reproduce the FV spectra, even after accounting for the $P$-wave contributions to the moving frame FV irreps. 

We have also investigated the origin of the observed energy shifts in the coupled channel $DD^*_s$-$D^*D_s$ system by artificially varying the strength ($\alpha$) of the evaluated cross-correlation data. The real simulated data ($\alpha=1$) have energy splittings that are very small, suggesting very weak interactions as well as weak coupling between the scattering channels, and is nearly consistent with decoupled scenario observed from the subsequent amplitude analysis. However, the correlation matrices with artificially enhanced cross-correlations suggest increase in the coupled channel effects. See Figure \ref{fig:coupling_dependence} and the discussion around it for details. We observe that the resultant energy shifts are more enhanced with increasing strength, suggesting the role of cross correlator data in rendering nonzero energy shifts. This is further supported by the observation of  energy shifts statistically consistent with zero in the variational analysis of the correlator data assuming a decoupled system of $DD^*_s$ and $D^*D_s$ scattering. Although the enhanced energy shifts in artificially strengthened cross correlator data suggest stronger coupled channel effects, a followup amplitude analysis of these artificial data leads to unstable fits and also introduces additional conceptual difficulties, with increasing negative energy shifts approaching the \lhc ~branch point. Future simulations at different quark masses would be beneficial in shedding more light on the interactions involved in this two-channel system.

Note that, our work is the first investigation of $D^{(*)}D^{(*)}_s$ scattering amplitudes. It is exploratory in nature and an investigation of the systematic uncertainties is not the main focus. While we have utilized bilocal meson-meson like interpolators in the basis, local diquark-antidiquark interpolators are not utilized in this entire work. The study on effects of these local interpolators in Ref. \cite{Cheung:2017tnt} and our own observations on similar systems \cite{Prelovsek:2025vbr} suggests nearly negligible effects on the low lying spectrum. All these finite-volume studies are performed at heavier-than-physical pion masses (and lighter-than-physical $K$-meson masses in our studies). Hence, it would be valuable to further investigate these effects on lighter pion masses to access the relevant dynamics that might dominate in the chiral limit. One of the most important systematics associated with charm physics are related to discretization effects. In Ref. \cite{Green:2021qol}, the significance of addressing the discretization effects in arriving at conservative estimates for the binding energy of a near-threshold hadron has been demonstrated. Our results and conclusions call for follow up studies of this system on multiple spatial volumes and with different lattice spacings to further constrain the amplitudes to reaffirm our findings. 

The scientific interest in doubly heavy tetraquark systems have spiked in the recent years, particularly triggered by the discovery of the $T_{cc}$ tetraquark in 2021. Despite a number of lattice studies in the isoscalar doubly bottom and doubly charm tetraquarks, investigations on the doubly charm strange system with valence content $cc\bar u\bar s$ have been limited \cite{Cheung:2017tnt,Junnarkar:2018twb} and did not go beyond the determination of the low lying FV spectra. While phenomenological expectations generally suggest the existence of a bound $T_{cc\bar u\bar s}$ state to be unlikely, first principles approaches such as ours using lattice QCD are crucial in identifying (or ruling out the existence of) hadronic pole features and confirming the predictions. To this end, this study makes an important step ahead and calls for more rigorous determinations in the future. 

\begin{acknowledgments}

We thank Navdeep Singh Dhindsa, Feng-Kun Guo, Eulogio Oset, Lu Meng, Raquel Molina, Alexey Nefediev, Bhabani Sankar Tripathy, and Ivan Vujmilovic for invaluable discussions. S.P. is supported by the Slovenian Research Agency (research core Funding No. P1-0035). M.P. gratefully acknowledges support from the Department of Science and Technology, India, SERB Start-up Research Grant No. SRG/2023/001235 and Department of Atomic Energy, India. P.J. acknowledges the support and hospitality provided by IISER Kolkata and IMSc Chennai during the completion of this work. We thank our colleagues in CLS for the joint effort in the generation of the gauge field ensembles which form a basis for the computation. We use the multigrid solver of Refs.~\cite{Heybrock:2014iga,Heybrock:2015kpy,Richtmann:2016kcq,Georg:2017diz} for the inversion of the Dirac operator. Our code implementing distillation is written within the framework of the Chroma software package \cite{Edwards:2004sx}. The simulations has been performed on the Regensburg Athene2 cluster. We thank the authors of Ref. \cite{Morningstar:2017spu} for making the~{\it TwoHadronsInBox} package public.  We acknowledge the use of computing clusters at IMSc Chennai. We also thank the HPC RIVR consortium (www.hpc-rivr.si) and EuroHPC JU (eurohpc-ju.europa.eu) for funding this research by providing computing resources of the HPC system Vega at the Institute of Information Science (www.izum.si). 

\end{acknowledgments}

\appendix

\section{parameterizations used in $S$-wave $DD_s$ scattering \label{app:elastic_fits}}

In this appendix, we present different functional forms utilized to parametrize the $S$-wave amplitude in $DD_s$ scattering. We present the parameterizations utilized within the L\"uscher's framework as well as the corresponding best fit estimates in dimensionless units built out of the energy of the $DD_s$ threshold leading to the fitting variables $\hat{k} = k/E_{DD_s}$ and $\hat{s}=s/E^2_{DD_s}$. With this convention, the parameterizations we utilize are as follows. 
\begin{align}
 \hat{k}~cot\delta_0 &= A_0+B_0\hat{k}^2+C_0\hat{k}^4 ~& [\mbox{EREk}^4], \label{EREp}\\
 \hat{k}~cot\delta_0 &= A_0+B_0\hat{s}+C_0\hat{s}^2 ~& [\mbox{sPol}], \label{sPoly}\\
 \hat{k}~cot\delta_0 &= A_0+B_0\hat{\Delta}+C_0\hat{\Delta}^2~& [\mbox{ERE}\Delta].\label{EREDel} \\
 \hat{k}~cot\delta_0 &= A_0\pm B_0\sqrt{\frac{\hat{k}^2-A'_0}{B'_0}+C'_0} ~& [\mbox{Hypb}], \label{Hypb} \\
 \frac1{\hat{k}~cot\delta_0} &= \frac{G^2}{m^2-\hat{s}}+\gamma~& [\mbox{BWs}].\label{BWs}
\end{align}
Here $\hat{\Delta}=\hat{s}-1$ is an alternate variable for an equivalent ERE, also used in Ref. \cite{BaryonScatteringBaSc:2023zvt,BaryonScatteringBaSc:2023ori}. The strings inside square brackets are names we use to refer to the generic form of the parameterizations. 

The summary of best fit parameters is given in table \ref{tab:Elastic_fits}. The generic form of the parametrization is indicated in the first column, whereas the second column identifies the fit with a label. We present the parameter values in dimensionless units in terms of $E_{DD_s}$ with errors quoted within the brackets estimated following bootstrap procedure and the corresponding $\chi^2/d.o.f$ values in last column. The parameters of the LSE potential are indicated as $c_q[l]$, where $q$ is the co-efficient index and $l$ is the partial wave, and the respective best fit values are presented in physical units. The irrelevant parameters for any given fit are indicated by a hyphen. 

We have demonstrated the consistency in the reproduced energy dependence by different parameterizations in Figure \ref{fig:scalar_pcotdel}, however comparing the respective best fit parameter values themselves across different functional forms is not justified. Yet one may investigate the variation or stability in best fit parameter values appearing in the leading terms in expansion based functional forms used, with varying number of terms used to parametrize the energy dependence. In the effective range expansion forms ($\text{EREk}^2$ and ERE$\Delta$), we observe that the leading $k^2$ independent term related to the $S$-wave scattering length remains intact with inclusion of the subleading terms in the fits. 

\begin{table*}[tbh!]
    \centering
\begin{tblr}{cccccccc}
\hline\hline
 & $A_0$ & $B_0$ & $C_0$ & $A_1$ & $B_1$ & - & $\chi^2/N_{dof}$\\\cline{2-8}
EREk$^{(0,2,4)}$ & $-0.25(^{+4}_{-9})$ & $4(^{+4}_{-2})$   &$-1077(^{+286}_{-864})$& 0 & $0.8(^{+8}_{-3})$ & - & 36/20\\\cline[dashed]{2-8}
     & $-0.27(^{+6}_{-2})$ & $3(^{+1}_{-21})$  &$-1004(^{+1151}_{-215})$& $0.0138(^{+15}_{-13})$ & - & - & 27.8/20\\\cline[dashed]{2-8}
     & $-0.25(^{+4}_{-7})$ & $4(^{+15}_{-9})$  & $-1161(^{+675}_{-1288})$ & - & - & -  & 16/11 \\\cline[dashed]{2-8}
     & $-0.22(^{+3}_{-4})$ & $-16(^{+3}_{-4})$ & - & - & - & -  &15.6/12\\
\hline\hline
 & $A_0$ & $B_0$ & $C_0$ & $A_1$ & $B_1$ & - & $\chi^2/N_{dof}$\\\cline{2-8}
ERE$\Delta$ & $-0.20(^{+2}_{-4})$ & $-2.73(^{+4}_{-1.84})$ & - & $0.016(^{+2}_{-4})$ & - & - & 29.7/21\\\cline[dashed]{2-8}
            & $-0.25(^{+4}_{-7})$ & $1(^{+4}_{-2})$ & $-73(^{+43}_{-79})$ & - & - & - & 16/11\\\cline[dashed]{2-8}
            & $-0.22(^{+3}_{-4})$  & $-4(^{+1}_{-1})$ & - & - & - & - & 15.6/12\\  
\hline\hline
 & $A_0$ & $B_0$ & $C_0$ & $A_1$ & $B_1$ & - & $\chi^2/N_{dof}$\\\cline{2-8}
sPol & $4(^{+2}_{-1})$ & $-4(^{+1}_{-2})$  & - & $0.014(^{+3}_{-9})$ & - & - & 26/21\\\cline[dashed]{2-8}
     & $4(^{+1}_{-1})$     & $-4(^{+1}_{-1})$       & - & - & - & - & 15.6/12\\ 
\hline\hline
 & $A_0$ & $B_0$ & $C_0$ & $A'_0$ & $B'_0$ &  & \\
 \cline{2-8}
 Hypb
      & $0.0759(^{+1693}_{-15})$ & $-0.0084(^{+8}_{-105})$ &$1117(^{+865}_{-726})$& $0.003(^{+1}_{-2})$ & $-0.00023(^{+5}_{-26})$  & - &15.4/9\\
\hline\hline
& $m$[S] & $G$[S] & $\gamma$[S] & $m$[P] & $G$[P] & $\gamma$[P] &\\
\cline{2-8}
BWs & $0.3(^{+6}_{-3})$ & $5(^{+1}_{-4})$  & $26(^{+5}_{-22})$ & - & - & - &17.8/11\\\cline[dashed]{2-8}
    & $0.32(^{+1}_{-1})$ & $7.7(^{+1}_{-4})$  & $-5.21(^{+6}_{-9})$ & $88(^{+5}_{-7})$ & $26.1(^{+9}_{-2})$ & $-68(^{+2}_{-2})$ &29/19\\
\hline\hline
& $c_0$[S] & $c_2$[S] & $c_0$[P] & $c_2$[P] &   &  &\\
& $[$GeV$]^{-2}$ & $[$GeV$]^{-4}$ & $[$GeV$]^{-4}$ & $[$GeV$]^{-6}$ &   &  &\\\cline{2-8}
LSE & $4.0(^{+5}_{-5})$ & $-4.4(^{+8}_{-8})$  & $-12.4(^{+6.2}_{-4.8})$ & $3.3(^{+6.4}_{-8.5})$ & - & - &31.4/18\\\cline[dashed]{2-8}
$\Lambda = 0.65$ GeV & $4.0(^{+5}_{-5})$ & $-4.4(^{+8}_{-8})$  & $-10.3(^{+1.5}_{-1.6})$ & - & - & - &31.4/19\\\cline[dashed]{2-8}
    & $4.2(^{+5}_{-5})$ & $-4.4(^{+8}_{-8})$  & -  & - & - & - &11.6/8\\\hline\hline
    & $c_0$[S] & $c_2$[S] & $c_0$[P] & $c_2$[P] &   &  &\\
    & $[$GeV$]^{-2}$ & $[$GeV$]^{-4}$ & $[$GeV$]^{-4}$ & $[$GeV$]^{-6}$ &   &  &\\\cline{2-8}
LSE & $3.6(^{+5}_{-5})$ & $-3.2(^{+6}_{-6})$  & $-12(^{+5}_{-4})$ & $5(^{+5}_{-5})$ & - & - &36.7/20\\ \cline[dashed]{2-8}
$\Lambda = 0.9$ GeV & $3.7(^{+5}_{-5})$ & $-3.2(^{+6}_{-6})$  & $-8.3(^{+1.0}_{-1.1})$ & - & - & - &37.3/21\\\cline[dashed]{2-8}
    & $4.1(^{+6}_{-3})$ & $-3.7(^{+5}_{-1.4})$  & -  & - & - & - &21.5/12\\\hline\hline
    \end{tblr}
    \caption{Summary of best fit parameter values for different parameterizations used for elastic $DD_s$ scattering in $S$-wave. The first column indicates the name of the generic parametrization used as indicated in Eqs. (\ref{EREp})-(\ref{BWs}). In the right most column, we present the $\chi^2/dof$ for different fits listed, whereas the columns in between are the best fit parameter values, all presented in units of $E_{DD_s}$. The power in EREk$^{(0,2,4)}$ refers to the highest power of $k$ included in the parametrization. The best fit parameter values for the LSE potentials are presented in physical units. See the discussion in Appendices \ref{app:elastic_fits} and \ref{app:elasticP} for details. The solid double horizontal lines separate results from different fit forms and procedures, whereas the dashed horizontal lines separate different fit results within the same generic fit form or procedure.}
    \label{tab:Elastic_fits}
\end{table*}

\section{$DD_s$ scattering in $P$-wave using L\"uscher-based fits \label{app:elasticP}}

\begin{figure}[h]
    \centering
    \includegraphics[width=1.0\linewidth]{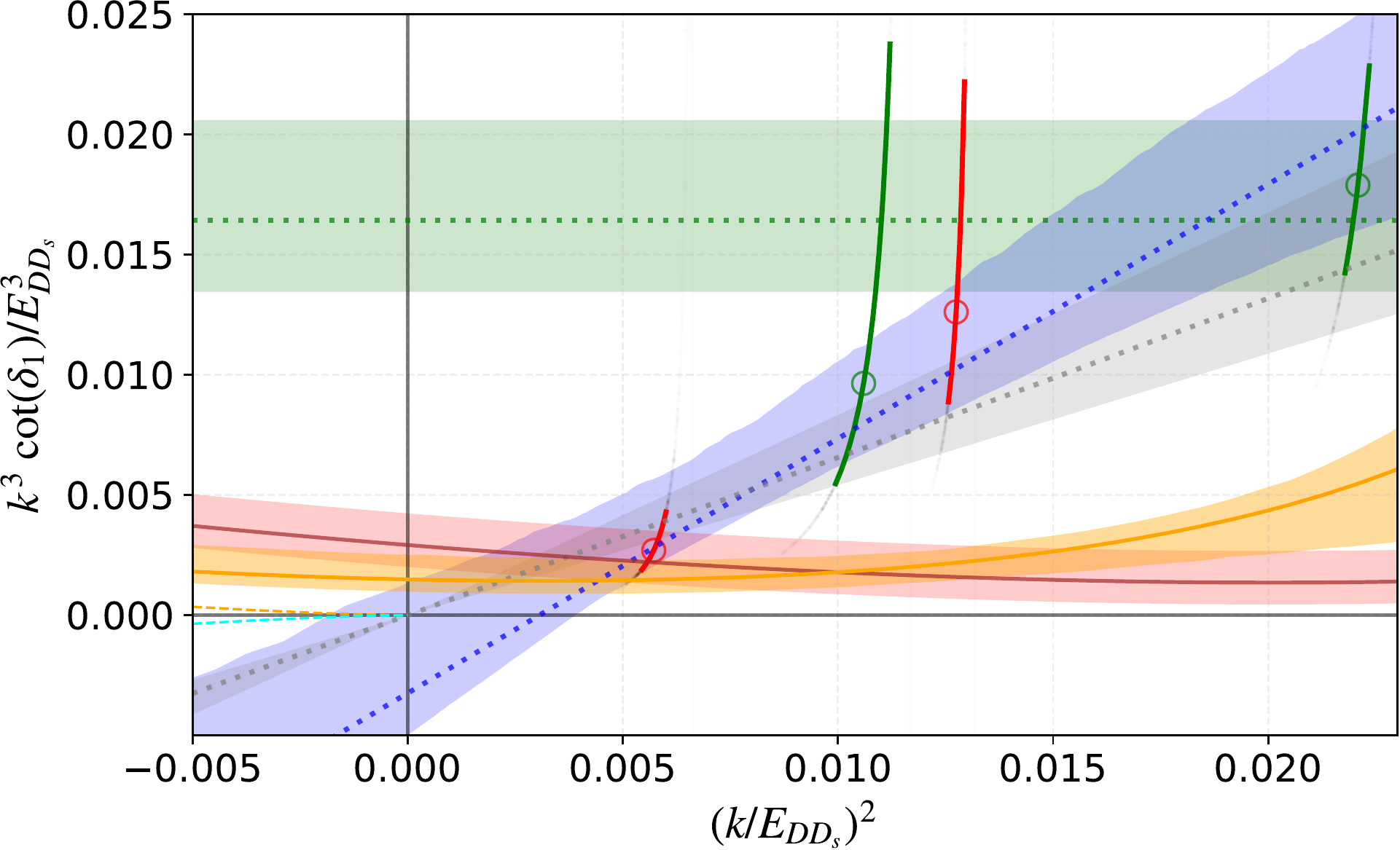}
    \caption{The momentum-squared dependence of the $P$-wave amplitude in $DD_s$ scattering.  The circle markers are obtained via L\"uscher's formalism from irreps where $l=1$ is the lowest contributing partial wave. The solid and dotted curves with bands represent fits with the LSE~(orange and red) and L\"uscher's~(green) approaches, respectively. The green band refers to the fit with constant parametrization, whereas the blue and gray band indicate fit results with a linear form in $k^2$ with a zero inverse scattering length assumed in the latter.}
    \label{fig:Pwave_amplitude_app}
\end{figure}

In this appendix, we discuss fits to elastic $DD_s$ scattering in $P$-wave based on L\"uscher's FV prescription. The three dotted curves and associated bands in Figure \ref{fig:Pwave_amplitude_app} are $P$-wave amplitudes extracted using L\"uscher's framework. It is evident from the circles that the level nearest to the threshold is also closest to the real axis and the momentum dependence of the amplitude suggests possibility of an $x$-axis crossing above threshold with a positive slope suggesting a pole in the physical sheet away from real axis. Those parameterizations that suggest such a crossing violate causality leading to physical Riemann sheet poles away from the real energy axis \cite{osti_4169822,Briceno:2017max,Briceno:2017qmb}. None of the FV eigenenergies at hand unfortunately constrain the $P$-wave amplitudes below this level and hence any parametrization  beyond a constant in $k^2$ can introduce a positive slope in the energy dependence leading to an inevitable pole in the physical sheet away from the real axis. This means any causal $P$-wave amplitude description goes beyond a constant in $k^2$ parametrization for $\tilde{K}_l^{-1}$ and would require more lattice data involving large spatial volume allowing access to near-threshold behavior of the $P$-wave amplitude. An example of fitted $k^3\cot{\delta_1}$ with a constant in $k^2$ (green) and a linear form in $k^2$ (blue band) is presented in the Figure \ref{fig:Pwave_amplitude}. The constant $\hat{k}^3\cot{\delta_1}=A_1$ or a zero-range approximation can be seen to miss the low energy amplitude leading to low quality fits, whereas the linear form can be clearly seen to cross the $x$-axis that will lead to a off-axis pole in the physical sheet, above the threshold. An ERE parametrization that does not lead to acausal poles is the one with infinite scattering length 
\beqa
\hat{k}^3\cot{\delta_1}&=B_{1}\hat{k}^2. \label{Pforma}
\eeqa
shown by gray band in Figure \ref{fig:Pwave_amplitude}, where the real-axis crossing is pushed to the threshold, as evident from the gray band in Figure \ref{fig:Pwave_amplitude}. 

We iterate that the $S$-wave amplitudes we extract for the $DD_s$ scattering remains robust irrespective of the $P$-wave parameterizations one utilize. Our main result for the $P$-wave, which based on solutions of LSE in \eqn{LSEfv}, is also free of any shallow bound states or near threshold resonance poles. 

\section{Parameterizations used in \texorpdfstring{$DD_s^*$-$D^*D_s$} scattering \label{app:inelastic_fits}}

In this Appendix, we present the summary of results and parameterizations used in the case of coupled channel $DD_s^*$-$D^*D_s$ scattering in $S$-wave. First, we present the parameterizations we have used for the amplitude fits based on L\"uscher's formalism. Similar to the ERE form presented in Eq. (\ref{ERE_inelastic}), we utilize an alternative form defined in terms of $\Delta_i=(s-E_{i,c}^2)/E_{i,c}^2$ \cite{BaryonScatteringBaSc:2023zvt,BaryonScatteringBaSc:2023ori} is given by 
\begin{align}
\tilde{K}^{-1}=\left[
\begin{array}{cc}
a_{11}+b_{11}\Delta_1 & a_{12}+b_{12}\Delta_1+c_{12}\Delta_2 \\
                      & a_{22}+b_{22}\Delta_2
\end{array}
\right].
\end{align}
Another parametrization used assumes polynomials in Mandelstam $s=E_{cm}^2$ to map the energy dependence as  
\begin{align}
\tilde{K}^{-1}=\left[
\begin{array}{cc}
a_{11}+b_{11}s & a_{12}+b_{12}s \\
                      & a_{22}+b_{22}s
\end{array}
\right].
\end{align}
For all the $S$-wave parameterizations utilized in the amplitude fits following L\"uscher's prescription, we limit to a three parameter form of the $P$-wave amplitudes, where an energy independent form 
\begin{align}
\tilde{K}^{-1}=\left[
\begin{array}{cc}
a_{33} & a_{34} \\
       & a_{44}
\end{array}
\right]. \label{3paramPwave}
\end{align}
is assumed for the amplitude in each scattering channel. 

A fourth form that we consider in our amplitude fits based on L\"uscher's formalism is following the standard Breit-Wigner type parametrization, where we typically consider one or no pole separately in the $S$ and $P$ waves coupled to either or both scattering channels, together with a constant term ($\gamma_{ij}$) or a linear term in Mandelstam $s$. The general form used can be expressed as
\begin{align}
 \tilde{K}_{ij} &= 
 \begin{cases} 
  \frac{G_i^2}{m^2-s}+\gamma_{ii}s^{[0,1]} & \text{if } i = j, \\
  \frac{G_iG_j}{m^2-s}+\gamma_{ij}s^{[0,1]} & \text{if } i \neq j.
 \end{cases}
\label{BW_inelastic}
\end{align}
In the case of no pole the first term in the above equation is dropped out. 

We also constrain the $DD_s^*$-$D^*D_s$ amplitudes following the solutions of LSE, for which the analytical reconstruction of the FV spectrum was demonstrated in the top pane of Figure \ref{fig:avector_chosen_spectrum}. We utilize an extension of the parametrization for pseudoscalar-vector scattering presented in Ref. \cite{Meng:2021uhz} for the coupled channel system being addressed. The form of the $S$ and $P$-wave potentials are chosen as
\beqa
\mathbb{V}^{[S]}(\vek,\vek') &=& [2c_0[S]+2c_2[S](k^2+k'^2)]~(\vec{\epsilon}\cdot\vec{\epsilon'}^*) \label{VinelasticS_LSE} \mbox{ ~and} \nonumber\\
\mathbb{V}^{[P]}(\vek,\vek') &=& 2c_2[P]~(\vek\cdot\vec{\epsilon})(\vek'\cdot\vec{\epsilon'}^*). 
\eeqa
With the form above, we utilize a two-dimensional potential matrix in the channel space
\begin{align}
V=\left[
\begin{array}{cc}
 (\mathbb{V}_{11}^{[S]}+\mathbb{V}_{11}^{[P]})(\vek_1,\vek'_1) & (\mathbb{V}_{12}^{[S]}+\mathbb{V}_{12}^{[P]})(\vek_1,\vek'_2) \\
                    & (\mathbb{V}_{22}^{[S]}+\mathbb{V}_{22}^{[P]})(\vek_2,\vek'_2)
\label{LSE_inelastic}
\end{array}
\right]
\end{align}
with a total of 9 parameters, while solving for the LSE in \eqn{LSEfv}. Here the subscripts 1 and 2 to $\mathbb{V}$, $\vek$, and $\vek'$ refer to the channel index.

In Table \ref{tab:inElastic_fits}, we present a summary of fits to the $S$-wave parameterizations and in Table \ref{tab:ERE_spwave_inelastic}, the fits constraining the $S$ and $P$-wave combined are presented. In the first column we indicate the form of parametrization.  Similar to the elastic case we present the parameter values in dimensionless units in terms of $E_{DD_s}$ with errors quoted within the brackets estimated following bootstrap procedure and the corresponding $\chi^2/N_{dof}$ values in last column. The parameters in the LSE potential are indicated as $c_{q,ij}[l]$, where $q$ is the co-efficient index, $i$ and $j$ are the channel indices and $l$ is the partial wave, and the respective best fit values are presented in physical units. The irrelevant parameters for any given fit are indicated by a hyphen. 

Assuming a decoupled scenario, in Figure \ref{fig:inelastic_pcotdel}, we have demonstrated that the various parameterizations suggest consistent energy dependencies. It can be observed that the $P$-wave contributions can be accounted in the moving frame irreps without changing the $S$-wave amplitudes. The successful reconstruction of the FV spectra using the $S$ and $P$-wave combined fits support this conclusion. Similar to the observation in the elastic scenario, the leading terms in the ERE-based parameterizations remain intact with the inclusion of the subleading terms related to shape parameters. The range parameters are generally observed to be subdominant indicating nearly energy independent behavior in the respective elastic amplitudes. Such arguments on leading terms are not immediate in other parameterizations used, as the threshold has no special relevance in the Breit-Wigner form (BWs) and in polynomials of $s$ (sPol). 

\begin{table*}[tbh!]
    \centering
    \begin{tblr}{cccccccc}
\hline\hline
    & $a_{11}$ & $b_{11}$ & $a_{12}$ & $b_{12}$ & $a_{22}$ & $b_{22}$ & $\chi^2/dof$\\ \cline{2-8}
    & $0.202(^{+21}_{-18})$ & -       & $-0.00083(^{+10}_{-5})$ & $0.20(^{+1}_{-2})$ & $-0.15(^{+1}_{-1})$ & -      & 26/18\\\cline[dashed]{2-8}
 EREk$^{(0,2)}$     & $0.202(^{+21}_{-19})$ & $-0.0002(^{+0}_{-0})$  & $-0.0018(^{+1}_{-1})$  & -       &  $-0.15(^{+1}_{-2})$ & $-0.24(^{+4}_{-1})$ &25.7/17\\\cline[dashed]{2-8}
    & $0.202(^{+21}_{-19})$ & $0.0026(^{+1}_{-4})$  & $-0.0012(^{+1}_{-1})$  & $0.096(^{+16}_{-4})$ &  $-0.15(^{+1}_{-2})$ & $-0.24(^{+3}_{-2})$ & $\underline{25.7/16}$\\
\hline\hline
    & $a_{11}$ & $b_{11}$ & $a_{12}$ & $b_{12}$ & $a_{22}$ & $b_{22}$ & $\chi^2/dof$\\ \cline{2-8}
sPol & $0.175(^{+11}_{-57})$ & $0.011(^{+5}_{-37})$  & $-0.0283(^{+11}_{-57})$ &  - & $-0.151$     &     -   &26.8/18\\\cline[dashed]{2-8}
    & $0.15(^{+4}_{-2})$  &  -      & $-0.01(^{+7}_{-5})$ & $0.1(^{+0}_{-1})$  &$-0.10(^{+4}_{-3})$ & $-0.04(^{+2}_{-5})$&        37.4/17\\\cline[dashed]{2-8}
    & $0.14(^{+2}_{-12})$ & $0.048(^{+43}_{-5})$  & $-0.003(^{+1}_{-1})$  & -       & $-0.10(^{+3}_{-5})$ & $-0.05(^{+1}_{-1})$        &$\underline{26.4/16}$\\
\hline\hline
    & $a_{11}$ & $b_{11}$ & $a_{12}$ & $b_{12}$ & $a_{22}$ & $b_{22}$ & $\chi^2/dof$\\ \cline{2-8}
    & $0.18(^{+3}_{-2})$  & - &  $-0.0020(^{+3}_{-2})$ & $0.10(^{+2}_{-1})$  &  $-0.19(^{+2}_{-2})$ & - & 25/18\\\cline[dashed]{2-8}
ERE$\Delta$& $0.17(^{+11}_{-3})$ &$0.9(^{+7}_{-9})$ &  $-0.14(^{+6}_{-5})$ & - &  $-0.15(^{+2}_{-3})$ & $0.1165(^{+3.8}_{-2.4})$ & $\underline{28.3/17}$\\\cline[dashed]{2-8}
    & $0.17(^{+2}_{-3})$ & $0.042(^{+59}_{-4})$ & $0.0025(^{+4}_{-5})$ & $0.039(^{+4}_{-9})$ & $-0.14(^{+2}_{-3})$ & $-0.032(^{+7}_{-6})$ & 33.2/16\\
\hline\hline
     & $m$ & $G_1$ & $G_2$ & $\gamma_1$ & $\gamma_2$ & $\gamma_3$  &   $\chi^2/dof$ \\\cline{2-8}
BWs  & $4.8(^{+2}_{-2})$ & $-0.024(^{+1}_{-7})$ & $-7.3(^{+0.4}_{-1.4})$& $5.0(^{+3}_{-7})$ & $-3.1(^{+2.9}_{-0.1})$ & $-9.0(^{+5}_{-9})$& 28.1/16\\\cline[dashed]{2-8}
     & $5.2(^{+6}_{-5})$ & $-0.196(^{+5}_{-30})$ & $-5(^{+1}_{-3})$&$5(^{+3}_{-0})$& - & $-8(^{+0}_{-2})$     & 26.7/17\\\cline{2-8}     
     & $m$ & $G_1$ & $G_1$ & $\gamma_1s$ & $\gamma_2s$ & $\gamma_3s$  & $\chi^2/dof$\\\cline{2-8}
     & $2.8(^{+7}_{-2})$ & $-5(^{+0.3}_{-3})$ & $-6(^{+1}_{-3})$&$1.5(^{+0.02}_{-2.8})$& $-0.012(^{+3}_{-5})$& $-11(^{+2}_{-3})$     & 30.6/16\\
\hline\hline
     & $c_{0,11}[S]$        & $c_{2,11}[S]$        & $c_{0,12}[S]$        & $c_{2,12}[S]$        & $c_{0,22}[S]$        & $c_{2,22}[S]$        &\\    
LSE  & $[$GeV$^{-2}]$    &$[$GeV$^{-4}]$     & $[$GeV$^{-2}]$    & $[$GeV$^{-4}]$    & $[$GeV$^{-2}]$    & $[$GeV$^{-4}]$   & $\chi^2/dof$\\ \cline{2-8}
     & $-2.7(^{+0.5}_{-0.7})$ & $-0.73(^{+1.57}_{-0.66})$ & $0.0001(^{+0.580}_{-0.001})$ & - &$7.2(^{+0.9}_{-1.1})$ & $-14.7(^{+3.0}_{-1})$ &18.2/18\\\cline[dashed]{2-8}
     & $-2.7(^{+7}_{-7})$ & $-0.7(^{+1.6}_{-0.7})$ & $0.5(^{+7}_{-1})$ & $-0.7(^{+1}_{-7})$ & $ 7.1(^{+0.7}_{-1.6})$ &$-14(^{+4}_{-1})$ &\underline{22.7/20}\\
\hline\hline
    \end{tblr}
    \caption{Summary of best fit parameter values for different $S$-wave amplitude parameterizations used for inelastic $DD_s^*$-$D^*D_s$ scattering in $S$-wave. The table follows the same conventions as used in Table \ref{tab:Elastic_fits}. The best fit parameter values from the L\"uscher-based fits are presented in dimensionless units with respect to $E_{DD^*_s}$. The power in EREk$^{(0,2)}$ refers to the highest power of $k$ included in the parametrization. The parameters in the LSE potentials are presented in the form $c_{q,ij}[l]$, where the indices $q$ refers to the co-efficient index in Eq. (\ref{VinelasticS_LSE}), $i$ and $j$ are channel indices, and $l$ refers to the partial wave. The LSE potential parameters are presented in physical units. See the discussion in Appendix \ref{app:inelastic_fits} for details. }
    \label{tab:inElastic_fits}
\end{table*}

\begin{table*}
    \centering
    \begin{tblr}{ccccc}
    \hline\hline
             & $a_{33}$             & $a_{34}$              & $a_{44}$             & $\chi^2/dof$\\\cline{2-5}
 EREk$^2$       & $0.0025(^{+4}_{-3})$ & $0.0001(^{+40}_{-1})$ & $-0.14(^{+1}_{-9})$  & 16/19\\\hline \hline 
             & $a_{33}$             & $a_{34}$              & $a_{44}$             & $\chi^2/dof$\\\cline{2-5}
 sPol        & $0.0025(^{+3}_{-5})$ & $0.032(^{+1}_{-15})$  & $-0.06(^{+1}_{-12})$ & 20.2/20\\\hline\hline
             & $a_{33}$             & $a_{34}$              & $a_{44}$             & $\chi^2/dof$\\\cline{2-5}
 ERE$\Delta$ & $0.0024(^{+4}_{-2})$ & $-0.0001(^{+2}_{-4})$ & $-0.0032(^{+2}_{-2})$& 28.9/20\\\hline\hline
             & $c_{2,11}$[P]          & $c_{2,12}$[P]           & $c_{2,22}$[P]          & $\chi^2/dof$\\
             & $[$GeV$]^{-4}$       & $[$GeV$]^{-4}$        & $[$GeV$]^{-4}$       &        \\\cline{2-5}
 LSE         & $-135(^{+21}_{-21})$ & $-2(^{+19}_{-14})$    & $56(^{+39}_{-35})$   & 47.6/19\\
 \hline\hline
    \end{tblr}
    \caption{Summary of best fit parameter values for a selection of $P$-wave amplitudes in the inelastic $DD_s^*$-$D^*D_s$ system that leads to a reasonable $\chi^2$ value in the respective fitting procedures. The $P$-wave parameters are fitted for the fixed $S$-wave parameter values taken from Table \ref{tab:inElastic_fits}. The chosen $S$-wave solutions are indicated by an underlined $\chi^2/dof$ in Table \ref{tab:inElastic_fits}. The table follows the same conventions as used in Table \ref{tab:inElastic_fits}.}
    \label{tab:ERE_spwave_inelastic}
\end{table*}

\bibliographystyle{apsrev4-1}
\bibliography{Tccus.bib}

\end{document}